\documentclass[usenatbib]{mnras}

\usepackage{newtxtext,newtxmath}

\usepackage[T1]{fontenc}

\DeclareRobustCommand{\VAN}[3]{#2}
\let\VANthebibliography\thebibliography
\def\thebibliography{\DeclareRobustCommand{\VAN}[3]{##3}\VANthebibliography}

\usepackage{graphicx}	
\usepackage{amsmath}	
\usepackage[version=4]{mhchem}
\usepackage[utf8]{inputenc}
\DeclareUnicodeCharacter{2212}{-}

\title[Non-Grey Hot Jupiter GCM Grid]{Hot Jupiter Diversity and the Onset of TiO/VO Revealed by a Large Grid of Non-Grey Global Circulation Models}

\author[A. Roth et al.]{
Alexander Roth$^{1}$\thanks{E-mail: Alexander.roth@physics.ox.ac.uk},
Vivien Parmentier$^{1,2}$ and
Mark Hammond$^{1}$
\\
$^{1}$Department of Atmospheric, Oceanic and Planetary Physics, University of Oxford, Oxford, OX1 3PU, UK\\
$^{2}$ Université Côte d’Azur, Observatoire de la Côte d’Azur, CNRS, Laboratoire Lagrange, France\\
}

\date{Accepted XXX. Received YYY; in original form ZZZ}

\pubyear{2024}

\begin{document}
\label{firstpage}
\pagerange{\pageref{firstpage}--\pageref{lastpage}}
\maketitle

\begin{abstract}
The population of hot Jupiters is extremely diverse, with large variations in their irradiation, period, gravity and chemical composition. To understand the intrinsic planet diversity through the observed population level trends, we explore the a-priori scatter in the population created by the different responses of atmospheric circulation to planetary parameters. We use the SPARC/MITgcm 3D global circulation model to simulate 345 planets spanning a wide range of instellation, metallicity, gravity and rotation periods typical for hot Jupiters, while differentiating between models with and without TiO/VO in their atmosphere. We show that the combined effect of the planetary parameters leads to a large diversity in the ability of atmospheres to transport heat from day-side to night-side at a given equilibrium temperature. We further show that the hot-spot offset is a non-monotonic function of planetary rotation period and explain our findings by a competition between the rotational and divergent parts of the circulation. As a consequence, hot-spot offset and phase curve amplitude are not necessarily correlated. Finally, we compare the observables from our grid to the population of Spitzer and Hubble observations of hot Jupiters. We find that the sudden jump in brightness temperature observed in the Spitzer secondary eclipse measurements can be naturally explained by the cold-trapping of TiO/VO at approximately 1800K. The grid of modelled spectra, phase curves and thermal structures are made available to the community, together with a python code for visualization of the grid properties, at \url{https://zenodo.org/doi/10.5281/zenodo.10785320} and \url{https://3dsim.oca.eu/hot-jupiters-3d-models}.
\end{abstract}



\section{Introduction}
As the number of high-quality exoplanet observation increases, with the James Webb Space telescope \citep{2016ApJ...817...17G} and upcoming Ariel mission \citep{2017EPSC...11..713T} providing wide wavelength coverage and ground-based telescopes supplying high resolution data \citep{2020Natur.580..597E}, hot Jupiters will continue to provide the best targets for atmospheric characterization. These planets are primarily classified based on their short-period orbits and high temperatures, meaning they offer optimal targets for transit observations. Many have been observed during secondary eclipse \citep{2020AJ....159..137G, 2020A&A...639A..36B,2023arXiv230103639D}, with a smaller number viewed over full orbits to provide phase curves \citep{2018haex.bookE.116P}. Strong gravitational interactions, due to the short-period orbits, are thought to cause tidal locking \citep{1996ApJ...470.1187R,1997ApJ...484..866L,2002A&A...385..166S}, leading to rotation periods synchronised with orbital periods. As all the incident stellar irradiation is deposited into the day-side hemisphere, steep day-night temperature contrasts form, making these planets intrinsically 3-dimensional in nature and driving strong winds \citep{2002A&A...385..166S}. These winds facilitate energy transport from the hot day-side onto the cold night-side \citep{2016ApJ...821...16K,2018RNAAS...2...36K}, in the form of either a super-rotating equatorial jet which can reach speeds of up to several km/s or direct day-to-night divergent flow \citet{2021PNAS..11822705H}. The winds and day-to-night flow have been measured in these atmospheres \citep{2010Natur.465.1049S,2016ApJ...817..106B}. The jets, first theorized by modelling \citep{2002A&A...385..166S,2005ApJ...629L..45C,2008ApJ...682..559S,2009ApJ...700..887M,2011MNRAS.418.2669H,2012ApJ...751...59P,2012ApJ...750...96R,2014A&A...561A...1M,2015ApJ...801...95S,2013MNRAS.435.3159D,2019ESS.....432603T}, have also been inferred through observation \citep{2007Natur.447..183K}. However, these strong winds are often not strong enough to transport energy from day-to-night before it is radiated back to space, leading to large horizontal contrasts in temperature \citep{2007Natur.447..183K,2010Natur.464.1161S}, chemistry \citep{2018A&A...617A.110P,2019A&A...631A..79H} and cloud coverage \citep{2021MNRAS.501...78P,2019A&A...631A..79H}.

The formation pathways responsible for the observed exoplanet population is one of the largest remaining unanswered questions. Planets form from the local material in the stellar disk. The location of formation, or subsequent migration after formation has occurred, could affect the atmospheres of giant planets \citep{2022A&A...665A.138B,2013ApJ...775...80F}. Consequently, measuring chemical abundances in planetary atmospheres can inform us on their formation and migration history \citep{2019A&A...627A.127C,2011ApJ...743L..16O,2014ApJ...794L..12M}. 
At the population level, however, it is not only the trend of chemical abundances with parameters, but also their scatter around the mean population value, that can inform us on the planetary formation mechanisms and their stochastic nature \citep{2013ApJ...775...80F}. However, abundance measurements in exoplanet atmospheres have to rely on a thorough understanding of their 3D thermo-chemical structure to avoid bias in atmospheric retrievals \citep{2018ApJ...855L..30A,2020MNRAS.493.4342T}. Particularly, at the population level, we wish to separate the part of the observed trends and scatter that is due to the diversity of the planetary parameters (e.g. temperature or rotation period) and the one that is due to intrinsic, formation related, parameters (such as metallicity).

\begin{figure*}
	\includegraphics[width=2\columnwidth]{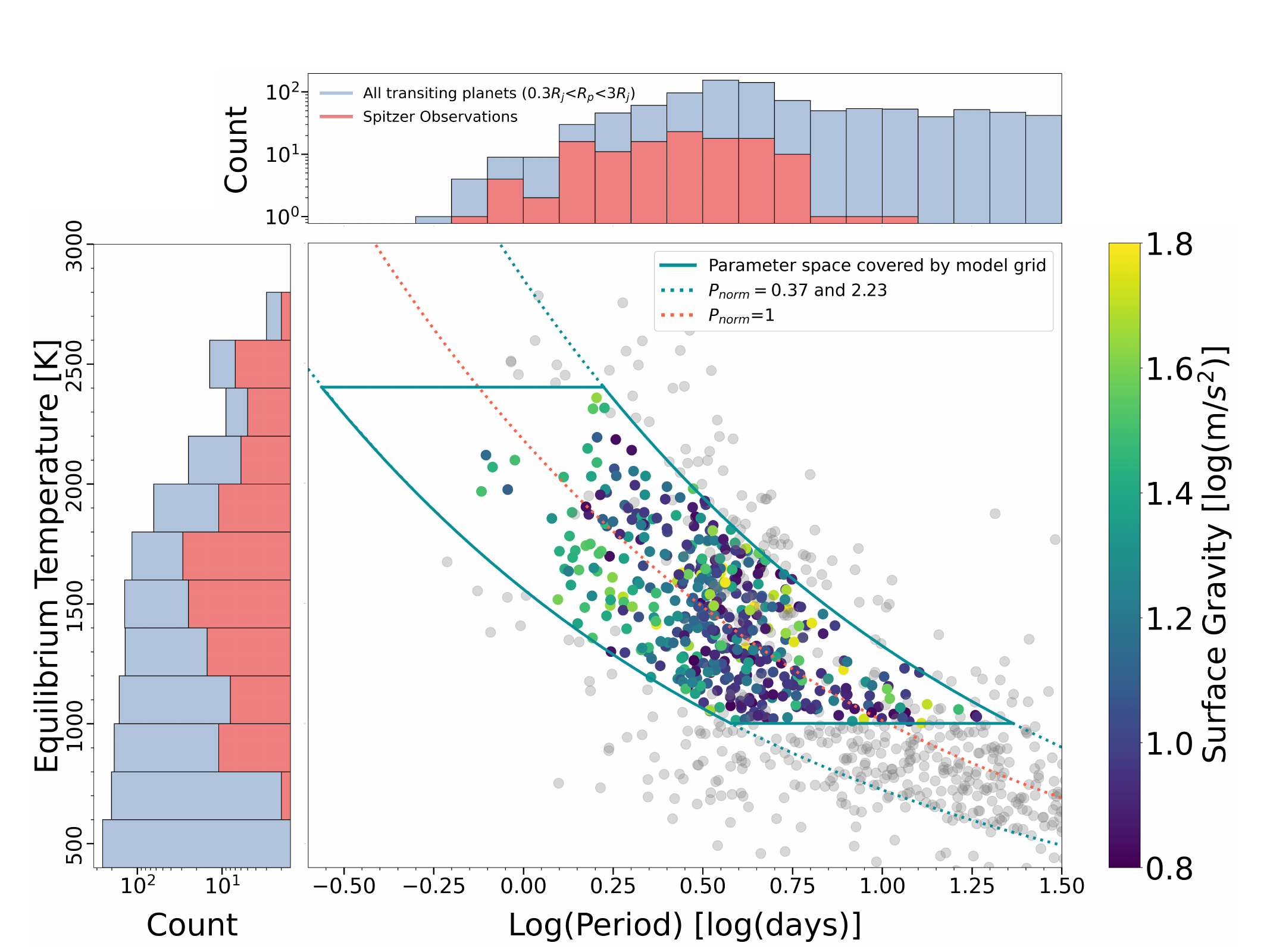}
    \caption{Distribution of known exoplanets (0.3<$R_{p}$<3$R_{j}$) with equilibrium temperature vs orbital period. Planets that fall within our grid boundaries are color-coded by their gravity. Planets that fall outside our grid boundaries are left grey. The blue dashed lines on the scatter plot represent the minimum and maximum values of constant period multiplicity for our model grid, as defined in Section \ref{sec:grid}. The outlined solid blue shape shows the area that is covered by our model grid in the equilibrium temperature vs log(period) space. The histogram to the left of the scatter plot displays the distribution of planets in the equilibrium temperature space (blue bars) when compared to the distribution of the measured Spitzer population from \citet{2023arXiv230103639D} (red bars). The histogram above the scatter plot shows the same population level comparison for the period distribution.}
    \label{fig:param_space}
\end{figure*}

The population of hot Jupiters is extremely diverse. As shown in Figure \ref{fig:param_space}, their equilibrium temperatures vary from $\approx$1000 to 2400K, which corresponds to irradiation varying by more than an order of magnitude. Cooler planets are often called "warm Jupiters" whereas hotter ones are called "ultra-hot Jupiters". Furthermore, gravity and period also vary by an order of magnitude within the population. Composition-wise, observations of their spectra have shown that atmospheric metallicity can also have an order of magnitude variation within the population \citep{2019ApJ...887L..20W}. There are multiple known correlations between these parameters at the population level which significantly boosts the spectral diversity we observe and introduce trends in their atmospheres. For example, because of tidal locking, equilibrium temperature and rotation are intrinsically linked. Additionally, the hottest hot Jupiters have an inflated radius \citep{2010SSRv..152..423F,2014prpl.conf..763B}, leading to a correlation between equilibrium temperature and surface gravity. Finally, a correlation between atmospheric metallicity and planet mass is observed in the solar-system and could also be present in giant exoplanets \citep{2019ApJ...887L..20W}.

The chemical composition of an atmosphere can strongly alter its thermal structure. In particular, the presence of TiO and VO, two strong optical absorbers that exist in M dwarfs, have been postulated to create a thermal inversion on the planet day-side \citep{2003ApJ...594.1011H,2008ApJ...678.1419F}, meaning that the temperature would increase with decreasing pressure. Such thermal inversions have been observed by the presence of emission features in the thermal emission spectra of the hottest planets \citep{2023arXiv230108192C,2020MNRAS.496.1638M,2023MNRAS.522.2145V,2022A&A...668A..53C}. For the cooler ones, clear absorption features have shown the definitive lack of a thermal inversion \citep{2016AJ....152..203L,2014ApJ...793L..27K}, which is expected as TiO and VO are supposed to condense out of the atmosphere at low temperatures. However, the exact point at which the population transitions from TiO/VO rich atmospheres with thermal inversions towards TiO/VO poor atmospheres without inversions is still unclear. Direct detection of TiO and VO through observation is complex as its spatial distribution is shaped by multiple processes. Condensation of these chemicals in the deep layers \citep{2009ApJ...699.1487S} or on the night-side \citep{2013A&A...558A..91P,2020A&A...636A.117M} could deplete these chemicals from the planet's day-side atmosphere \citep{2018ApJ...860...18P}. Furthermore, atmospheres that are hot enough to suppress condensation can be too hot for TiO and VO to be stable against thermal dissociation \citep{2018A&A...617A.110P}.

Additionally, the 3D structure of hot Jupiters makes the measurement of chemical abundances from their spectra challenging. Indeed, observations, either in transit or eclipse, are usually sensitive to the spatially averaged spectra. Information from different parts of the atmosphere, which can have very different properties, are mixed together. 1D retrieval models can have strong biases when applied to hot Jupiter data sets, for both emission \citep{2016ApJ...829...52F,2017ApJ...848..127B,2020MNRAS.493.4342T} and transmission \citep{2016ApJ...820...78L,2017MNRAS.469.1979M,2019A&A...623A.161C,2020A&A...636A..66P,2020ApJ...893L..43M,2020ApJ...905..131L}. This is due to the spatially varying cloud coverage \citep{2016ApJ...820...78L}, thermal structure \citep{2016ApJ...829...52F,2020MNRAS.493.4342T, 2019A&A...623A.161C} or chemical abundance maps \citep{2020ApJ...893L..43M,2020A&A...636A..66P}. This problem becomes even greater at higher spectral resolutions, as the spectral line shapes, depths and positions, which are heavily dependent on the chemical, thermal and dynamical structure in the atmosphere, become increasingly resolved \citep{2013ApJ...762...24S,2012ApJ...751..117M,2014ApJ...795...24K,2017ApJ...851...84Z,2021AJ....161....1B,2021MNRAS.502.1456K,2021MNRAS.506.1258W}. This has been shown to cause heavily asymmetric features in transmission \citep{2020Natur.580..597E,2021ApJ...908L..17K,2021MNRAS.506.1258W,2022AJ....163..107K,2022ApJ...926...85S}. These effects can lead to biases when inferring abundances because of the non-linear nature of radiative transfer: the spatially averaged spectra is different from the spectra of the spatially averaged thermal structure of the planet. This is particularly true when spatial correlations between physical quantities are present in the planet (e.g. clouds forming in the cooler parts of the atmosphere, chemical abundances being determined by the local temperature etc...). Consequently, to obtain robust chemical abundances from observed spectra, we must understand how planetary parameters influence the physical processes in hot Jupiter atmospheres. Therefore, to draw conclusions on observational trends arising from planet formation processes, we must first understand expected trends due to planetary properties. With rapidly increasing numbers of high quality exoplanet observations, large scale population studies (e.g \citep{2020A&A...639A..36B,2020AJ....159..137G,2022arXiv220315059M,2023arXiv230103639D}) are now possible. The goal of this paper is to estimate how the 3D atmospheric dynamics can shape the observational properties of hot Jupiters across the observed population of planets.

Such work has been done in 1D and 2D by the use of radiative/convective model grids \citep{2021MNRAS.505.5603B,2022MNRAS.512.4877B,2020MNRAS.498.4680G,2021ApJ...923..242G, 2021NatAs...5.1224M,2019ApJ...883..194M}. \citet{2021MNRAS.505.5603B,2022MNRAS.512.4877B} explore how chemical mixing and photochemistry affect atmospheres in a pseudo-2D framework. \citet{2020MNRAS.498.4680G,2021ApJ...923..242G} generate a 1D-2D model grid with a range of re-circulation factors, metallicities, and C/O ratios and use this to explore population trends in the observed Spitzer thermal emission. In a similar way, \citet{2021NatAs...5.1224M} use a 1D model grid to explore trends in the water feature strength of HST/WFC3 observations.

A number of 3D Global Circulation Model (GCM) grids have also been used to identify the importance of planetary parameters in shaping the atmospheric circulation in hot Jupiters. Day-to-night temperature contrast has been shown to depend heavily on irradiation received, rotation period and atmospheric composition \citep{2015ApJ...801...95S,2016ApJ...821....9K,2016ApJ...821...16K,2017ApJ...835..198K,2012ApJ...751...59P,2024MNRAS.528.1016T}. Super-solar metallicities are also found to increase the day-to-night temperature contrast at the photosphere and reduce eastward hot-spot offset \citep{2009ApJ...699..564S,2015ApJ...801...86K,2016ApJ...821....9K,2018A&A...612A.105D}. Additionally, \citet{2016ApJ...821....9K,2022ApJ...941L..40K} show how surface gravity plays an important role when determining the photospheric pressure in the atmosphere. However, until now no study has utilised non-grey 3D modelling to investigated the impact of systematically and jointly varying all of these intrinsic planetary properties on an atmosphere, which is the ultimate goal for this study.

We begin by describing the dynamical, radiative transfer and chemical modelling in Section \ref{sec:methods}. In Section \ref{sec:grid}, we then detail the setup of the model grid. This includes a brief discussion on each of the free parameters, why certain values have been selected and how we expect them to affect an atmosphere. Then, in Section \ref{sec:discuss} we show how the wide range of planetary free parameters chosen influence the atmosphere of hot Jupiters. Starting with an overview of the thermal structure, then a discussion on the wide variation in day-to-night heat redistribution within the grid. We then move on to an overview of the secondary eclipse spectra and phase curves, before discussing the phase curve offset in more detail. Comparisons with current planetary observations are then detailed in Section \ref{sec:obs}. We provide an explanation for the sharp jump in brightness temperature that is seen by Spitzer observations of the population of hot Jupiter atmospheres \citep{2023arXiv230103639D}, inferences from investigating spectral indices for water in HST/WFC3 observations \citep{2021NatAs...5.1224M}, and compare predictions from our grid to the Spitzer phase curve observations before summarizing our conclusions in Section \ref{sec:conc}.

\section{Methods}
\label{sec:methods}
\subsection{Dynamics}
\label{sec:methods_dynamics}
We use the non-grey SPARC/MITgcm global circulation model \citep{2009ApJ...699..564S}, which has been widely applied to modelling the atmospheric dynamics for hot, Jupiter sized planets in tidally locked orbits \citep{2009ApJ...699..564S,2015ApJ...801...95S,2015ApJ...801...86K,2016ApJ...821....9K,2019ApJ...880...14S,2017arXiv170600466L,2013A&A...558A..91P,2016ApJ...828...22P,2018A&A...617A.110P,2021MNRAS.501...78P,2024MNRAS.528.1016T}. This model couples the dynamical code from the MITgcm \citep{2004MWRv..132.2845A} with the plane-parallel radiative transfer code of \cite{1999Icar..138..268M} and solves the primitive equations on a cube-sphere grid.

The setup of the GCM here is very similar to that described in \citet{2018A&A...617A.110P,2021MNRAS.501...78P}. All models assume an internal heat flux corresponding to a temperature of 100K. Although it has been shown that internal heat flux probably increases with equilibrium temperature \citep{2019ApJ...884L...6T}, the effect of the internal heat on the wind and thermal structure at the photosphere is not yet fully understood \citep{2022ApJ...941L..40K}. We therefore leave such study for future work and choose to keep the internal heat constant here. All of our models also use a fixed planetary radius of 1.3$R_{j}$, where $R_{j}$ is the equatorial radius of Jupiter at the 1 bar pressure level ($71,492 km$). Despite the radii of hot Jupiter's varying by around a factor of 2 across the population, this was done in order to minimize the number of parameter dimensions within this study. We assume a specific heat capacity of the atmosphere $C_{p} = 1.3\times 10^{4} Jkg^{1}K^{-1}$, with a specific heat capacity ratio of $\gamma = 1 + 2/7$ and a mean molecular weight of $\mu = 2.3mH$ - which is valid for the H2-dominated atmospheres we are simulating. Other prescribed model parameters are changed on a model to model bases, including: equilibrium temperature, atmospheric metallicity, rotation period (changed with stellar parameters), surface gravity and the inclusion of certain molecular species, see Section \ref{sec:grid} for details. Our models purposely do not add a Rayleigh drag parameterization as has been done in the past by e.g. \citet{2016ApJ...821...16K} and most dissipation is due to the 4th order Shapiro filter applied to the temperature and velocity fields, which dissipates kinetic energy by smoothing the horizontal gradients between grid points \citep{2018ApJ...853..133K}. We use a fixed dissipation timescale of $\tau_{\rm shap}$=50s for all models, meaning that the numerical dissipation is proportional to the kinetic energy of each model \citep{2018ApJ...853..133K}. This provides us with the least dissipative models that can be obtained at the spatial resolution we are using while being numerically stable.

The atmosphere was allowed to evolve for 150 days, 300 days for the 1000-1200K models, with all quantities being averaged over the final 50 days for the output. This is enough integration time for a pseudo-steady state to be reached at the photospheric pressure level \citep{2009ApJ...699..564S}, although not enough to reach a fully converged state which would take several thousand days \citep{2018AJ....155..150M,2020MNRAS.491.1456M,2020ApJ...891....7W,2017A&A...604A..79M}. As shown in Appendix \ref{sec:convergence}, running for much longer times does not 
significantly change the observable properties of the planets we are interested in. Given that the interactions between the deep interior and upper atmosphere are still not fully understood for hot Jupiter exoplanets, we are assuming that the deep layers are quiet as they don't have time to develop winds, giving us the same solution as a photosphere equilibriated with a quiet deep interior and meaning that our choice of deep boundary condition has no effect on the final result. The models were run at a horizontal resolution of C32, which corresponds to an approximate resolution of 128 longitudinal and 64 latitudinal cells. The atmosphere is split into 53 pressure levels, between $2\times 10^{-6}$ - 200 bars. For low metallicity models the time-step was set to 25 s, which was decreased to 12.5 s at higher metallicities in order to stabilize the simulations.

\subsection{Radiative Transfer}
\label{sec:methods_radtrans}
In both the 3D GCM simulations and subsequent spectral calculations, radiative transfer is handled via the plane-parallel radiative transfer code of \citet{1999Icar..138..268M}. Although this code was first developed to study Titan's atmosphere \citep{1989Icar...80...23M}, it has previously been used to study both hot Jupiters \citep{2005ApJ...627L..69F,2008ApJ...678.1419F,2021MNRAS.501...78P} and ultra-hot Jupiters \citep{2018A&A...617A.110P}.

For the opacities, we use those detailed in \citet{2008ApJS..174..504F}, including updates from \citet{2014ApJS..214...25F}. When coupled to the GCM a lower spectral resolution including only 11 frequency bins is used in order to speed up calculations \citep{2013ApJ...767...76K}. Within each of these bins, thousands of spectral lines are modelled through the correlated-k distribution method \citep{1989artb.book.....G,1991JGR....96.9027L}, allowing us to compress the information down to just 8 k-coefficients.

We then use the 3D thermal structure from the GCM and post process using a much higher spectral resolution of 196 wavelength bins, ranging from 0.2679 - 227.5313 $\mu m$. This is done using the two-stream radiative transfer equations along the line of sight for each atmospheric column in the cube-sphere grid at each planetary phase, taking into account emission, absorption and scattering. By using this method we naturally take into account geometrical effects such as limb darkening. For more details see \citet{2021MNRAS.501...78P,2016ApJ...828...22P,2006ApJ...652..746F}. For the stellar spectra we use the NextGen stellar models \citep{1999ApJ...512..377H}. Spectra from the grid produced using this method can be seen in Section \ref{sec:atmstruc}.

To additionally speed up post-processing of the grid, the P-T structure from the models was recalculated with a lower spatial resolution of C16, equivalent to 64 longitude and 32 latitude cells. This is done as an insignificant portion of the spectral information is lost when decreasing the spatial resolution in this manner, whilst a significant amount of time is saved in the calculations \citep{2022ApJ...930...93R}. 

The opacity affects the radiative timescale, and therefore temperature structure, of an atmosphere. Opacity itself is highly dependent on the temperature structure due to pressure and temperature broadening of spectral lines and changing chemistry \citep{2018A&A...617A.110P,2018ApJ...855L..30A}. Using non-grey opacity is therefore important for realistic modelling to ensures that energy is conserved when we calculate the spectrum of the planet from the modelled temperature field.

\subsection{Chemistry}
\label{sec:methods_chem}
In our models we assume local chemical equilibrium with local rain-out of condensate materials \citep{2006ApJ...648.1181V,2010ApJ...716.1060V} and a solar C/O ratio of 0.46 \citep{2009LanB...4B..712L}. Both molecular and atomic chemical abundances are calculated using a version of the NASA CEA Gibbs minimisation scheme \citep{GMB1994}, assuming solar elemental abundances and local chemical equilibrium, developed to study gas and condensate equilibrium chemistry under different atmospheric conditions \citep{2013ApJ...777...34M,2016ApJ...817..166S,2015ApJ...801...86K,2017MNRAS.464.4247W,2017AAS...23031507M,2018A&A...617A.110P}. Chemical equilibrium is far from a safe assumption in hot Jupiter atmospheres, where the extreme conditions cause effects such as photochemistry and chemical quenching which smooths out the very large day to night chemical gradients \citep{2006ApJ...649.1048C,2018ApJ...855L..31D,2018ApJ...869..107M,2019ApJ...880...14S}. However, although non-equilibrium abundances (through quenching) can change the overall heat transport and the observables in specific band-passes (i.e temperature changes of $\sim$ 100-200K can lead to a $\sim$ 20\% change in the spectra \citep{2019ApJ...880...14S}), the effects of disequilibrium chemistry on the overall atmospheric energy balance is small compared to the population trend \citep{2019ApJ...880...14S, 2016A&A...594A..69D}.

For the models with increased metallicity, we multiply the elemental abundances of all elements apart from hydrogen and helium by the metallicity factor and adjust the H/He abundance to ensure that the sum of all mixing ratios is equal to 1. To the first order, atmospheric metallicity increases the abundance of all molecules present in the atmosphere, overall increasing the opacity of the atmosphere. At second order, atmospheric metallicity can change the balance between molecules in chemical equilibrium and, for example, favour CO instead of CH4 in cool planets \citep{2023Natur.614..649J}. Additionally, we include models both with and without TiO and VO in the grid, more details on these species can be found in Section \ref{sec:tiovo}.

\section{Grid Parameters}
\label{sec:grid}
In the grid, we have five free planetary parameters which are varied: equilibrium temperature ($T_{\rm eq}$), atmospheric metallicity (Log(M/H)), rotation period (determined via the stellar parameters), surface gravity (Log(g)) and the presence of strongly absorbing chemicals TiO and VO. The full range of values used can be seen in Table \ref{tab:gridparams}. Any other necessary model parameters (i.e C/O ratio, planet radius, internal temperature etc..) remain consistent across the grid. The values used for these fixed parameters can be seen in Section \ref{sec:methods}. Combinations of these free parameters were used to run a total of 345 simulations. The extent of our model grid when compared to the observed exoplanet population can be seen in Figure \ref{fig:param_space}. From this, we note that the population scatter is well covered within the grid, with the vast majority of planets within our equilibrium temperature range having a closely analogous model.
\begin{table*}
	\centering
	\caption{Free parameters used in the 3D GCM grid.}
	\vspace{-5pt}
	\fontsize{10}{13}\selectfont
		\begin{tabular}{ccccc}
			\hline
			\hline
			\textbf{T$_{eq}$ [K]} & \textbf{Log(M/H)} & \textbf{Normalised Period Multiplier} & \textbf{log(g[SI])} & \textbf{Molecular Species}\\ \hline \hline
			1400-2400 (200K intervals) & 0.0, 0.7, 1.5 & 0.37, 1, 2.23 & 0.8, 1.3, 1.8 & \ce{C2H2}, \ce{C2H4}, \ce{C2H6}, \ce{CaH}, \ce{CH4}, \ce{CO},\\
            &&&& \ce{CO2}, \ce{CrH}, \ce{FeH}, \ce{H2O}, \ce{H2S}, \ce{HCN},\\
            &&&& \ce{K}, \ce{Li}, \ce{LiCl}, \ce{MgH}, \ce{N2}, \ce{Na}, \ce{NH3},\\
            &&&& \ce{OCS}, \ce{PH3}, \ce{SiO}, \ce{TiO}, \ce{VO}\\
            \hline
			1000-2400 (200K intervals) & " & " & " & No \ce{TiO}, \ce{VO}\\ \hline \hline
	    \end{tabular}
	\label{tab:gridparams}
\end{table*}

\subsection{Equilibrium Temperature}
The equilibrium temperature, $T_{\rm eq}$, is a measure of the stellar irradiation received by an atmosphere. It is defined as the brightness temperature that a homogeneous planet with zero albedo would have. This temperature is incredibly important for the atmospheric structure, as it affects the dynamical, radiative and chemical timescales significantly. In our grid we have selected a wide range of equilibrium temperatures, from 1000K (which is on the lower end for hot Jupiter exoplanets) to 2400K. We sample every 200K giving us eight values. We don't use temperatures higher than 2400K as above this molecular dissociation \citep{2018A&A...617A.110P}, transport of heat through H2 dissociation \citep{2018ApJ...857L..20B,2021MNRAS.505.4515R,2018RNAAS...2...36K} and atomic iron opacities \citep{2018ApJ...866...27L} start to cause significant complications to the planetary energy balance. Magnetic drag may also start to disrupt the equatorial jet \citep{2022AJ....163...35B,2017NatAs...1E.131R,2014ApJ...794..132R}. On the other end, we don't use temperatures lower than 1000K as for these planets the assumption of tidally locked orbits starts to break down \citep{2014arXiv1411.4731S,2015ExA....40..481P}.

The distribution of observed equilibrium temperatures is displayed in Figure \ref{fig:param_space}. From this we can see that the majority of hot Jupiters have equilibrium temperatures between $\sim 1000-1800$K, with a lower number at higher temperatures, and that our grid range covers most of the observed population.

\subsection{Orbital Period}
In order to stay close to the observed population, we decided to run models for planets orbiting three different stellar types. A K2 dwarf, similar to HD189733, a G0V dwarf, similar to HD209458 and a F6 dwarf similar to HAT-P-7. The influence of different stellar spectra on planet atmospheres, and particularly the sensitivity of temperature inversions to the spectral type, have been previously studied using 1D models for both hot \citep{2015ApJ...813...47M} and ultra-hot \citep{2019ApJ...876...69L} Jupiters. Associated stellar spectra come from the NextGen stellar model archive \citep{1999ApJ...512..377H}. As such, $M_{\rm star}$, $T_{\rm star}$ and $R_{\rm star}$ all vary between the different model tracks. Even though the exact stellar spectrum surely has an effect on the radiative energy balance of the atmosphere, the main effect of changing stellar parameters that we are interested in is the related change of orbital period. Because we assume that our planets are tidally locked, their rotation period is directly determined by the equilibrium temperature and stellar parameters:

\begin{equation}
P = \left(\frac{R_{\rm star}}{R_{\rm sun}}\right)^{\frac{3}{2}}\left(\frac{T_{\rm star}}{T_{\rm sun}}\right)^{3}\left(\frac{M_{\rm sun}}{M_{\rm star}}\right)^{\frac{1}{2}}\left(\frac{1991.5\rm K}{T_{\rm eq}}\right)^{\frac{1}{3}}\rm days.
\end{equation}

As shown in Table \ref{tab:starparams}, our three stars correspond to three model tracks with periods shifted by 0.37 and 2.23 compared to our nominal case. Thus, at a given equilibrium temperature our grid has rotation period varying by a factor of 6. The goal of varying the orbital period in this way is to compare how the atmosphere varies when looking at different planets around the same stellar type. The distribution of observed orbital periods for the hot Jupiter population is displayed in Figure \ref{fig:param_space}. We note that for high temperature models (>2000K), we do model some orbital periods that are lower than observed in the population, which was done to maintain the completeness of the grid.
\begin{table*}
	\centering
	\caption{Parameters used for the stellar models, we use the NextGen stellar spectra \citep{1999ApJ...512..377H} for these stars.}
	\vspace{-5pt}
	\fontsize{10}{13}\selectfont
		\begin{tabular}{lllll}
			\hline
			\hline
			\textbf{$M_{\rm star}$ [M$_{\rm sun}$]} & \textbf{$ R_{\rm star}$ [R$_{\rm sun}$]} & \textbf{$T_{\rm star}$ [K]} & \textbf{normalised Period Multiplier} & \textbf{Absolute Period Range [days]} \\ \hline \hline
			0.8 & 0.8 & 4875 & 0.37 & 0.27 - 3.79 \\ \hline
			1.1 & 1.18 & 5920 & 1 & 0.75 - 10.36 \\ \hline
			1.5 & 2 & 6259 &  2.23 & 1.67 - 23.13 \\ \hline \hline
	    \end{tabular}
	\label{tab:starparams}
\end{table*}

\subsection{Metallicity}
We define the planetary metallicity here as the atomic molar mixing ratio of all other chemical elements to Hydrogen and Helium in the atmosphere. Elemental abundances are scaled from solar composition \citep{2009LanB...4B..712L}, with all atoms other than H and He multiplied by the metallicity value. For the grid we have selected 3 values to sample in the metallicity: Log(M/H)=0.0, 0.7 and 1.5. These range from solar to roughly thirty-times solar metallicity.

We do not explore very high values of metallicity, ensuring that our approximation of a constant mean molecular weight and heat capacity throughout the population stay correct (see \citet{2017ApJ...836...73Z}). The main effect of metallicity is therefore to increase the molecular opacities of the atmosphere, which leads to lower photospheric pressures.

\subsection{Surface Gravity}
The surface gravity of a planet is dependent on the mass and radius, and displays order of magnitude variations in the hot Jupiter mass regime. For example WASP-18b \citep{2009Natur.460.1098H} is a very high density planet with a surface gravity of log(g)=$\sim$2.2 $ms^{-2}$, whilst there are also very low density planets, such as WASP-39b \citep{2016AcA....66...55M} which has a surface gravity of log(g)=$\sim$0.63 $ms^{-2}$. We therefore sample three values for the surface gravity, representing close to the extremes in the hot Jupiter population: log(g[SI])=0.8, 1.3, 1.8 (g=6.31, 19.95, 63.10 $ms^{-2}$). As shown in \citet{2017ApJ...836...73Z}, gravity naturally cancels out in the dynamic equations. However, the role of gravity is important in setting the pressure of the infrared photosphere and thus the opacities at the photosphere.

\subsection{TiO \& VO}
\label{sec:tiovo}
If strong UV/optical absorbers are present in the upper atmosphere, a portion of the incident stellar irradiation will be intercepted creating an area of local heating which causes a temperature inversion. In hot Jupiter atmospheres, the primary chemicals thought to be responsible for these temperature inversion are gaseous titanium oxide (TiO) and vanadium oxide (VO) \citep{2003ApJ...594.1011H,2008ApJ...678.1419F}. These species are predicted by chemical equilibrium to be present in sufficient quantities to produce a strong thermal inversion above temperatures of $\sim 1400K$, creating two distinct regimes in which hot Jupiter atmospheres can be classified \citep{2008ApJ...678.1419F,2015A&A...574A..35P,2020MNRAS.496.3870P}

TiO and thermal inversions have indeed been detected in the hottest hot Jupiters, such as WASP-33b ($T_{\rm eq}=2781$K) \citep{2022A&A...668A..53C}, WASP-121b ($T_{\rm eq}=2450$K) \citep{2016ApJ...822L...4E}, WASP-189b ($T_{\rm eq}=2636$K) \citep{2022NatAs...6..449P} and WASP-18b ($T_{eq}=2505$K) \citep{2023arXiv230108192C}, and VO was recently detected in WASP-76b ($T_{\rm eq}=2182$K) \citep{2023arXiv230608739P}. However, TiO, VO nor thermal inversions have been detected in cooler planets, such as HD209458b ($T_{\rm eq}=1477$K) \citep{2016AJ....152..203L} or WASP-43b ($T_{\rm eq}=1379$K) \citep{2014ApJ...793L..27K}, even though these planets are hot enough to have gaseous TiO in their atmospheres. It is therefore not yet clear at which equilibrium temperature the population of planets transitions from TiO/VO dominated atmospheres to atmospheres devoid of TiO/VO \citep{2021NatAs...5.1224M}. The main mechanisms through which to deplete TiO/VO from the day-side atmosphere are condensation processes that can either happen in the deep layers of the atmosphere \citep{2009ApJ...699.1487S,2016ApJ...828...22P} or in the planet's night-side \citep{2016ApJ...828...22P,2017AJ....154..158B,2023arXiv230608739P}. The actual strength of these cold-trap processes depends on poorly constrained microphysical parameters (such as the condensates size) and the strength of the atmospheric mixing \citep{2019ApJ...887..170P}. Here, we decide to be agnostic about the transition from an atmosphere containing TiO/VO to one without. We therefore run both a set of models at chemical equilibrium (and thus containing TiO/VO) between 1000-2400K, and a set of models where the elemental abundances of Ti and V have been artificially set to zero, mimicking the disappearance of TiO/VO due to non-local condensation processes, between 1400-2400K.

\section{Atmospheric Characteristics for a Range of Planetary Parameters}
\label{sec:discuss}
\subsection{Atmospheric Structure}
\label{sec:atmstruc}
\begin{figure*}
	\includegraphics[width=2\columnwidth]{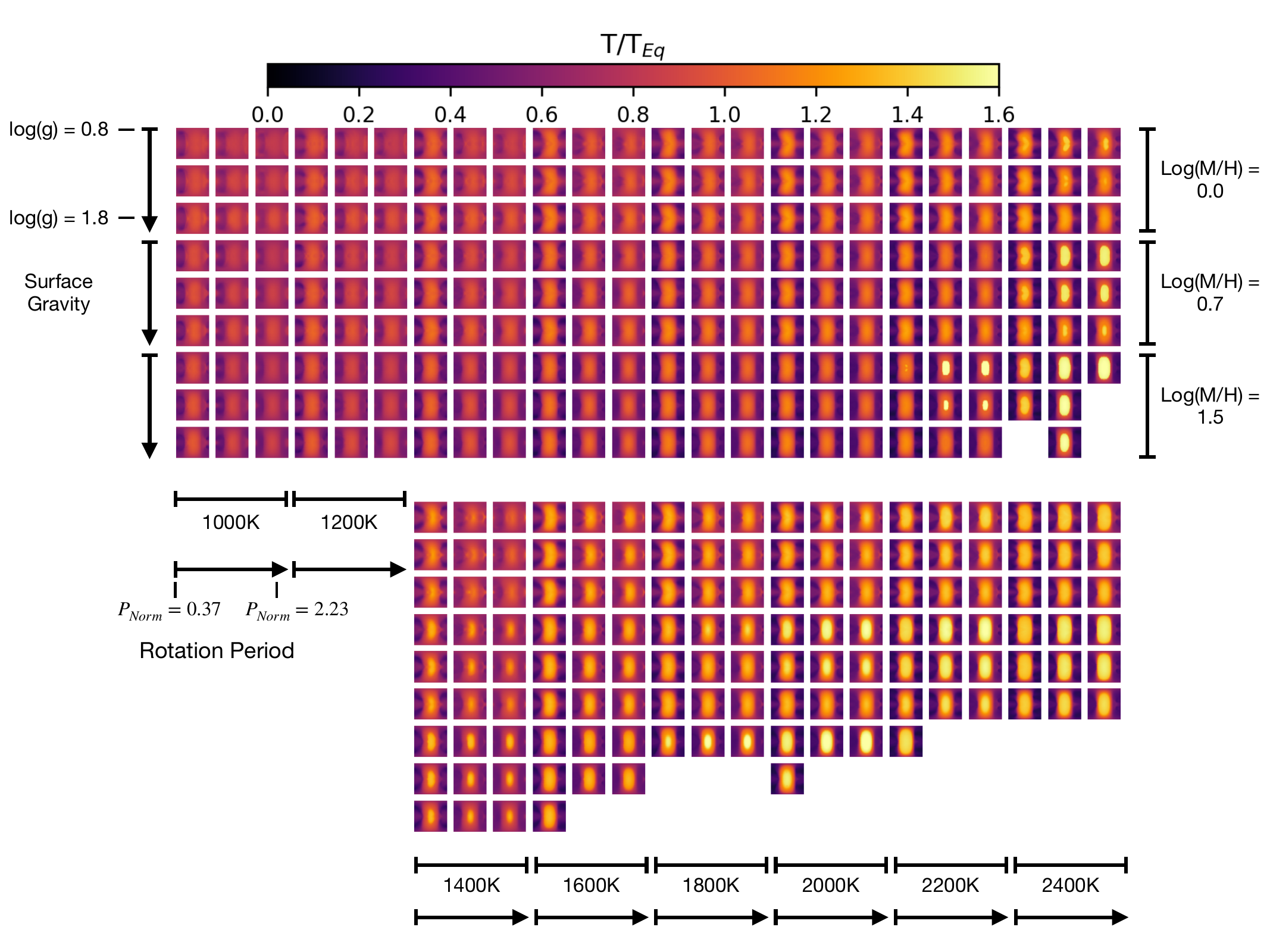}
    \caption{Latitude vs longitude thermal structure, temperature normalised by equilibrium temperature, for all models within the grid, at the pressure level closest to the 1.4 micron photospheric pressure. Equilibrium temperature is sorted into columns and atmospheric metallicity into rows. The surface gravity increases between the values seen in Table \ref{tab:gridparams} down each bracket of rows and the normalised orbital period increases in the same manor along columns. Models excluding TiO/VO are separated into the top half, with models including the two species in the bottom half.}
    \label{fig:fullgrid}
\end{figure*}

For much of the analysis in this investigation we will be boiling the atmosphere down into a series of characteristic observable features. However, we show in Figure \ref{fig:fullgrid} the wide range of diversity displayed by our models. This shows the temperature structure, normalised by the equilibrium temperature, at the pressure level closest to the 1.4 micron photospheric pressure for all the models within our grid. The 1.4 micron photospheric pressure can be calculated as \citep{2018A&A...617A.110P},

\begin{equation}
P_{\rm phot,1.4\mu m}=\frac{2}{3}\frac{g\mu}{\sigma_{\rm H_2O,1.4\mu m}A_{\rm H_2O}},
\end{equation}

\noindent where $\sigma_{\rm H_2O,1.4\mu m}$ is the absorption cross section of water at 1.4$\mu m$, which we assume to be constant, $A_{\rm H_2O}$ is the volume mixing ratio of water, $g$ is the gravity and $\mu$ is the mean molecular weight. this can be scaled by our model parameters as,

\begin{equation}
P_{\rm phot,1.4\mu m}=\left(\frac{g}{2046.2ms^{-2}}\right)\left[\frac{M}{H}\right]^{-1}\rm bars.
\end{equation}

\noindent This method of calculating photospheric pressure works well for lower temperature planets, however it does become less accurate for higher temperatures as it does not include any treatment for water dissociation. Each model is sorted into columns and rows based on their grid parameters. Gaps in the grid shown in Figure \ref{fig:fullgrid} are due to models that are unable to converge. This usually happens at high temperature, high metallicity and fast rotation rates due to numerical instabilities.

The temperature structure in our models varies drastically based on the prescribed model parameters. Surface gravity and metallicity alter the location of the photospheric pressure level, orbital period changes the width and strength of equatorial jet circulation and TiO/VO create temperature inversions. From these temperature maps we can already start to see some trends among the models. The equatorial jet structure is clearly visible in many. This jet becomes more narrow in latitude for models with higher temperatures, which are rotating much faster than their cooler counterparts in the grid, and for models with lower metallicity. High latitude Rossby waves are also visible for many of the models. The day-to-night temperature contrast can clearly be seen to increase with the equilibrium temperature.

\begin{figure*}
	\includegraphics[width=2\columnwidth]{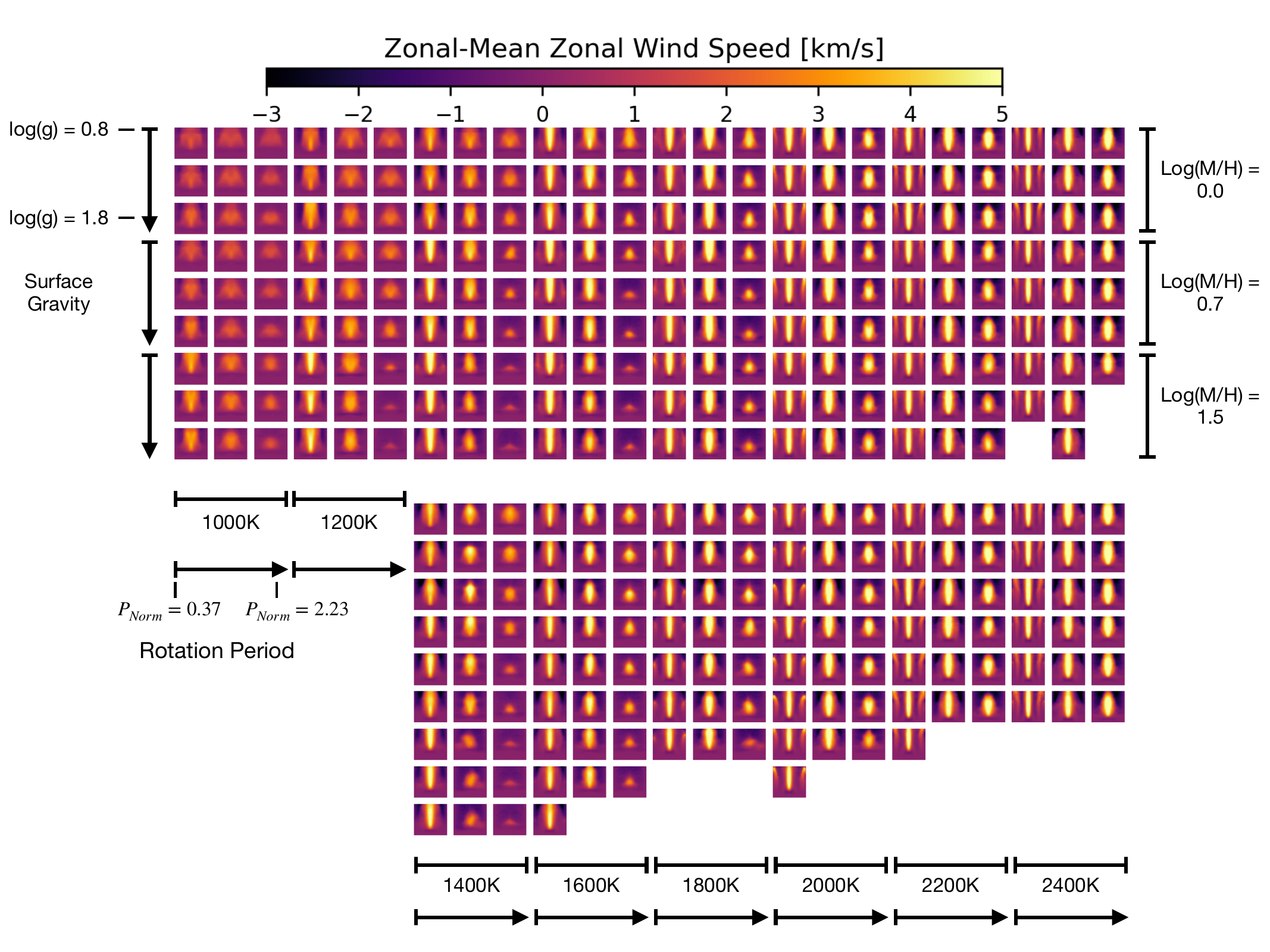}
    \caption{Zonal-mean zonal wind speed for all models within the grid. Within each subplot, the x-axis shows latitude and the y-axis pressure. Equilibrium temperature is sorted into columns and atmospheric metallicity into rows. The surface gravity increases between the values seen in Table \ref{tab:gridparams} down each bracket of rows and normalised orbital period increases in the same manor along columns. Models excluding TiO/VO are separated into the top half, with models including the two species in the bottom half.}
    \label{fig:fullgrid_zmzws}
\end{figure*}

Figure \ref{fig:fullgrid_zmzws} shows the zonal-mean zonal wind speed with pressure for the grid arranged in the same format. Again, there are some clear trends that emerge with our model parameters. The strength of the super-rotating jet is clearly increased with the equilibrium temperature. The jet also extends to deeper pressures in the atmosphere as the surface gravity is increased. As the normalised rotation period is increased, the jet also clearly becomes wider, and the flow structure transitions from multiple jets to a single one for higher temperature models.

\begin{figure*}
	\includegraphics[width=2\columnwidth]{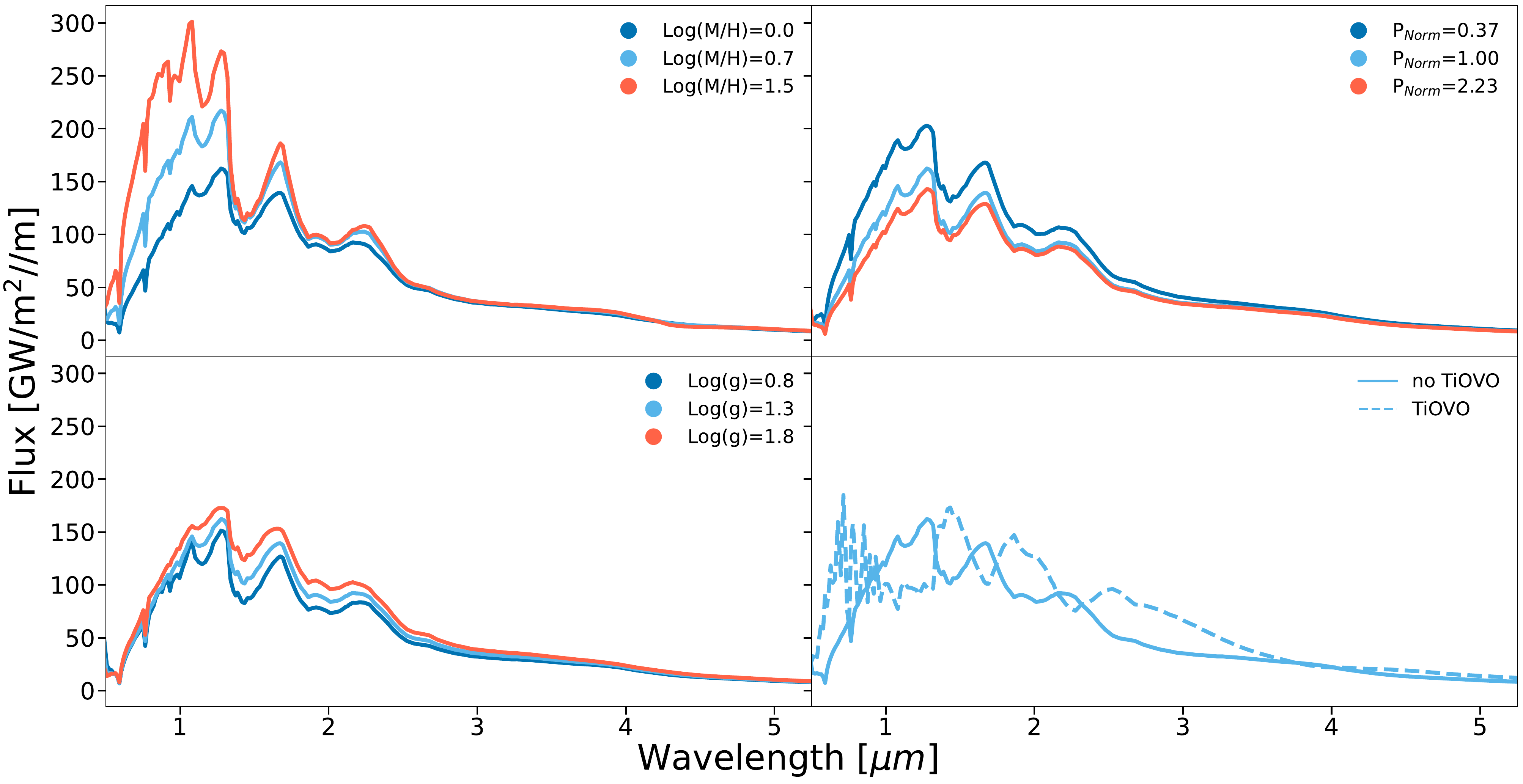}
    \caption{Secondary eclipse spectra for a number of 1800K T$_{\rm eq}$ models within the grid. \textbf{Upper Left Panel:} Shows the variation in atmospheric metallicity. \textbf{Upper Right Panel:} Shows the variation in normalised orbital period. \textbf{Lower Left Panel:} Shows the variation in atmospheric surface gravity. \textbf{Lower Right Panel:} Shows the difference between the spectra for a model containing and not containing TiO and VO.}
    \label{fig:spectra_comp}
\end{figure*}

To show how the secondary eclipse spectra vary with our model parameters, we include Figure \ref{fig:spectra_comp}. Here we can see that increasing metallicity and surface gravity act to increase the emission flux, with increasing normalised period having the opposite effect. This is caused by the effect of parameters on the heat redistribution, which is discussed in detail throughout Section \ref{sec:redist}. It is interesting to note here that the metallicity appears to have a larger affect on the flux within the emission bands, surface gravity has a larger affect outside of the bands and the normalised period has a fairly constant affect across the spectra. The bottom right panel of Figure \ref{fig:spectra_comp} shows how the presence of TiO and VO affect the emission spectra.

\subsection{Heat Redistribution}
\label{sec:redist}
The incident stellar flux on a planet's atmosphere has, essentially, two main pathways. It is either reflected back into space, the portion of which can be determined by the bond albedo, or it is deposited in the day-side atmosphere where it can either be re-radiated or partially advected onto the night-side before being re-radiated \citep{2011ApJ...729...54C}. This can be described as the heat redistribution efficiency. Ideally, measurements of the emitted light from a planet at all phases is necessary to estimate the heat redistribution efficiency. However, for planets with a small albedo, this efficiency can be determined by comparing the incoming radiation ($\sigma T_{\rm eq}^{4}$) to the outgoing longwave radiation from the planetary day-side ($\sigma T_{\rm day}^{4}$). Thus we define the heat redistribution efficiency, or $f$-factor, as \citep{2018ApJ...855L..30A},
\begin{equation}
f = \left(\frac{T_{\rm day}}{T_{\rm eq}}\right)^4.
\label{eq:redist}
\end{equation}

\noindent $T_{\rm day}$ is the temperature of the equivalent blackbody emitting as much energy in our direction as the planetary day-side. Following this definition, the redistribution factor ($f$-factor) has a minimum value of $f=1$, corresponding to a case of maximum planet wide heat redistribution where the atmospheric temperature is fully homogenized between the day and night-sides, and a maximum value of $f=2.66$ where there is zero heat redistribution between the two hemispheres. These limits only apply when considering the bolometric flux. A value of $f=2$ corresponds to a theoretical case where there is day-side only heat redistribution. These $f$-factor values are related to the $\epsilon$ values defined in \citep{2011ApJ...729...54C} by $\epsilon=1+0.6(1-f)$.
Equation \ref{eq:redist} is used to calculate the heat transport in our cloudless models, using the same method as described in \citet{2021MNRAS.501...78P}, since the albedos are always small (see Appendix \ref{sec:albedo} for our model albedos). This will further enable a direct comparison with secondary eclipse data.
\begin{figure*}
	\includegraphics[width=2\columnwidth]{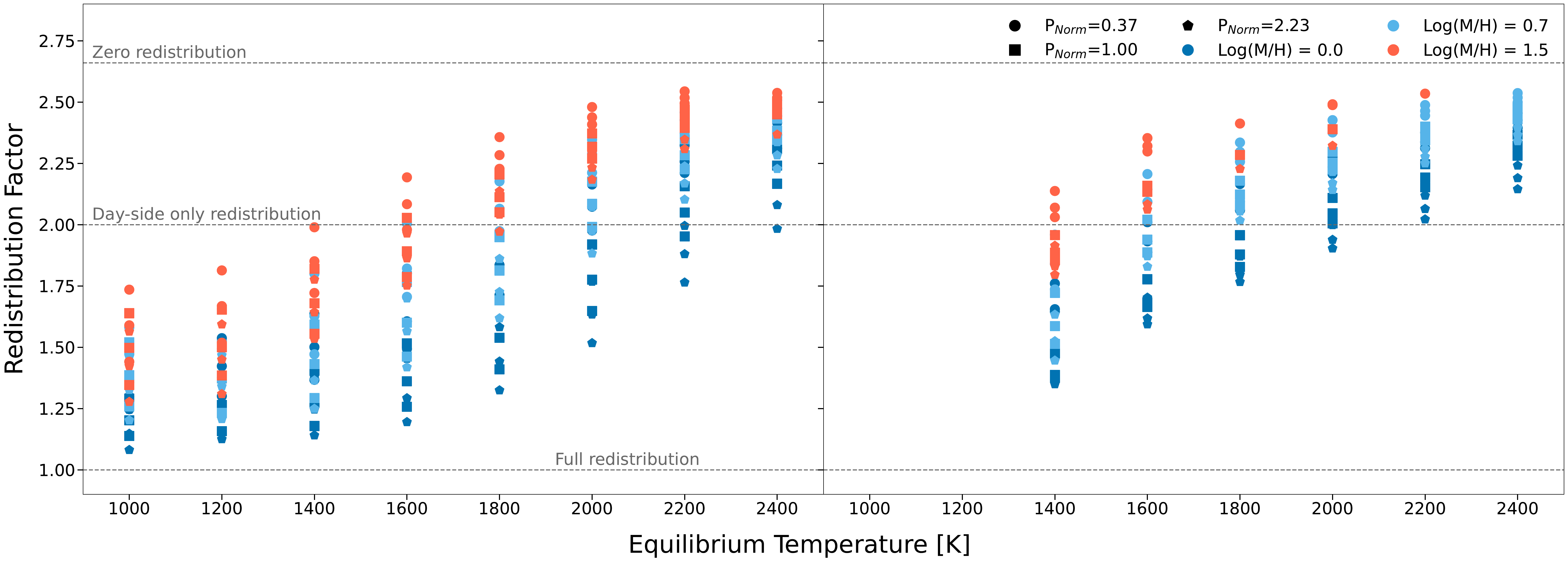}
    \caption{Bolometric redistribution factor ,$f$-factor, a measure of the day-night atmospheric heat redistribution, for the model grid vs equilibrium temperature. \textbf{Left Panel:} Shows models with TiO/VO removed. \textbf{Right Panel:} Shows models containing TiO/VO. The shape of each point depicts the orbital period, and the color depicts the metallicity of the model. The grey dashed lines are drawn at key values corresponding to specific redistribution regimes.}
    \label{fig:ffactorsall}
\end{figure*}
\begin{figure*}
	\includegraphics[width=2\columnwidth]{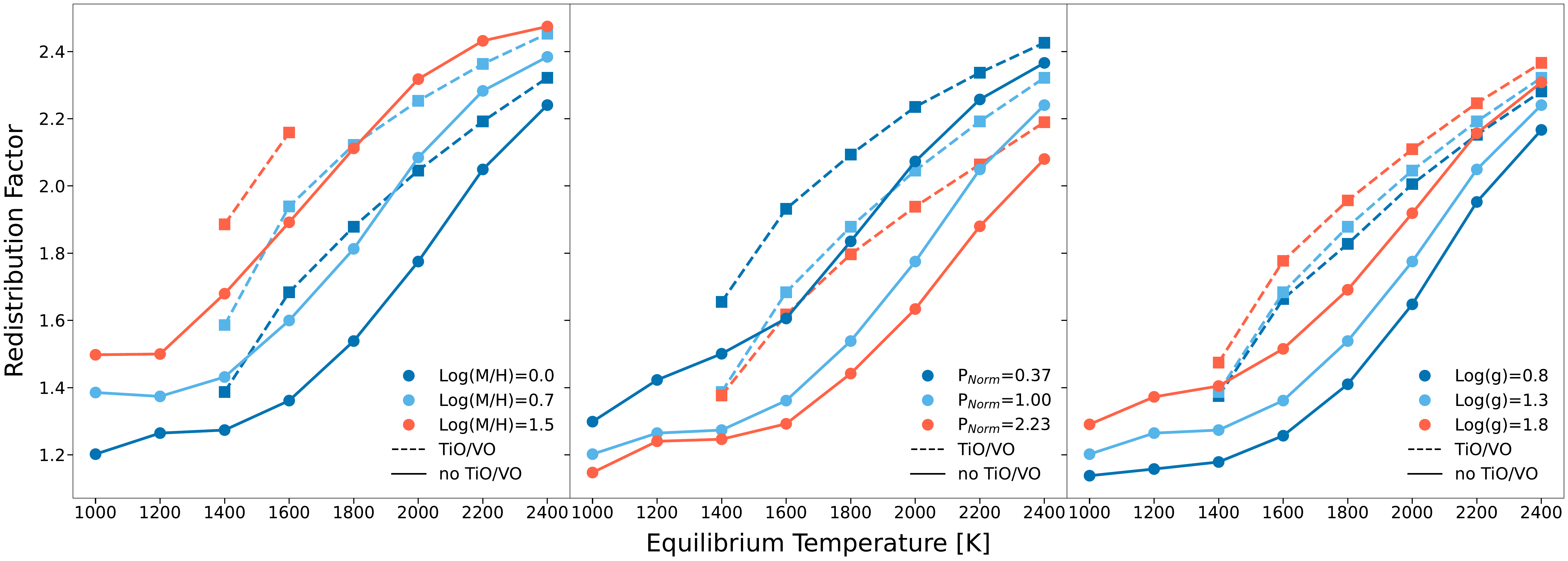}
    \caption{Calculated bolometric $f$-factor values for a subset of models within the grid. Each panel displays the variation in $f$-factor with $T_{\rm eq}$ due to one of: metallicity, orbital period and surface gravity, whilst the other two parameters are kept constant (when constant Log(M/H)=0.0, P$_{\rm Norm}$=1, log(g)=1.3). Cross-terms between parameters are not included. Models excluding and including TiO/VO are displayed with solid and dashed lines respectively. \textbf{Left Panel:} Variation in $f$-factor with metallicity. \textbf{Middle Panel:} Variation in $f$-factor with different normalised orbital period. These tracks corresponds to planets tidally-locked around different stellar types. \textbf{Right Panel:} Variation in $f$-factor with surface gravity.}
    \label{fig:f_bol_tracks}
\end{figure*}

\subsubsection{Expected scaling for the heat redistribution factor}
\label{sec:effects_of_g&met}
Heat is transported in hot Jupiter atmospheres via two mechanisms: direct advection of heat by the winds and wave adjustment mechanisms. As the atmosphere tries to move heat from day to night-side, energy is lost by the gas through radiation. The final efficiency of the heat transport is therefore determined by competition between the advective or wave driven transport and radiative losses \citep{2002A&A...385..166S,2013ApJ...776..134P,2016ApJ...821...16K,2018AJ....155...83Z}. Heat advection itself occurs, primarily, via two different and competing methods: rotational circulation, which is dominated in hot Jupiters by the super-rotating equatorial jet, and divergent overturning circulation between the day-side and night-side hemispheres. \citet{2021PNAS..11822705H} show that these two regimes can both contribute to the heat circulation efficiency equally for certain planets. As the jet circulation is highly dependent on a host of external factors such as the orbital period of the planet, these components will also have variable relative magnitudes. This means that their are essentially three ways to transport heat: wave adjustment mechanisms, rotational circulation and divergent circulation. Because of this, heat transport varies widely over the observed hot Jupiter population as it is highly dependent on atmospheric thermal and chemical structure. \citet{2008ApJ...678.1419F}, for example, suggested that the presence of stratospheric temperature inversions could explain some variations in observed redistribution efficiency, which has driven much of the investigation into low pressure optical absorbers in hot Jupiter atmospheres. Additionally, heat redistribution is thought to be negatively effected with increasing temperature \citep{2016ApJ...821...16K,2013ApJ...776..134P} and metallicity \citep{2015ApJ...813L...1C}.

The wave timescale can be defined as as the timescale for gravity waves to travel across a hemisphere of the planet in the isothermal case \citep{2013ApJ...776..134P,2021MNRAS.501...78P}:
\begin{equation}
    \tau_{\rm wave} \sim \frac{\pi R_p}{\sqrt{k_b T_{\rm photo}/\mu}} \propto \frac{1}{\sqrt{T_{\rm photo}}},
    \label{eq:tauwav}
\end{equation}

\noindent where $R_p$ is the planetary radius, $\mu$ the mean molecular weight, $k_b$ the Boltzmann constant and $T_{\rm photo}$ the photospheric temperature. 

The advective timescale can be defined as the timescale it takes for a parcel of gas to be advected across one hemisphere by the equatorial zonal jet \citep{2021MNRAS.501...78P}:
\begin{equation}
    \tau_{\rm adv} \sim \frac{\pi R_p}{U_{\rm jet}} \propto \frac{1}{T_{\rm eq}}.
    \label{eq:tauadv}
\end{equation}

\noindent where $U_{\rm jet}$ is the mean wind speed at the equator and $T_{\rm eq}$ the equilibrium temperature. \citet{2021MNRAS.501...78P} find $U_{\rm jet}$ to increase linearly with $T_{\rm eq}$ in their cloudless GCM models, which we also find for our models (see Appendix \ref{sec:zmws_appendix}). $U_{\rm jet}$ also increases linearly with increasing surface gravity. Increasing atmospheric metallicity slightly increases $U_{\rm jet}$ at lower $T_{\rm eq}$, but reduces it at higher $T_{\rm eq}$. Additionally, increasing rotation period decreases $U_{\rm jet}$, although this relationship is non-linear. However, order of magnitude variations in these parameters change $U_{\rm jet}$ by at most a factor of 2.

The radiative timescale, the time it takes for a parcel of gas to lose its energy to space, at the photosphere of an atmosphere can be approximated as \citep{2002A&A...385..166S}:
\begin{equation}
    \tau_{\rm rad} \sim \frac{P_{\rm photo}c_{p}}{4\sigma gT_{\rm photo}^3},
    \label{eq:taurad}
\end{equation}

\noindent where $\sigma$ is the Stefan-Boltzmann constant, $g$ the surface gravity, $P_{\rm photo}$ the photospheric pressure and $c_p$ the specific heat capacity at constant pressure. This timescale can range from a few seconds to thousands of years at deeper pressure levels \citep{2002A&A...385..166S}. Given the radiative timescale has a much stronger dependency on the temperature and gravity parameters than the advective and wave timescales, we expect adjustments to the radiative timescale to dominate changes in the heat redistribution. 

The opacity will increase with atmospheric metallicity, as a higher ratio of metals to hydrogen will increase the atomic weighting of the atmosphere. From Equation \ref{eq:taurad}, the radiative timescale at the photosphere will therefore reduce with increasing metallicity \citep{2015ApJ...801...86K,2018A&A...612A.105D}, leading to less efficient heat redistribution. However, as $P_{\rm phot}$ also depends on gravity, the scaling with $g$ is not clear until we rewrite Equation \ref{eq:taurad}. The radiative timescale can equally be expressed in terms of the opacity, $\kappa$. By first taking the hydrostatic equation and the definition of optical depth,
\begin{align}
    dP &= -\rho g dz, \\
    d\tau&=\kappa dm,
\end{align}
 
\noindent where $\tau$ is the optical depth. Then noticing that $\rho dz=dm$, we can rewrite the pressure as,
\begin{equation}
    dP = \frac{gd\tau}{\kappa},
\end{equation}

\noindent where $\tau\sim2/3$ is the optical depth at the photosphere. Assuming constant gravity and opacity, the photospheric pressure can then be described as.
\begin{equation}
    P_{\rm photo} \sim \frac{2g}{3\kappa}.
\end{equation}

\noindent Because pressure and optical depth vary exponentially in our atmosphere, and because the opacity usually increases with pressure, the opacity value that matters to determine the photospheric pressure is the opacity at the photosphere, $\kappa_{\rm photo}$. Substituting this back into equation \ref{eq:taurad}, the radiative timescale then becomes,
\begin{equation}
    \tau_{\rm rad} \sim \frac{c_{p}}{6\sigma\kappa (P_{\rm photo}(g),T_{\rm photo})T_{\rm photo}^3}.
    \label{eq:tauradnew}
\end{equation}

\noindent As $\kappa$ increases with pressure, due to pressure broadening of spectral lines, $\tau_{\rm rad}$ will decrease with increasing surface gravity. The exact scaling relation between opacity and pressure can be found in \cite{2014ApJS..214...25F}. In conclusion, we expect planets with high metallicity and high gravity to be the less efficient at transporting heat from day to night.

\subsubsection{Overall model distribution}
The bolometric redistribution factor values for all models in the grid are shown in Figure \ref{fig:ffactorsall}, with those models containing TiO/VO on the right and those without on the left. There is a trend towards higher $f$-factor values, meaning weaker day-night heat redistribution, at higher equilibrium temperature models in our grid. This is a well documented result that has been previously shown in both non-grey \citep{2021MNRAS.501...78P} and semi-grey modelling \citep{2017ApJ...835..198K, 2012ApJ...751...59P}, and also predicted with energy balance models \citep{2011ApJ...729...54C}. The scaling is caused by the strong temperature dependency of the radiative timescale ($T^{3}$ from equation \ref{eq:tauradnew}) compared to the scaling of the wind and wave speeds ($T$ and $\sqrt{T}$ from equations \ref{eq:tauadv} and \ref{eq:tauwav} respectively). Overall, hot planets do not move heat around fast enough compared to the time needed for a parcel of gas to cool down radiatively. This naturally explains why the night-sides of hot Jupiters are almost independent of the equilibrium temperature \citep{2021MNRAS.501...78P}.

There is also a noticeable kink in the distribution of $f$-factors for models not containing TiO/VO. This is due to the competition between the radiative, advective and wave timescales. Below $\sim 1600K$ the photospheric pressure stays roughly constant (see Figure 4 in \citep{2021MNRAS.501...78P}), as opacity is increasing with the temperature but the Plank function is shifting to shorter wavelengths. Over this range of temperature, $f$-factor does not vary strongly with $T_{\rm eq}$. Over the same range of $T_{\rm eq}$, the radiative and advective timescales stay very close to each other. This is because the advective timescale is not inversely linear but inversely \emph{affine} (linear minus constant alpha, see \citet{2021MNRAS.501...78P} equation 6). As a consequence, when both radiative and advective timescales cross each other they do not scale as $T_{\rm}^{2}$ but $T_{\rm}^{3}/(T_{\rm eq}-\alpha)$, which shows a much weaker variation with $T_{\rm eq}$.

When TiO and VO are present, right panel of Figure \ref{fig:ffactorsall}, the heat redistribution is worse, making the $f$-factor larger. The reason is multi-fold. First, the opacities do not decrease with decreasing wavelength anymore, meaning that, contrary to the case without TiO, as the temperature increases the photospheric pressure decreases, leading to shorter radiative timescales. Second, a portion of the stellar energy is absorbed at very low pressure and re-emitted directly from the stratosphere, rather than transported to the IR-photosphere, hence decreasing the overall rate of day-night heat transport.

The main takeaway here, however, is that at each equilibrium temperature we find that there is a significantly large spread in the heat redistribution based on the other planetary parameters. When gravity, period and metallicity are varied together, they can alter the efficiency of the heat redistribution as much as the equilibrium temperature itself. If we compare the spread in these values to that found in \citet{2021MNRAS.501...78P}, showing the change in redistribution factor between cloudless and cloudy atmosphere models when altering only the equilibrium temperature, we find that the intrinsic properties of the planet are having a comparable effect on the atmosphere to that of night-side clouds. Furthermore, the spread is also equivalent to that found in \citet{2017ApJ...835..198K} between models with very low drag timescales ($\tau_{\rm drag}=10^{3}s$) and infinite drag timescales. 

The spread in heat redistribution factor at a given equilibrium temperature in our models, due to the combined variability of rotation period, gravity and metallicity, is well in line with the spread observed by \citet{2022arXiv220315059M} who measure a heat redistribution factor varying between 0.6 and 2.6 for planets between 1100-2400K (in the Spitzer bands the $f$-factor can be below 1). \cite{2022arXiv220315059M} also display the heat redistribution to have a very wide range within a narrow variation of equilibrium temperatures ($f$=0.6-2.4 between 1100-1700K). This is an important point to note, as due to the wide range of effects that different parameters have on the redistribution, there is a large amount of degeneracy between the efficiency of heat redistribution between cases. This greatly complicates the characterisation of hot Jupiter atmospheres, as the intrinsic properties of the planet, chemical constituents of the atmosphere and the presence of night-side clouds all act to effect the redistribution in a wide range of potentially offsetting ways. 

\subsubsection{Specific trends with parameters}
We now look into more detail at the trends with each parameter and discuss whether it is in agreement with simple timescale analysis. To take a closer look at how each planetary parameter is individually affecting heat redistribution, we can plot the $f$-factor as a function of temperature for a subset of models within our grid (ignoring cross-terms between the parameter space for now). This can be seen in Figure \ref{fig:f_bol_tracks}.

The change in $f$-factor with atmospheric metallicity is found in the left panel of Figure \ref{fig:f_bol_tracks}. Here, we find the calculated $f$-factor to increases with higher atmospheric metallicity. This is caused by higher metallicity increasing the opacity and therefore shifting the photosphere to a lower pressure (as explained in Section \ref{sec:effects_of_g&met}). This reduces the radiative timescale at the photosphere, which increasingly outweighs the dynamical timescale in the atmosphere and therefore reduces the day-night heat redistribution efficiency. 

In the right panel we see that the $f$-factor increases with surface gravity. From the re-written radiative timescale equation (see Equation \ref{eq:tauradnew}), we can see that, as the atmospheric opacity increases with surface gravity, the radiative timescale at the photosphere decreases with surface gravity therefore leading to a decrease in heat redistribution (again see Section \ref{sec:effects_of_g&met}). 

At a given equilibrium temperature, a longer orbital period is shown to increase the heat redistribution efficiency (decrease the $f$-factor) in the middle panel Figure \ref{fig:f_bol_tracks}. This is caused by the increased strength of the combined rotational and divergent components of the circulation, as both transport heat from the day to the night-side and are both dependent on rotation period \citep{2021PNAS..11822705H,2022ApJ...941..171L}. As we will see in more details in Section \ref{sec:offset}, the opposite is true for the offset, for which it is the \emph{relative} strength of divergent and rotational components of the circulation that matter.

Comparing the models excluding and including TiO/VO shows us that the presence of these species reduces redistribution across the board. Strong absorption in the near UV/optical range by these chemical in the high altitude day-side atmosphere, where the radiative timescale is short, reduces the efficiency of energy redistribution and significantly increases the temperature compared to the night-side. The average effect of the TiO/VO on the heat redistribution is $\Delta f=0.216$, but this varies significantly with temperature.
\begin{figure*}
	\includegraphics[width=2\columnwidth]{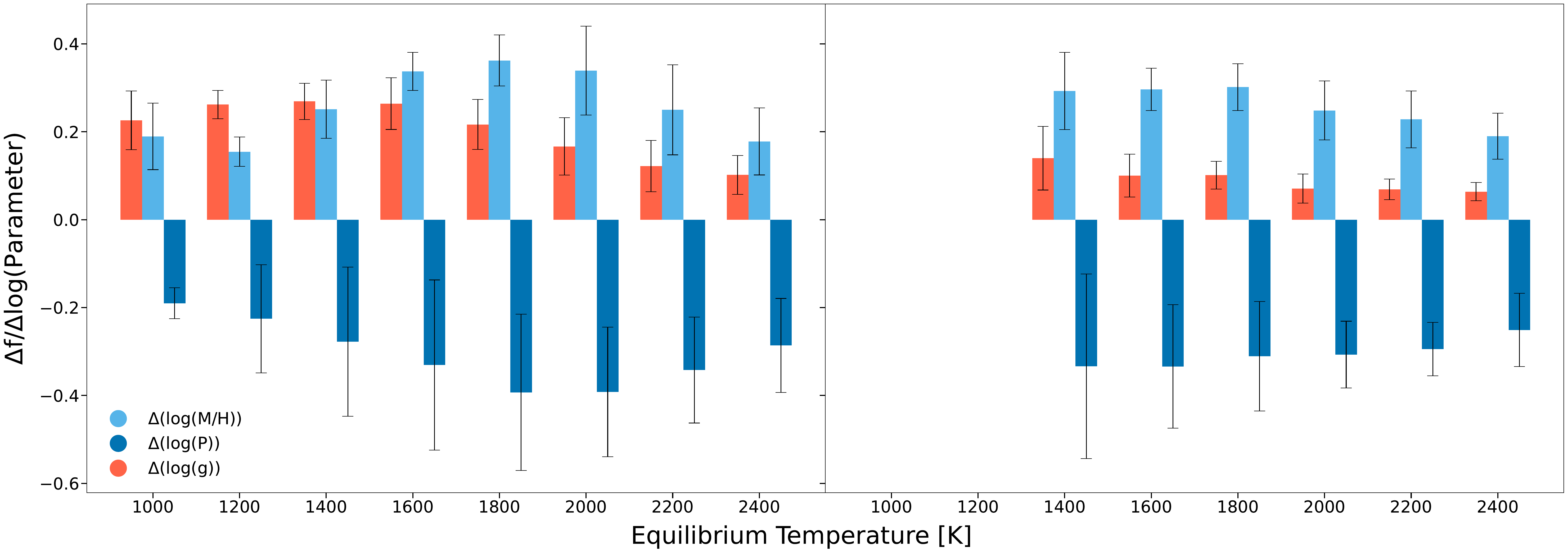}
    \caption{Relative impact of each planetary parameter on heat redistribution. The y-axis of each plot is the delta in the observable feature divided by the delta in the log of each planetary parameter. The x-axis is the Equilibrium temperature. Light blue denotes the change due to metallicity, dark blue the change due to period and red the change due to surface gravity. The error bars display the standard deviation in the impact of each parameter. \textbf{Left Panel:} Models with TiO/VO removed. \textbf{Right Panel:} Models containing TiO/VO.}
    \label{fig:param_delta_f}
\end{figure*}

\subsubsection{Combined parameter trends: heat redistribution}
\label{sec:delta_params}
We now discuss the quantitative effect that each parameter has on the heat redistribution. Because the change of heat transport with, say, period, depends on the other parameters, such as temperature, we propose to look at both the average effect of the parameters and the variation of this average effect within the grid. For this, we measure the difference in heat redistribution between two models when varying only one parameter and keeping the other ones constant: 

\begin{equation}
\Delta f_{P_{\rm 1,2}(T_{\rm eq},\log(M/H),\log(g))}=\frac{f_{P_2,\rm all}-f_{P_1,\rm all}}{\log(P_2)-\log(P_1)}.
\end{equation}

\noindent Here $P_1$ is our model with a period normalization factor of $P_{\rm norm}=0.37$ and $P_2$ the model with $P_{\rm norm}=1$. We then take the statistical mean between all the $\Delta$ values for the combinations of $T_{\rm eq}$, $\rm log(\rm M/H)$ and $\rm log(\rm g)$. This gives us the statistical average for the change we would expect to see in the heat redistribution when transitioning between consecutive grid points for the period, i.e $\Delta f_{P_{1,2}}$. This process is then repeated for each of the other parameters. The standard deviation in the $\Delta f$ values is also calculated, i.e $\sigma f_{P_{1,2}}$. The mean and standard deviation for each parameter $\Delta f$ is shown in Figure \ref{fig:param_delta_f}.

The size of the bar is the averaged change of the heat redistribution parameter when changing metallicity, pressure or gravity. The size of the error bar is the variation of this change when evaluated at different values of the other parameters. For example, one can see that the impact of period on the heat redistribution is similar for all metallicities and gravities. Hence it has a small variance. On the contrary, the change of heat redistribution due to a change of period depends strongly on the specific metallicity and gravity of the planet, hence the larger variance.

Overall, we see for the case without TiO/VO that gravity, period and metallicity have a quantitatively similar effect on the heat redistribution at a given equilibrium temperature. For the case with TiO/VO (right panel), the variation with gravity has a much smaller effect than the other parameters but both metallicity and period contribute in a similar way. However, atmospheric metallicity is largely unconstrained in comparison to period and gravity, which vary by no more than an order of magnitude within the population (see Figure \ref{fig:param_space}). If the metallicity variation is strong enough across the population, its heat redistribution trend could therefore dominate the observed population trend.

\subsection{Bolometric Phase Curves}
\label{sec:pc}
The phase curve of a planet, characterised as the change in brightness temperature observed during a full orbit, contains information about the longitudinal structure of the planet's atmosphere. As the phase curve amplitude and shape are determined by longitudinal inhomogeneities for transiting planets, they contain useful information about the 3D nature of planets. This includes the contrast in brightness temperature between day and night, which is directly linked to the phase curve amplitude, and the asymmetry of the brightness distribution, which is directly obtained by the measurement of the phase curve offset.

\subsubsection{Offset}
\label{sec:offset}
The phase curve offset is a measure of the longitudinal offset of the brightest hemisphere from the substellar point ($0^{\circ}$). This has a positive value if offset eastward (respective to the direction of planetary rotation) from the substellar point, meaning that the brightest hemisphere would be observed before secondary eclipse. Over the population of planets, phase curve offset is firstly determined by equilibrium temperatures, with offsets varying between 50$^{\circ}$ in the colder cases to a few degrees for the hottest planets. Within each temperature bin, however, the variation of the phase curve offset is extremely large. For example, for planets at 1400K it ranges between 50$^{\circ}$ for low gravity, low metallicity cases to 10$^{\circ}$ for high metallicity, high gravity cases. Additionally, the phase curve offset is predicted to vary in a non-monotonic way with temperature \citep{2016ApJ...821...16K,2018AJ....155...83Z}.
\begin{figure*}
	\includegraphics[width=2\columnwidth]{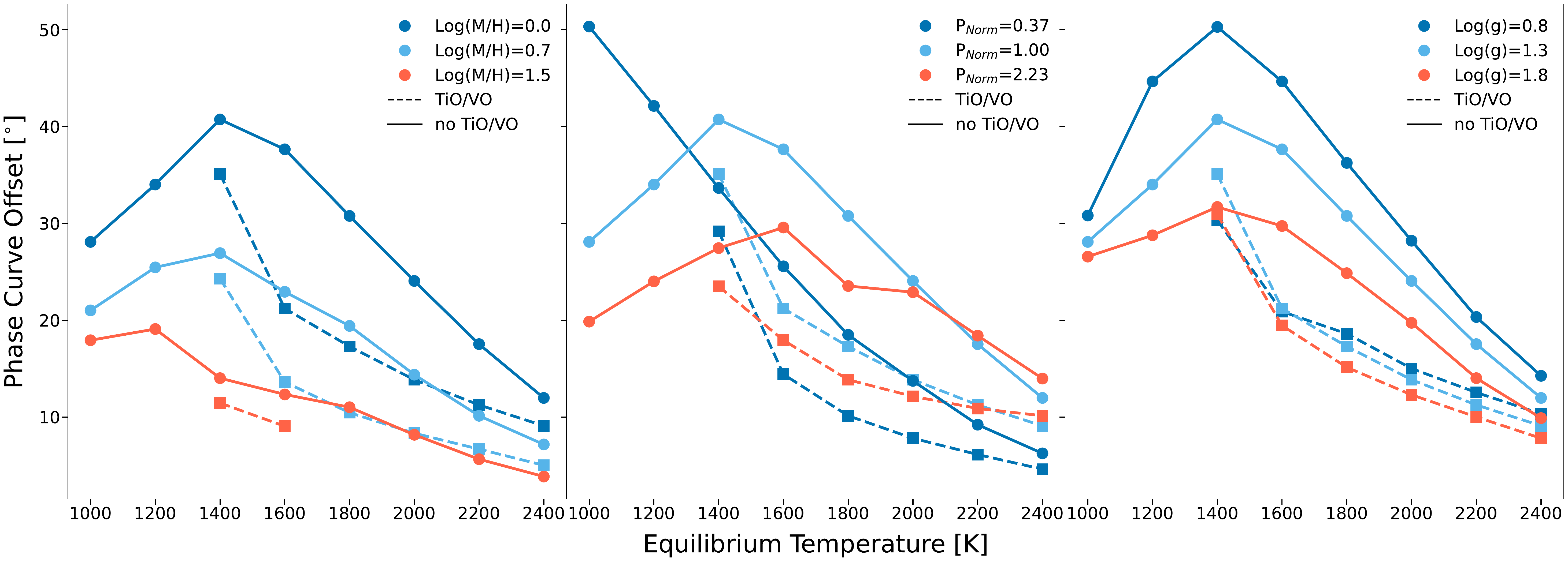}
    \caption{Phase curve offset for a subset of models within the grid. Each panel displays the variation in offset with $T_{\rm eq}$ due to one of: orbital period, surface gravity and metallicity, whilst the other two parameters are kept constant (when constant Log(M/H)=0.0, P$_{\rm Norm}$=1, log(g)=1.3). Cross-terms between parameters are not included. \textbf{Left Panel:} Variation in offset whilst changing metallicity. \textbf{Middle Panel:} Variation in offset with different normalised orbital period. These tracks corresponds to planets tidally-locked around different stellar types. \textbf{Right Panel:} Variation in offset whilst changing surface gravity.}
    \label{fig:offset_bol_comp}
\end{figure*}

The bolometric phase curve offsets for a subset of our models are displayed in Figure \ref{fig:offset_bol_comp}. For the most part the trends we find can be explained using the same logic as that of the heat redistribution. A reduction in the efficiency of heat redistribution naturally leads to lower phase offsets at higher temperatures and the trends in both increasing metallicity and surface gravity can again be explained by the reduced pressure of the photosphere, where a shorter radiative timescale will result in a lower hot-spot shift from the substellar point. The presence of TiO and VO significantly reduce the phase offset in all cases, again due to the reduction in heat redistribution efficiency caused by the strong near UV/optical absorption in the upper atmosphere of the day-side. As mentioned in Section \ref{sec:redist}, the presence of TiO/VO causes a portion of the energy to be re-emitted directly from the stratosphere instead of the IR-photosphere, which leads to a smaller hot-spot offset.

The phase offset appears to peak between equilibrium temperatures of 1400-1600K, with models at lower temperatures having reduced offsets. This behaviour is also found in \citet{2017ApJ...835..198K} (see their Figure 9). This is due to homogenisation of temperature at these lower equilibrium temperatures, caused by longer radiative timescale, which can be seen in Figure \ref{fig:fullgrid}. Because of this, their phase offset is determined by small, wavenumber-2 variations of the thermal structure. The affect of rotation period on the phase offset (via the implied planetary rotation rate) is more complex, which we discuss in the next section.

\subsubsection{Offset trend with period}
\begin{figure*}
\includegraphics[width=2\columnwidth]{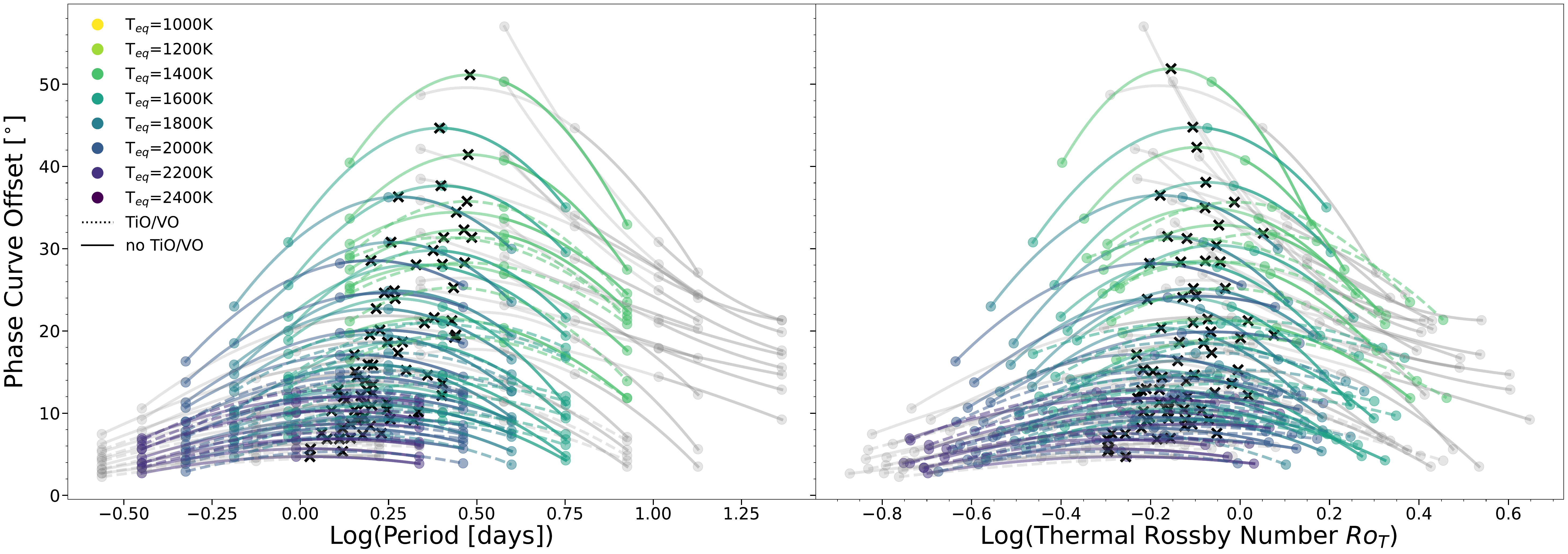}
\caption{\textbf{Left Panel:} Phase curve offset for all models plotted against log(period). \textbf{Right Panel:} Phase curve offset for all models plotted against log($Ro_{T}$). In both plots, models with a larger offset at P$_{\rm Norm}$=1 than at P$_{\rm Norm}$=0.37 or 2.23 are color coded in equilibrium temperature. The black crosses display the maxima of the curve fitted to these models. Any models that do not satisfy this criteria are in grey. Models with TiO and VO are shown in dashed lines and those without in filled lines.}
\label{fig:offset_periodandrossby}
\end{figure*}

As we saw in Section \ref{sec:redist}, larger rotation period leads to more efficient heat redistribution (smaller $f$-factor). Based on simple energy balance models \citep{2011ApJ...729...54C}, we would therefore expect phase curve offsets to increase with increasing period, as was proposed by \citet{2022arXiv220315059M} based on Spitzer phase curve observations. However, as shown in Figure \ref{fig:offset_periodandrossby}, this is only true for small enough rotation periods. Indeed, our models systematically show a turning point, where, past a critical rotation period, the phase curve offset starts \emph{decreasing} with increasing period. This turning point, however, appears to be a function of equilibrium temperature.

We can explain this trend, and the fact that heat redistribution and phase curve offsets have different behaviour, by considering the competition between the divergent and the rotational component of the circulation (as in \citet{2021PNAS..11822705H,2022ApJ...941..171L}) and their impact on the Doppler shifting of Rossby waves. Whereas both components contribute to the heat redistribution, only the rotational component contributes to the phase curve offset. Therefore, planets with a strong divergent component can have a small phase curve offset despite having a significant heat transport efficiency. The position of the peak due to the rotational circulation is approximated by Equation 12 of \citet{2018ApJ...869...65H}, from which we derive the following expression for the longitudinal shape $R(\theta)$ of the wavenumber-1 rotational circulation:
\begin{equation}
\label{eq:phase_shift_approx}
    R(\theta)=\operatorname{\mathbb{R}e} \left\{ \frac{\cos(\theta) + i\sin(\theta)}{(1/\tau_{\mathrm{rad}})-i\left(\omega_m-\bar{U}\right)}\right\}
\end{equation}

\noindent where $\theta$ is longitude (with $\theta=0$ at the substellar point), $\tau_{\mathrm{rad}}$ is the radiative timescale, $\omega_{m}$ is the Rossby wave frequency (which increases with increasing $\Omega$), and $\bar{U}$ is the jet speed.

This equation is valid only in the rotationally dominated regime (e.g. for short rotation periods). It shows the balance between the radiative damping ($1/\tau_{\mathrm{rad}}$ term), that tends to drag the offset towards zero, the Rossby wave ($-i \omega_m$ term), that tends to drive the offset towards negative values, and the jet speed ($+i \bar{U}$ term), that tends to increase the hot spot offset. For a given equilibrium temperature (and thus a given radiative timescale), a change in rotation period would change both terms in the imaginary part of the denominator. In our grid, as the rotation period increases from $P_{\rm norm}=0.37$ to $P_{\rm norm}=1$, the coriolis force decreases and $\omega_m$ decreases. At the same time, the strength of the wind speed stays roughly constant, but the size of the equatorial jet increases (see Figure \ref{fig:fullgrid_zmzws}), so the term $+i \bar{U}$ increases. Overall, this means that the eastward shift from the jet becomes increasingly larger than the westward shift from the Rossby waves, leading to an increase of the eastward shift with increasing period. 

When the normalised period goes from $P_{\rm norm}=$1 to 2.23, both the jet speed and the Rossby wave speed reduce significantly. Therefore the divergent circulation \citep{2021PNAS..11822705H,2022ApJ...941..171L}, which by definition has no offset, becomes more and more dominant compared to the rotational circulation. This leads to an offset that decreases with increasing period. 

In summary we find that for a given equilibrium temperature, at short orbital periods the circulation is dominated by the rotational component, which, because the Rossby wave speed decreases with increasing period while the jet becomes wider, leads to an offset that increases with period. Then, for larger rotation periods, the divergent circulation becomes more and more dominant, leading to an offset that decreases with increasing period.

\begin{figure*}
\includegraphics[width=2\columnwidth]{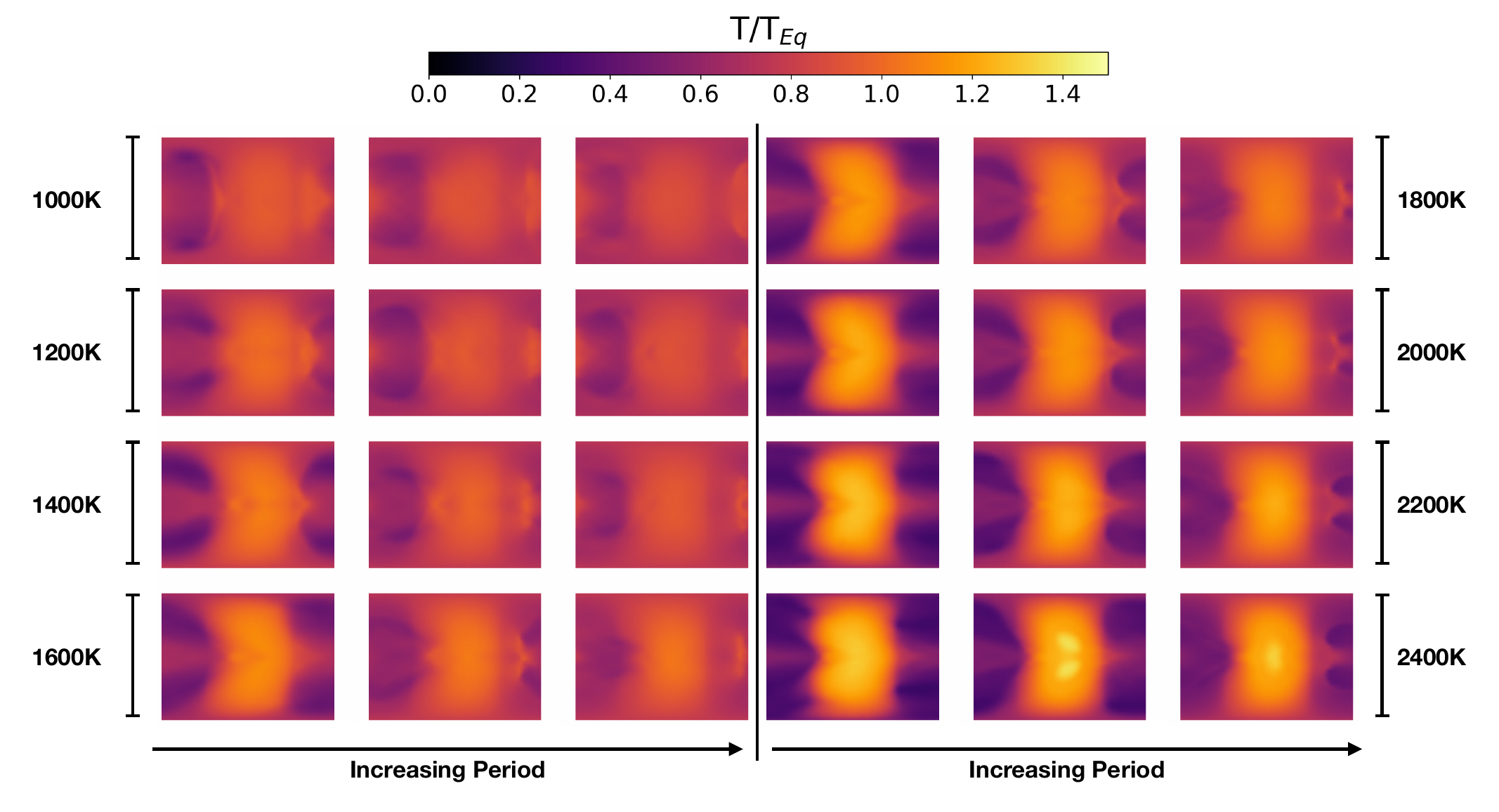}
\caption{Latitude vs longitude thermal structure at the 1.4 micron photosphere for the subset of models displayed with solid lines in the middle panel of Figure \ref{fig:offset_bol_comp}. Equilibrium temperature increases down columns and orbital period increases along rows.}
\label{fig:thermal_rossby_explain}
\end{figure*}

To illustrate this behaviour, the thermal structure at the photosphere for our models at differing periods is plotted in Figure \ref{fig:thermal_rossby_explain}. Taking the 1800K row, in the leftmost plot we can clearly see the jet Doppler-shifting the Rossby wave towards the evening terminator, with the divergent circulation present at higher latitudes. In the centre column, the Doppler shift has increased, shifting the overall hot-spot further eastward. In the rightmost column, the magnitude of the rotation has now decreased to the point where the divergent circulation is now pulling the hot-spot back towards $0^{\circ}$. 

Figure \ref{fig:offset_periodandrossby} shows that the maximum phase offset occurs at different orbital periods for different equilibrium temperatures. We can explain this by estimating that the peak phase shift will occur when $\bar{U}$ is larger than $\omega_m$ in Equation \ref{eq:phase_shift_approx}. We estimate the ratio of these two quantities as the thermal Rossby number \citep{2010JGRE..11512008M}, which at the equator is
\begin{equation}
\label{eq:rot}
    Ro_{T}=\frac{U}{2\Omega L},
\end{equation}

\noindent where $\Omega$ is the rotation rate, $L$ is the characteristic length scale (for which we use the planetary radius), and $U$ is the characteristic velocity scale (for which we use the jet speed).

The right panel of Figure \ref{fig:offset_periodandrossby} displays the phase curve offset plotted against $Ro_{T}$. For those sets of models which display turning points, the maxima of the fitted curves are all found between values of $Ro_{T}\sim0.5-1$, supporting the argument that the rotational phase shift depends on the balance of the jet speed and Coriolis parameter. This trend breaks down for lower equilibrium temperatures. However, as explained in the previous section, these planets have a very uniform brightness temperature due to their long radiative timescale, and so their phase offset is determined by small variations in the thermal structure.

It should be noted that although all of the offset values calculated for our model grid are positive, negative values of the phase curve offset have been previously observed in certain wavelength bands \citep{2015ApJ...804..150E,2018NatAs...2..220D}. The mechanisms causing westwards offsets are currently unclear, although MHD has been suggested as a potential reason \citep{2014ApJ...794..132R,2021ApJ...922..176H}. 

\subsubsection{Combined parameter trends: offset}
\begin{figure*}
	\includegraphics[width=2\columnwidth]{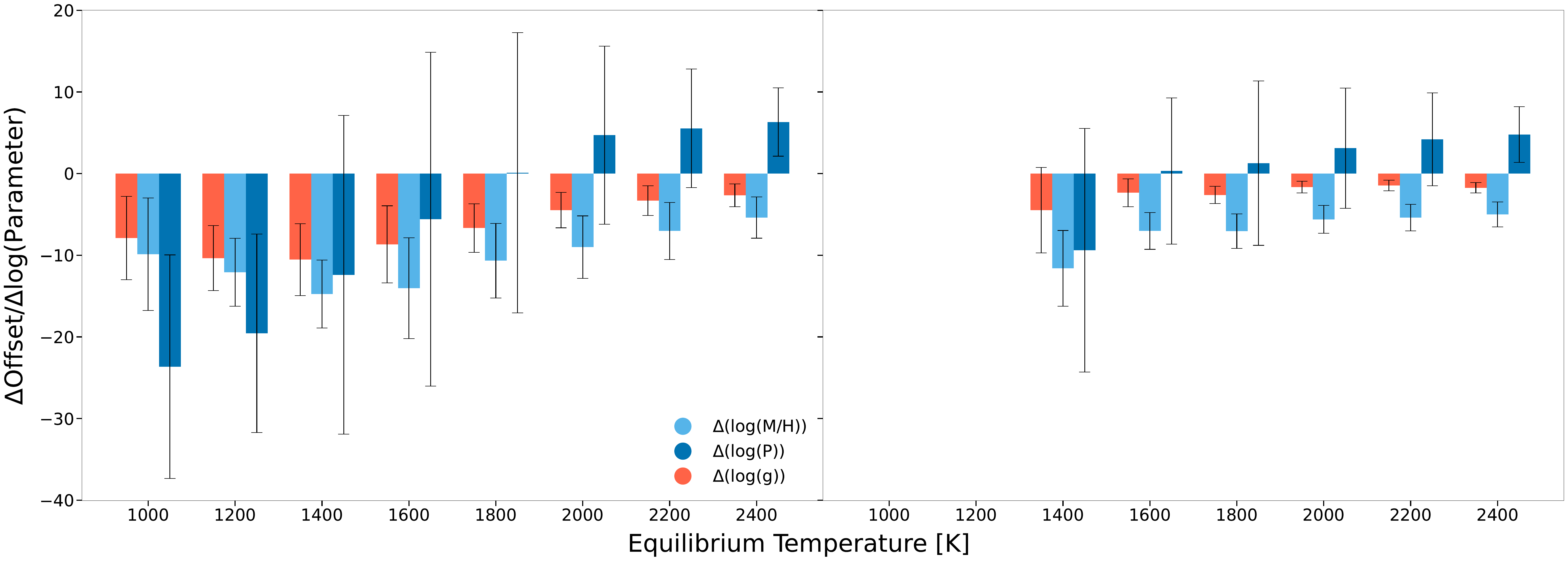}
    \caption{Relative impact of each planetary parameter on phase curve offset. The y-axis of each plot is the delta in the observable feature divided by the delta in the log of each planetary parameter. The x-axis is the Equilibrium temperature. Light blue denotes the change due to metallicity, dark blue the change due to period and red the change due to surface gravity. The error bars display the standard deviation in the impact of each parameter. \textbf{Left Panel:} Models with TiO/VO removed. \textbf{Right Panel:} Models containing TiO/VO}
    \label{fig:param_delta_offset}
\end{figure*}

To investigate the combined effect of the parameters we perform the same calculation from Section \ref{sec:delta_params} for the phase curve offset. The mean and standard deviation for each parameter $\Delta$ is shown in Figure \ref{fig:param_delta_offset}. The offset, as expected, has a much more complex dependency on the parameters. Increasing both surface gravity and metallicity reduce the offset across the board, with the metallicity having the larger impact of the two. Whereas increasing period first increases the offset then reduces it, as explained in the previous section. However, the standard deviation in the impact of period is large, indicating that cross parameters can drastically affect this relationship.

Phase curve offsets are most sensitive to gravity, metallicity and periods in the 1000-1600K range. The same trends can be seen in the models containing TiO/VO, although notably here the offset is much less significantly affected by the parameters, due to the substellar temperature inversion having a dominating effect on the hotspot position.

\subsubsection{Amplitude and offset are set by different mechanisms}
\begin{figure}
	\includegraphics[width=\columnwidth]{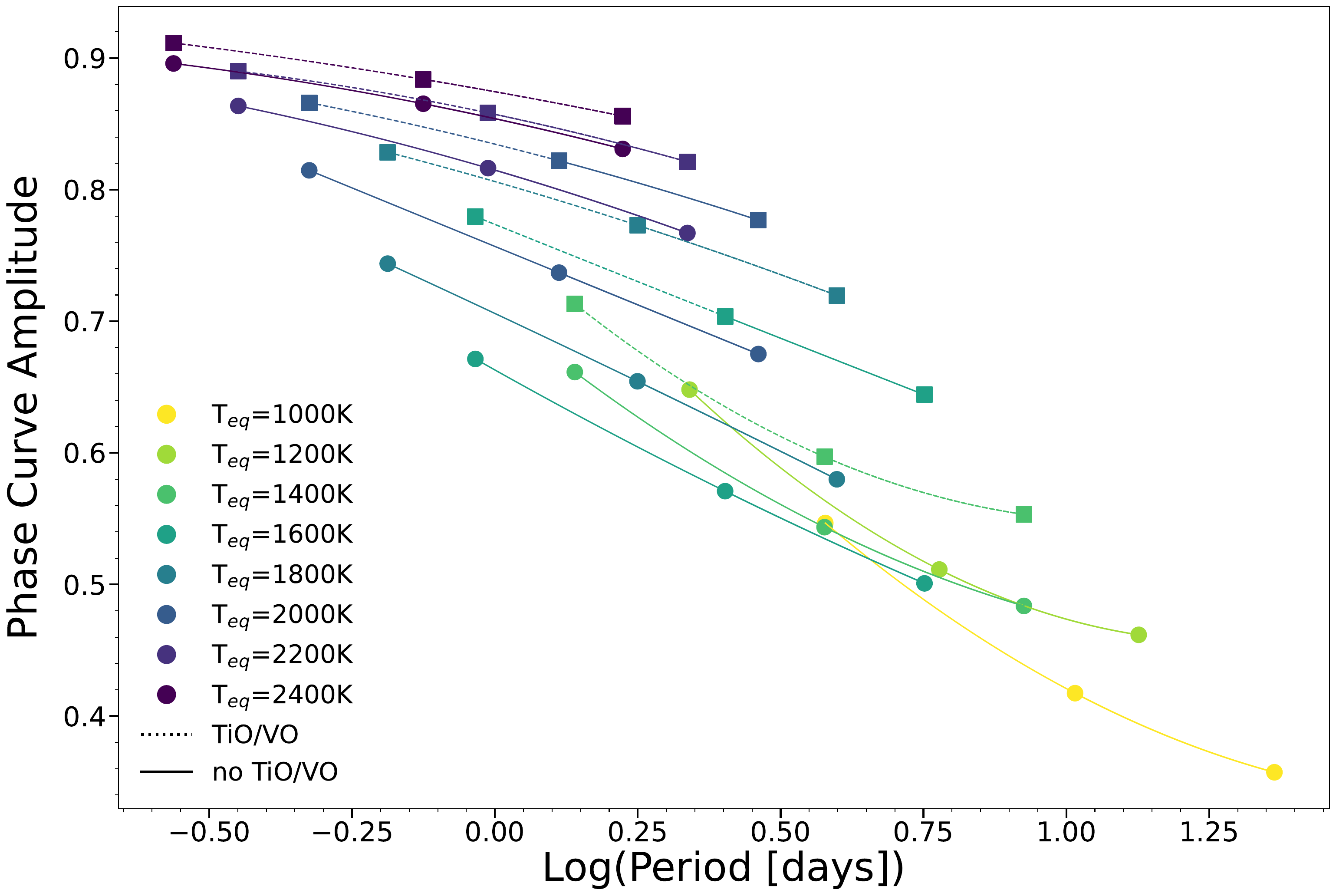}
    \caption{Phase curve amplitude for the subset of models shown in Figure \ref{fig:thermal_rossby_explain} plotted against log(period).}
    \label{fig:ampvp}
\end{figure}
\begin{figure}
	\includegraphics[width=\columnwidth]{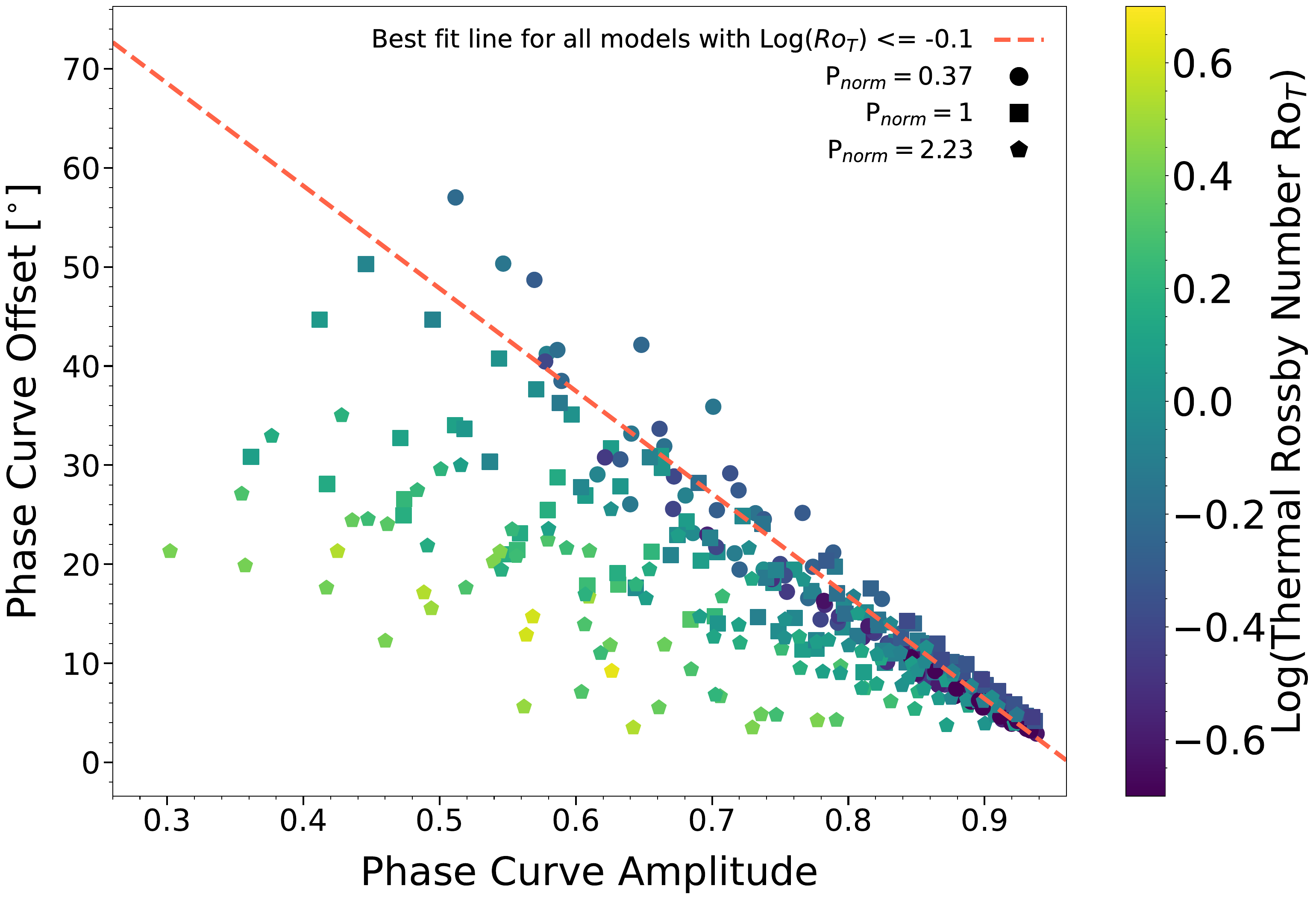}
    \caption{Phase curve offset plotted against amplitude for all models within the grid. Colour denotes the thermal Rossby number of the model and shape denotes the normalised period. The red dashed line displays the best fit to the models with value of log($Ro_{T}$)$\leq$-0.1.}
    \label{fig:ampvoff}
\end{figure}
The relative amplitude is a measure of the maximum brightness contrast between hemispheres, and so is intrinsically linked to the heat redistribution which we discuss in Section \ref{sec:redist}. It has been defined in a few different ways throughout literature. When investigating atmospheric energy balance (i.e \citep{2012ApJ...747...82C,2015MNRAS.449.4192S,2016ApJ...821...16K,2017ApJ...835..198K}) it is often defined as the day-night brightness temperature contrast. Here we define the relative amplitude in terms of the phase curve fluxes \citep{2013ApJ...776..134P,2018haex.bookE.116P},
\begin{equation}
A = \frac{F_{p,\mathrm{max}}-F_{p,\mathrm{min}}}{F_{p,\mathrm{max}}}.
\end{equation}

The less efficient that day-to-night heat redistribution is, the larger the phase curve amplitude is expected to be. Because of this, it follows the same trends as seen in Figure \ref{fig:f_bol_tracks} for the $f$-factor: increasing with temperature \citep{2016ApJ...821...16K,2020A&A...636A..66P}, metallicity and surface gravity, but decreasing with orbital period. A figure displaying this behaviour can be found in Appendix \ref{sec:amp_appendix}.

Energy balance models based on advection of heat by a jet would lead to the conclusion that phase curve offset and amplitude should correlate with each other, with high winds expected to produce efficient heat redistribution and a large phase shift. However, as seen in Figure \ref{fig:ampvp}, our models do not see a systematic correlation between phase curve amplitude and phase curve offset

This highlights that phase curve offset and phase curve amplitude are not set by the same component of the atmospheric circulation. The amplitude of the phase curve increases when heat redistribution becomes less efficient. As both the jet and the day-to-night flow contribute to the heat transport, the phase curve amplitude is sensitive to the sum of the rotational and divergent components of the circulation. On the contrary, rotational and divergent parts of the circulation affect the phase curve offset in opposite directions. The rotational part tends to increase the offset whereas the divergent part tends to reduce it. As a consequence, the offset depends on the \emph{relative} strength of divergent and rotational flow whereas the amplitude depends on the \emph{combined} strength of divergent and rotational circulation.

This means that the relationship between phase curve offset and amplitude should not necessarily be linear, which we can see in Figure \ref{fig:ampvoff} where there are two clear regimes in the amplitude vs offset trend based on the thermal Rossby number (discussed in Section \ref{sec:offset}. For planets with short periods ($Ro_{T}\leq1$) the amplitude increases when offset decreases, For those with long periods ($Ro_{T}>1$) it is the opposite. We do see that a linear relationship holds for fast rotating planets, the low ($Ro_{T}\leq1$) branch in Figure 
\ref{fig:ampvoff}, with offset=$-103.5^{\circ}\times$amplitude$+99.6^{\circ}$. However, this relationship does not hold for larger rotation periods (those with $Ro_{T}>1$). The models deviating from the linear trend are those with $P_{\rm norm}=1$ (planets around $\sim$G0 stars, Table \ref{tab:starparams}) between $T_{\rm eq}=1000-1600K$ and those with $P_{\rm norm}=2.23$ (planets around $\sim$F6 stars, Table \ref{tab:starparams}) between $T_{\rm eq}=1000-2200K$. However, we expect the linear relationship between amplitude and offset to hold for planets around $\sim$K2 dwarfs.

There are, however, some processes our models omit which could alter the relationship between phase curve offset and amplitude. Our model does not include H2 dissociation, which can have a non-negligible impact on the phase curve offset and day-night temperature contrast below 2400K \citep{2019ESS.....432603T}. H2 dissociation reduces the day-night contrast and resulting phase curve amplitude, by transporting additional heat from the day to the night-side \citep{2018ApJ...857L..20B,2018RNAAS...2...36K,2021MNRAS.505.4515R}. This can reduce the speed of winds at a constant rotation period and reduce the phase-offset. Including this additional heating source could change the non-monotonic trend found between offset and rotation period to a linearly increasing trend for high temperature models. However, the effects of the H2 dissociation are greatly lessened in the presence of TiO/VO, which are likely to be present in the atmosphere at these high temperatures. Additionally, \citet{2022AJ....163...35B,2017NatAs...1E.131R,2014ApJ...794..132R} show that MHD can affect hot Jupiter atmospheres from $\sim$1500K. However, these effects do not become very strong until $\sim$2000K when there is sufficient coupling between the atmospheric magnetic field and circulation. As the magnetic drag timescale decreases the flow structure can be altered, initially providing pure drag then moving to oscillatory mean flows and eventually stationary westward mean flows. This could cause time variable winds and result in variable, or even westward, phase curve offsets. For the high temperature models, this would likely reduce the phase curve offsets, potentially altering the location of the turning point. Our models also do not include any clouds, which have been shown capable of drastically changing the observable features of hot Jupiter atmospheres \citet{2021MNRAS.501...78P}. The presence of clouds, particularly if mostly contained on the night-side, can increase the phase curve amplitude and decrease phase curve offset.

\subsection{Temperature Contrast Between the Limbs}
\begin{figure*}
	\includegraphics[width=2\columnwidth]{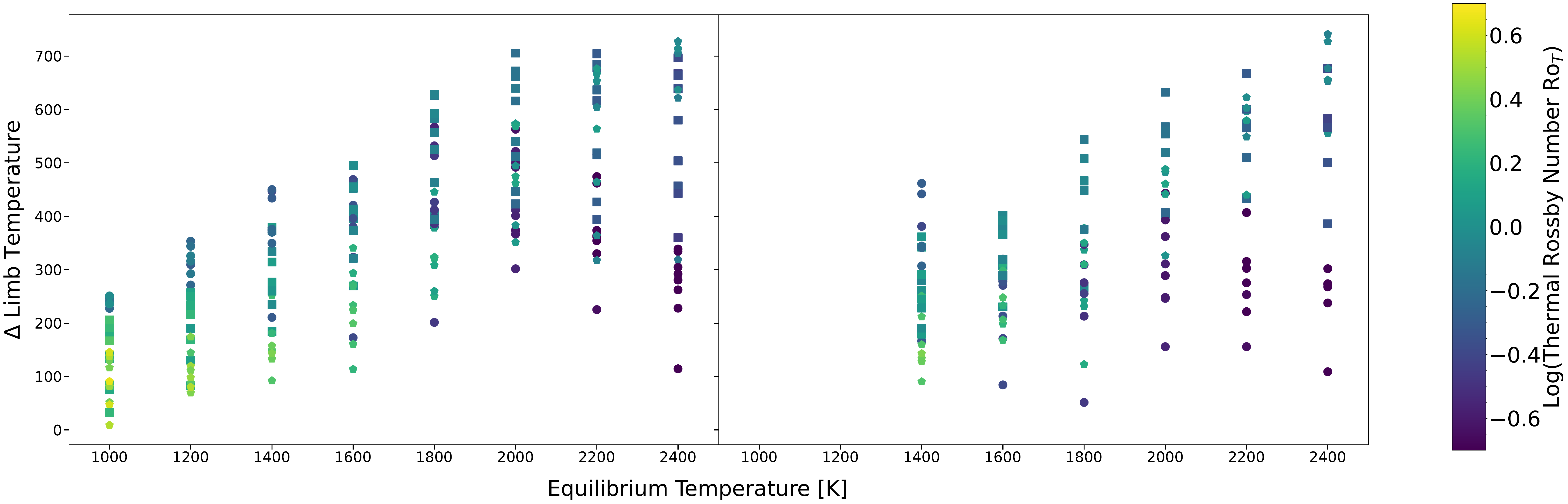}
    \caption{Temperature difference between the evening and morning limb pressure-temperature profiles at the transit photospheric pressure level plotted against equilibrium temperature. \textbf{Left Panel:} Without TiO/VO. \textbf{Right Panel:} With TiO/VO.}
    \label{fig:limbtempdif}
\end{figure*}
During transit spectroscopy one probes the limb of the planet. Limb-to-limb variation in temperature can lead to variations in chemistry and cloud coverage. All of these lead to potential bias when interpreting transmission spectra \citep{2016ApJ...820...78L,2017MNRAS.469.1979M,2020A&A...636A..66P} and can have specific signatures in the transit light curve both at low spectral resolution \citep{2016ApJ...820...78L} and at high spectral resolution \citep{2020Natur.580..597E,2021MNRAS.506.1258W}.

Figure \ref{fig:limbtempdif} shows the temperature contrast between the east and west limbs of our models at the photospheric pressure level, corrected for the transit chord optical depth by the geometric factor $\sqrt{2\pi R_{p}/H}$ \citep{2005MNRAS.364..649F}. The temperature is calculated as the average over all latitudes and $\pm 20^{\circ}$ of longitude, in order to take into account the line-of-sight effect and the planet rotation during transit \citep{2022MNRAS.510..620W}. A positive value here denotes an evening limb that is hotter than the morning limb. We can see that the temperature contrast does become increasingly large, although the upwards trend begins to flatten out at higher equilibrium temperatures. This is due to the combined affects of both the heat redistribution and hotspot offset. In Section \ref{sec:redist} we see that heat redistribution becomes increasingly less efficient at higher equilibrium temperatures, which will naturally lead to a hotter evening limb and colder morning limb. Then, in Section \ref{sec:pc} we see that the phase curve offset, which tracks the hotspot offset in the case of a cloudless atmosphere, reduces with equilibrium temperature. This will naturally lead to a more symmetrical structure and thus a smaller east/west contrast.

As the phase offset peaks for models with $Ro_{T}\sim1$, models with $Ro_{T}<1$ or $>1$ will have lower phase offset and therefore more symmetrical atmospheres. At low equilibrium temperatures ($T_{\rm eq}<$1600K) the heat redistribution efficiency is strong and most models have $Ro_{T}>1$, so the limb temperature contrast is small. As the equilibrium temperature increases, the efficiency of heat redistribution reduces, but the offset also reduces. In our models we find the limb temperature difference starts to increase. Implying that the reduction to heat redistribution with equilibrium temperature must be more impactful on the limb temperatures in this regime than the offset reducing with equilibrium temperature. This behaviour is maintained between 1000-2000k. Above 2000K, the heat redistribution trend with temperature flattens out as it reaches close to its minimum (e.g large $f$-factor). However, the offset is still reducing with $T_{\rm eq}$, so the limb temperature difference stops increasing for models with $Ro_{T}\sim1$ and even decreases for those with $Ro_{T}<1$.

In the presence of TiO/VO, the heat redistribution is much worse at lower temperatures. The offset is also significantly reduced. From the right panel of Figure \ref{fig:limbtempdif}, we can see that in this case the relationship between the limb temperature contrast appears to flatten at much lower temperatures, which is likely due to the decrease in variation for both the heat redistribution and offset when TiO/VO are present.

\section{Comparison with Observations}
\label{sec:obs}
We now look at two different types of population level eclipse observations: the population of planets observed with the Spitzer Space Telescope and the one observed with the Wide Field Camera 3 on board the Hubble Space Telescope and compare them to our population of modelled planets. 

\subsection{Spitzer Band Secondary Eclipse}
\label{sec:demtransition}
\begin{figure*}
	\includegraphics[width=2\columnwidth]{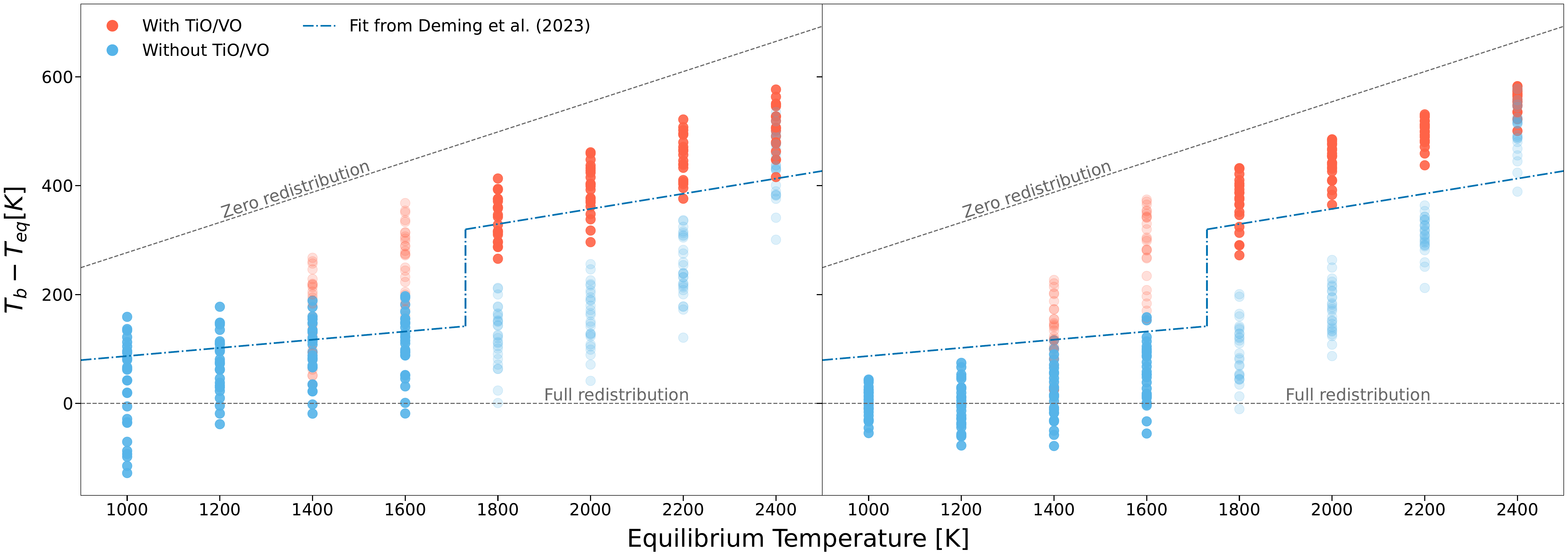}
    \caption{Comparison between the Spitzer band-pass brightness temperatures and model equilibrium temperatures. The opacity of points containing TiO/VO with equilibrium temperatures below the transition temperature found in \citet{2023arXiv230103639D}, and points not containing TiO/VO above the the transition temperature, have been reduced to emphasise the transition. The blue dashed line correspond to the fits seen in Figure 3 of \citet{2023arXiv230103639D}. \textbf{Left Panel:} Spitzer 1 (3.6 microns). \textbf{Right Panel:} Spitzer 2 (4.5 microns).}
    \label{fig:tdem}
\end{figure*}
\begin{figure*}
	\includegraphics[width=2\columnwidth]{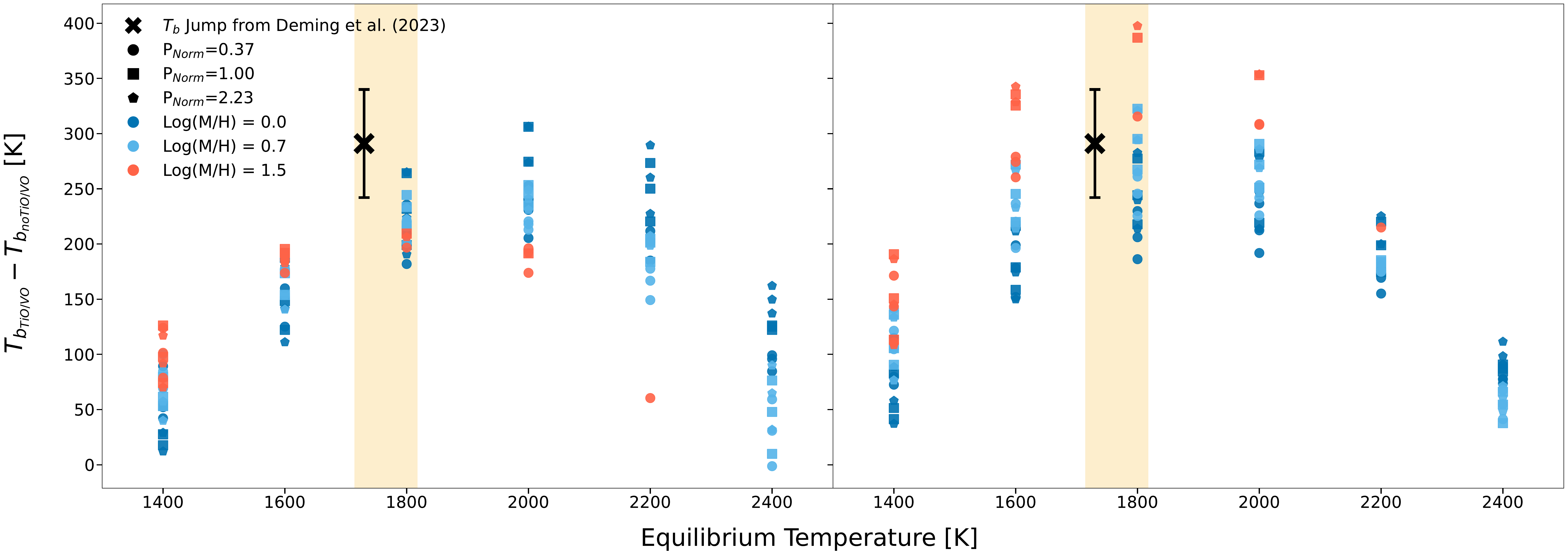}
    \caption{Difference between the day-side brightness temperature for models containing and not containing TiO/VO. The yellow shaded region displays the temperature range of the abrupt increase in emergent flux from \citet{2023arXiv230103639D}. The black cross shows the exact value found in \citet{2023arXiv230103639D}. \textbf{Left Panel:} Spitzer 1 (3.6 microns).\textbf{Right Panel:} Spitzer 2 (4.5 microns).}
    \label{fig:tdem_comp}
\end{figure*}

The band-averaged measurements provide the brightness temperature of the planet day-side in a given band-pass, which can be used to measure a "band averaged" heat redistribution parameter. Observations of exoplanet atmospheres do not generally probe the bolometric quantities described above, but band averaged or spectrally resolved properties of the atmosphere. Because different band-passes probe different atmospheric depth, the band-averaged quantities can be different from the bolometric ones. Thus we will now look at the model predictions for the Spitzer eclipse measurements.

Over its lifetime Spitzer observed 457 eclipses of 122 hot Jupiters in it's 3.6 and 4.5 micron observing modes \citep{2023arXiv230103639D}. Eclipses from both channels spectra, where CH4 and CO features are expected to be present depending on the planets equilibrium temperature, have been re-analysed in homogeneous ways by several authors \citep{2020A&A...639A..36B,2020AJ....159..137G,2023arXiv230103639D}. The distribution of planet parameters observed matches well with the overall transiting planet population, as seen in Figure \ref{fig:param_space}, making it a good representation for diversity in population level studies. \citet{2020A&A...639A..36B} find that there is a transition between seeing CO in absorption to CO in emission at $\rm T_{eq}=1600\pm100K$, and are the first to attribute this switch to a transition between non-inverted and inverted temperature profiles. We hereafter use the recent re-analyses of \citet{2023arXiv230103639D}, because it is the most recent and comprehensive one. \citet{2023arXiv230103639D} identify a statistically significant unexpected and sudden rise in the brightness temperatures of the population, with the brightness temperature increasing by 291$\pm$41 K between equilibrium temperatures of 1714-1818 K. \citet{2023arXiv230103639D} propose a few explanation to explain this sudden rise in brightness temperature, such as the onset of magnetic drag which would inhibit longitudinal heat redistribution or the rapid dissipation of day-side clouds at these temperatures.

Here we propose a novel mechanism to explain this rise in temperature: that the sudden appearance of TiO and VO changes the ability of the atmosphere to redistribute heat. Indeed, as shown in Figure \ref{fig:ffactorsall}, the presence of TiO and VO in a planetary atmosphere reduces the heat transport and thus increases the heat redistribution factor ($\Delta f_{\rm avg}=0.216$). We will now look at the specific effect of TiO/VO on the brightness temperature of the Spitzer band-pass.

For this, we calculate the brightness temperature in the Spitzer band-pass in our grid of models by first fitting a blackbody to the spectral region of the Spitzer band-passes, we can then use the temperature of this blackbody as the brightness temperature and compare with the equilibrium temperature of our models. 

We show the difference between day-side brightness temperature and equilibrium temperature in Figure \ref{fig:tdem}. This difference would be zero for a planet with full heat redistribution and would be $(2.66^{1/4}-1)*T_{\rm eq}$ for a planet with no heat redistribution. In between these two extremes, the band averaged redistribution value depends on both the day/night heat transport and on the spatially varying chemical composition of the planet, thus sometimes leading to negative values \citep{2017ApJ...851L..26D}. Figure \ref{fig:tdem} shows two tracks for the brightness temperature versus equilibrium temperature, corresponding to models in chemical equilibrium and to models where TiO/VO have been artificially removed from the atmosphere. Both tracks have a brightness temperature that increases relative to the equilibrium temperature as the equilibrium temperature gets larger. This is a direct consequence of the heat redistribution parameter that increases with equilibrium temperature, as shown in Figure \ref{fig:ffactorsall}. 

Additionally, we can see that the with and without TiO/VO track are offset by, approximately, 50-300K depending on the equilibrium temperature. This was discussed in Section \ref{sec:redist} and is due to TiO/VO intercepting and re-radiating part of the stellar flux at very low pressures, where the radiative timescales are very short. Overall, this reduces the heat transport to the planetary night-side and increases the planet brightness temperature. 

Figure \ref{fig:tdem} also compares our population of models with the best fit to the population of planets proposed by \citet{2023arXiv230103639D}, which consist of a rising slope at low equilibrium temperature, a jump around 1714-1818 K and a gentle slope at high equilibrium temperature. Without any adjustment, we can see that the best fit to the observed population of planets falls right inside our population of models, with the low temperature fit corresponding to the no TiO/VO track and the high temperature fit corresponding to the TiO/VO track. We do note, however, that towards very high temperatures the \citet{2023arXiv230103639D} fit lies slightly below the TiO/VO track. Given that the parameter distribution of the Spitzer observations is representative of the wider hot Jupiter population (see Figure \ref{fig:param_space} for this comparison), and not weighted towards planets with parameters that would give lower values here, one possible cause is partial depletion of TiO/VO in these atmospheres, which we discuss with more detail in Section \ref{sec:specindex}.

We now show in Figure \ref{fig:tdem_comp} the brightness temperature difference between models with and without TiO/VO, with all other parameters being constant (e.g. size of the "jump" in Figure \ref{fig:tdem} between equivalent models). We see that the position and amplitude of the jump observed by \citet{2023arXiv230103639D} (black cross in Figure \ref{fig:tdem_comp}) aligns very well with the model predictions if the population of planets jumps from the no TiO/VO track to the TiO/VO track around 1800K. We therefore conclude that the appearance of TiO/VO between 1714 K and 1818 K can explain the jump in brightness temperature observed in the population of hot Jupiters. 

This would align well with observational evidence suggesting the appearance of gaseous TiO/VO in atmospheres at around this temperature range. \citet{2016ApJ...828...22P}, for instance, predict the transition to occur at $\leq$1900K based on the large apparent albedos for Kepler-76b and HAT-P-7b, and better fitting of Kepler light-curves for planets above this temperature. This transitional temperature is consistent with cold trap models \citep{2017AJ....154..158B,2016ApJ...828...22P,2013A&A...558A..91P} and the temperature predicted for deep cold traps to possibly occur from micro-physical cloud models \citep{2018ApJ...860...18P}.

We further note that, although our modelled planets have a large scatter along the temperature trend ($\sigma=\sim104$K, which increases to $\sigma=\sim166$K when the error bars from the observed Spitzer population are included), this scatter is significantly smaller than the one seen in the population of Spitzer data points ($\sigma\sim221$K). We conclude that either additional mechanisms lead to an additional scatter around the mean value compared to our expectation (e.g. chemical disequilibrium, presence of night-side clouds, atmospheric drag), or that the Spitzer IRAC instruments have additional noise that increases the scatter of the population \citep{2014MNRAS.444.3632H}.
\begin{figure}
	\includegraphics[width=1\columnwidth]{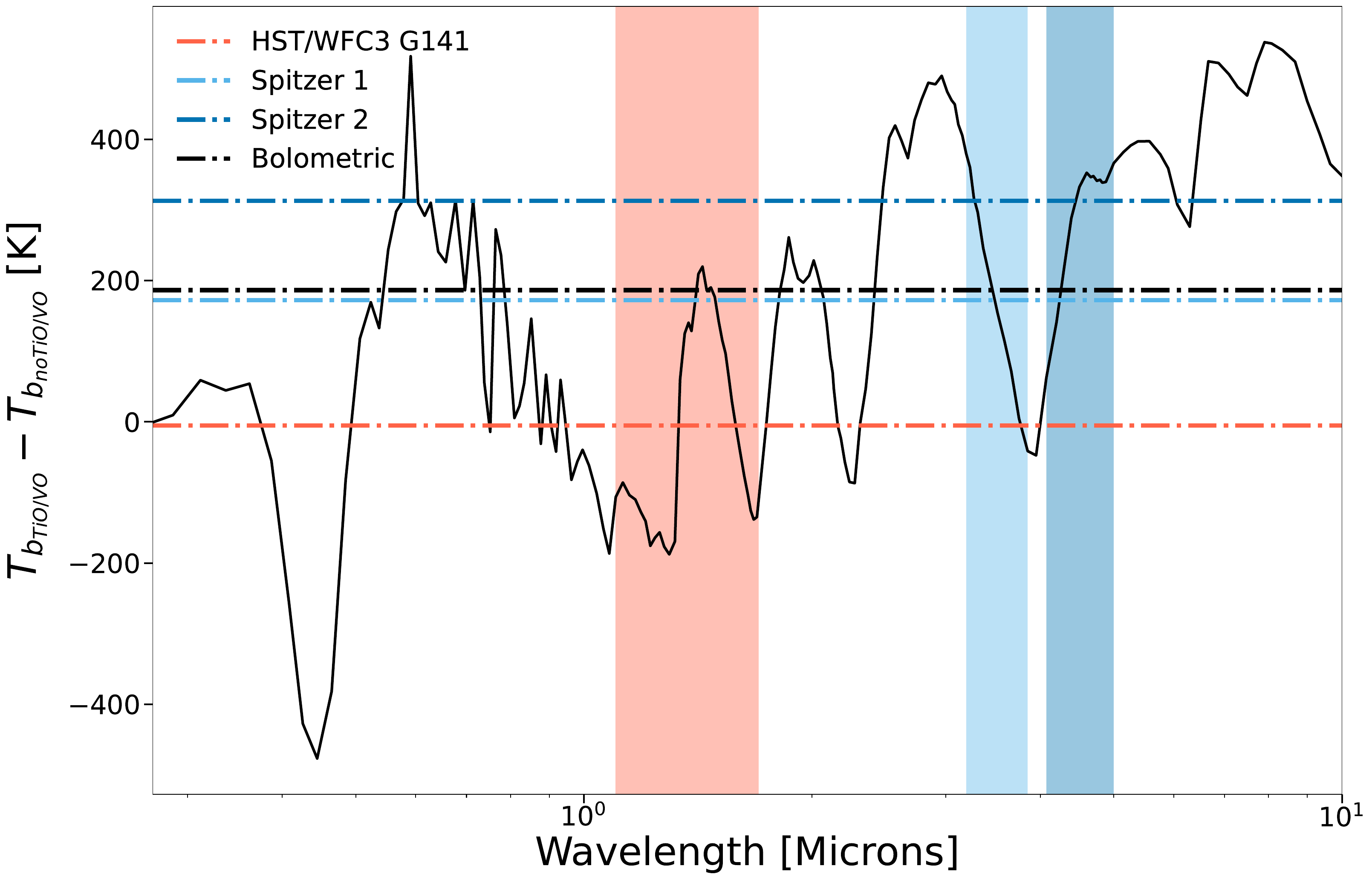}
    \caption{This figure shows $T_{b_{\rm TiO/VO}}-T_{b_{\rm noTiO/VO}}$ for an 1800K model over all wavelengths. The red, light blue and dark blue shaded regions show the HST/WFC3, Spitzer 1 and Spitzer 2 band-passes respectively. Dashed lines of the same colors represent the average value within each band-pass.}
    \label{fig:tdem_wl_single}
\end{figure}

We now wonder whether this jump should be seen with other instruments, such as HST/WFC3. We show in Figure \ref{fig:tdem_wl_single} the difference between the brightness temperature minus the equilibrium temperature for a single model at 1800K with and without TiO/VO as a function of wavelength. We can see that the size of the jump is highly dependent on the wavelength. This is because different wavelengths probe different depths in the atmosphere. Because TiO/VO changes both the heat redistribution and the vertical structure of the thermal profile, at some wavelengths (or some pressure levels) the reduction of the heat redistribution due to the presence of TiO/VO is compensated by the cooling of the atmosphere below the thermal inversion due to the anti-greenhouse effect \citep{2014A&A...562A.133P}. This is particularly prevalent in the WFC3 band-pass, where no specific jump in the population would be expected.

We would therefore predict that if a similar population study using HST/WFC3 observations were to be carried out, then the brightness temperature increase would be absent altogether. However, over a wavelength range such as that probed by the JWST/NIRSPEC/G395 instrument, 2.87–5.14$\mu m$, the signature of this transition would be larger.

\subsection{Spectral Feature Strength}
\label{sec:specindex}
\begin{figure*}
	\includegraphics[width=2\columnwidth]{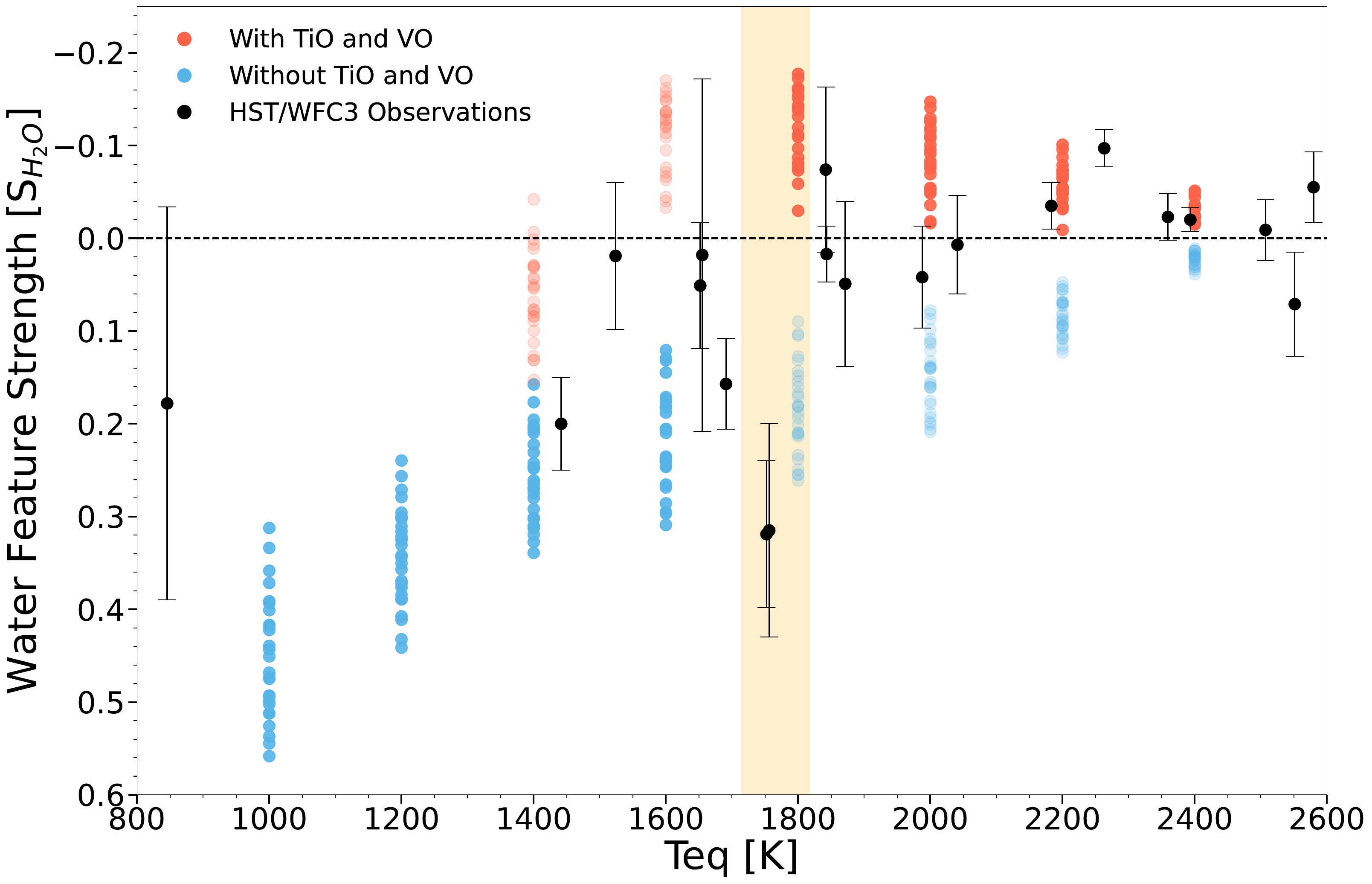}
    \caption{This figure shows a comparison between the water feature strength calculated for the 3D GCM grid and observational data. The opacity of points containing TiO/VO with equilibrium temperatures below the transition temperature found in \citet{2023arXiv230103639D}, and points not containing TiO/VO above the the transition temperature, have been reduced to emphasise the transition. The blue points are models without TiO/VO, the red points are the models with TiO/VO and the black points with error-bars are the observational data from \citet{2021NatAs...5.1224M,2022AJ....163..261M,2022ApJ...925L...3F}. The yellow shaded region displays the temperature range of the abrupt increase in emergent flux from \citet{2023arXiv230103639D}}
    \label{fig:sh2o}
\end{figure*}
The thermal flux emitted by a planet's day-side can be measured by observing its eclipse behind the star. When the spectroscopic eclipse is measured, the spectra can provide insights into the thermal profile of the planet. In particular, the spectra can be used to determine whether the thermal profile is increasing with decreasing pressure (an inverted atmosphere that produces emission features) or whether it is decreasing with decreasing pressure (a non-inverted atmosphere with absorption features). Therefore, another way to probe the TiO/VO to no TiO/VO transition at the population level is to look at the spectral variation of the water feature within the HST/WFC3 band-pass, where water has a strong molecular feature between 1.35-1.48$\mu m$. When this feature transitions from absorption to emission, it indicates that a temperature inversion is present in the atmosphere. For hotter planets the combination of thermal dissociation and the appearance of H- opacities reduces the spectral feature of water and brings the spectrum close to a blackbody. If the transition from TiO/VO to no TiO/VO happens at cooler temperatures than the onset of dissociation, we would expect a sharp transition between planets with inverted thermal profile and thus emission features (such as \citet{2022A&A...668A..53C,2016ApJ...822L...4E,2023arXiv230108192C}) to planets with non-inverted thermal profile and thus absorption features (such as \citet{2016AJ....152..203L,2014ApJ...793L..27K}). 

The Hubble Space Telescope observed the eclipse of 21 planets with the WFC3 instruments. A spectral index was determined by \citet{2021NatAs...5.1224M} in order to reduce these points (Initially 19 in \citet{2021NatAs...5.1224M}, with later additions in \citet{2022AJ....163..261M,2022ApJ...925L...3F}) into a single index value. The WFC3 index measured the departure of the 1.4 micron feature (usually due to water) to the underlying continuum. To calculate it we first fit a blackbody to two 'out-of-band' regions between 1.22-1.33$\mu m$ and 1.53-1.61$\mu m$, where water absorption in the band-pass is minimal. The temperature of this blackbody is then referred to as the observed day-side temperature, T$_{\rm day}$. In \citet{2021NatAs...5.1224M}, this temperature is plotted against the water feature strength. However, for ease of comparison to Section \ref{sec:demtransition}, we use the equilibrium temperature here. The water feature strength in the 1.35-1.48$\mu m$ 'in-band' region is then calculated by \citep{2021NatAs...5.1224M}:
\begin{equation}
S_{H_{2}O}=\log\left(\frac{F_{\rm blackbody,in-band}}{F_{\rm planet,in-band}}\right).
\end{equation}

\noindent Based on this definition, positive values correspond to a water feature observed in absorption and negative values to a water feature observed in emission.

In Figure \ref{fig:sh2o} the water feature strength for all models within the grid can be seen when compared to observations from HST/WFC3 seen in \citet{2021NatAs...5.1224M,2022AJ....163..261M,2022ApJ...925L...3F}. In our models we see that, when TiO/VO are absent from the models, the water feature is seen in absorption (negative index value) and the strength of the water feature decreases with increasing temperature as the day-side thermal profile becomes more isothermal for more irradiated planets. When TiO/VO are present, the water feature is initially seen in absorption below $\sim$ 1500K, where the TiO/VO begin to condense out of the atmosphere in chemical equilibrium. The feature rapidly swaps to emission above this temperature, with the strength of emission decreasing towards higher temperature models. Both models containing TiO/VO and those without are seen to converge towards the zero line at high day-side temperatures. This occurs as increasing quantities of water are thermally dissociated in the atmosphere at high temperatures and H- opacities start filling the gap in the water bands \citep{2021MNRAS.501...78P}. 

Figure \ref{fig:sh2o} further shows that our models are able to encompass all the observed data-points. Particularly, we see that the spread of the water feature strength index becomes smaller at high equilibrium temperature, where the dissociation of water and the H- opacity reduce the water feature strength in all planets.

Looking in more details, we see that the observations seem to lie in between the TiO/VO and no TiO/VO model tracks for the 1400-2000K equilibrium temperature range. Particularly, we do not see the large emission features that should be prominent for planets in the 1600K-1800K range. Similarly, we do not see the non-inverted water features expected by the no TiO/VO model in that specific range of temperatures.

One possible explanation is that TiO/VO is indeed present in these atmospheres, but at a smaller abundance than in the chemical equilibrium prediction. This could happen due to deep or day/night partial cold trap mechanisms \citep{2009ApJ...699.1487S,2016ApJ...828...22P,2017AJ....154..158B,2023arXiv230608739P}, which are omitted from our 3D models, or to original elemental Ti and V depletion. \citet{2023arXiv230608739P}, for instance, recently detected VO but no TiO in WASP-76b ($T_{\rm eq}=2228$K), and suggest condensation of TiO to be responsible. Such a partial depletion of TiO/VO would naturally bring the models between the TiO/VO and no TiO/VO track in Figure \ref{fig:sh2o}, allowing a closer match to the data.

In Figure \ref{fig:sh2o}, we also highlight the transition temperature between TiO/VO and no TiO/VO expected from the Spitzer data in Section \ref{sec:demtransition}. The current HST data does not show a steep difference in water feature strength around this temperature, however, the step might be too small to be measured given the error bars on the data.

We note that the partial reduction of TiO/VO that we argue can explain the HST data set does not necessarily rule out our explanation for the sharp increase in brightness temperature in the Spitzer data set (see Section \ref{sec:demtransition}). Indeed, the TiO/VO model track in Figure \ref{fig:tdem} lies above the observations and partial depletion of TiO/VO could easily lower the model track and lead to a better agreement with the observations. To determine if this works quantitatively would require models with partial TiO/VO abundances, which are out of the scope of the current paper. Nonetheless, we believe that the current Spitzer data provides compelling argument for future ground-based optical hi-res observations in order to determine the TiO and VO abundances in these atmospheres.

Another possible explanation for the damped features in the HST bandpass would be that the H- abundance is larger than expected by our models. This could be due to an increase in alkali abundances, which are the prime source of electrons for the H-. As for the partial TiO/VO abundance scenario, we leave this for a future study.

\subsection{Spitzer Phase Curve offset and amplitude}
\label{sec:spitzer2_offsetandamp}

Spitzer observed 21 phase curves of hot Jupiters in its 3.6 $\mu m$ and 4.5 $\mu m$ bands. For those planets within the equilibrium temperature bounds of our grid, we take the values of phase curve offset and amplitudes from \citet{2018haex.bookE.116P} with a correction for HAT-P-7b found in \citet{2021MNRAS.501...78P}. For phase curve data published after this we use the values summarised in \citet{2022arXiv220315059M} for Qatar-1b, Qatar-2b, WASP-34b, WASP-52b and WASP-104B, \citet{2019AJ....158..166B} for KELT-1b and \citet{2021MNRAS.504.3316B} for KELT-16b and MASCARA-1b. For CoRoT-2b, we take the offset value from \citet{2018NatAs...2..220D}, but use the re-analysis of \citet{2021MNRAS.504.3316B} to calculate an amplitude. We also exclude HD149458b from our data set, as done in \citet{2018AJ....155...83Z}. For WASP-43b we use the most resent re-analysis of \citet{2020AJ....160..140M}.

\begin{figure*}
	\includegraphics[width=2\columnwidth]{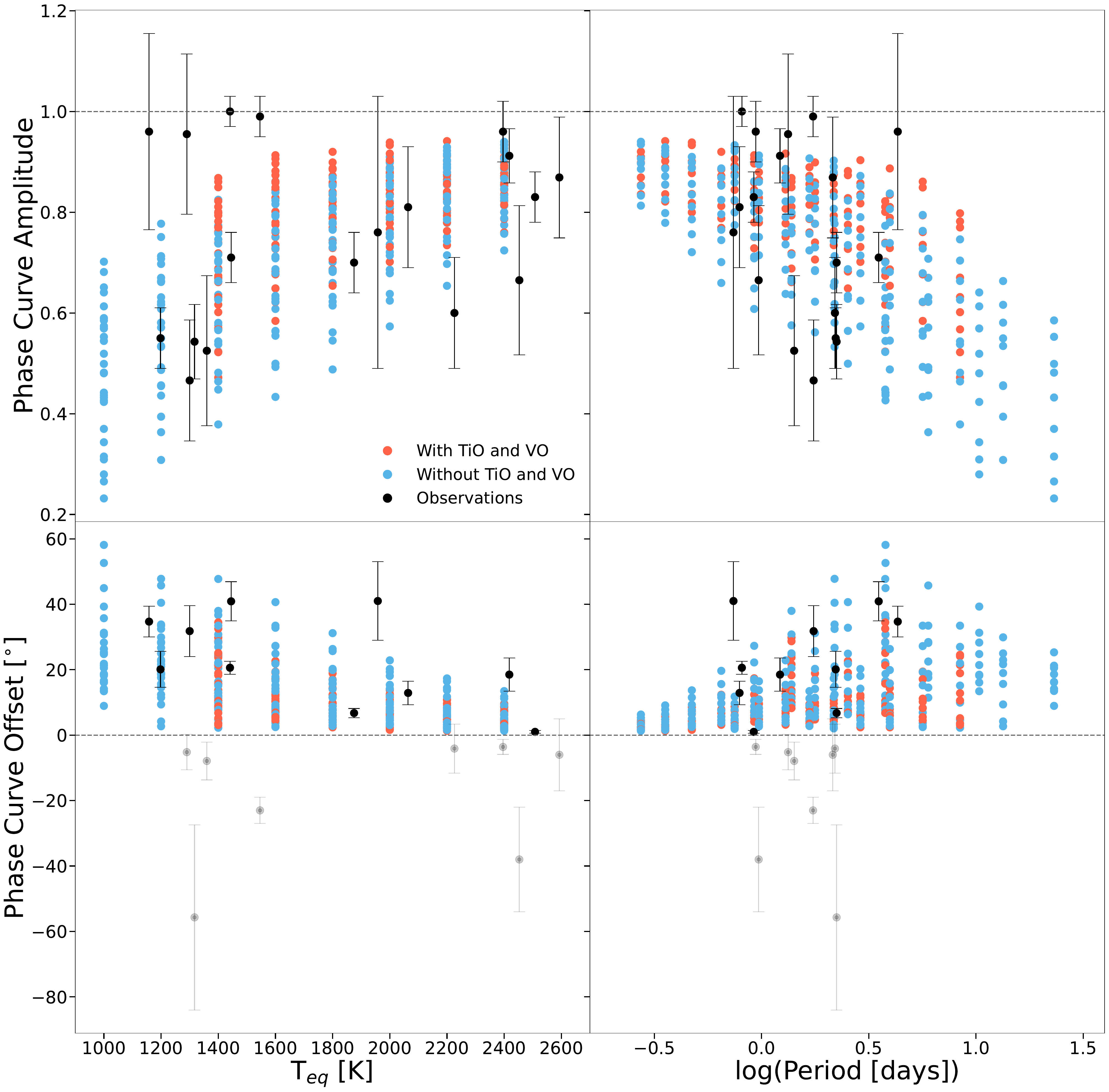}
    \caption{This figure shows a comparison between the phase curve offset and amplitudes in the Spitzer/IRAC $4.5\mu m$ band calculated for the 3D GCM grid and the observational data. The blue points are models without TiO/VO, the red points are the models with TiO/VO and the black points with error-bars are the observational data. \textbf{Left Panels:} shows the offset (lower) and amplitude (upper) vs equilibrium temperature. \textbf{Right Panels:} shows the offset (lower) and amplitude (upper) vs orbital period.}
    \label{fig:pccomp_4.5}
\end{figure*}

Figure \ref{fig:pccomp_4.5} displays the phase curve offsets and amplitudes calculated from the model grid compared to the Spitzer/IRAC 4.5$\mu m$ band observations. The 3.6 $\mu m$ comparison can be found in Appendix \ref{sec:spitzer1_offsetandamp}. The model grid shows the amplitude to increase with the equilibrium temperature and decrease with orbital period. Overall the trend with equilibrium temperature matches the observations quite well, with a few outliers at lower temperatures. These outliers (Qatar-2b, WASP-34b \citet{2022arXiv220315059M}, CoRoT-2b \citet{2021MNRAS.504.3316B} and WASP-43b \citet{2020AJ....160..140M}), with amplitudes close to unity, are likely due to night-side clouds, which are omitted from our models and would significantly reduce the night-side brightness temperature and increase observed amplitude. In particular, for WASP-43b night-side clouds have been postulated to be the reason for the observed amplitude \citep{2015ApJ...801...86K,2020ApJ...890..176V,2021MNRAS.501...78P}. One feature that seem to be present in the data and not in the models is the sudden jump to lower amplitudes for planets with orbital periods larger than 1 day. However we do not have an explanation for this potential jump.

As our model grid omits all processes which could cause negative phase curve offsets, we will focus on comparison to only ones with positive values. The trend in offset with equilibrium temperature is remarkably consistent between the models and observations, decreasing for higher equilibrium temperatures. The observational data also matches the model grid well when compared by period. In our models we clearly see the turning point behaviour detailed in Section \ref{sec:offset}. However, there is no observed data beyond periods of $\sim$4 days for which the offset would begin to noticeably decrease if following the model grid trend.

Overall, from this comparison we can see that the spread of values observed for both phase curve amplitude and (positive) offset can be induced by the varying planetary parameters within our model grid.

\section{Conclusions}
\label{sec:conc}
In this paper, we have created the largest library (345 simulations) of 3D hot Jupiter models to date. These models were produced using the non-grey SPARC/MITgcm global circulation model \citep{2009ApJ...699..564S}, under the assumption of local chemical equilibrium with local rain-out of condensate materials \citep{2006ApJ...648.1181V,2010ApJ...716.1060V}. Phase curves and spectra were further calculated via the plane-parallel radiative transfer code of \citep{1999Icar..138..268M}, at increased spectral resolution. The parameters of equilibrium temperature, orbital period, metallicity, surface gravity were all varied throughout the grid, including cross-variation. In addition, models with the strong photo-absorbing molecules TiO and VO artificially removed from chemical equilibrium were also run. These intrinsic planetary properties were all varied systematically and jointly over the parameters space.

We show that the day-night heat redistribution in hot Jupiter atmospheres is as sensitive to the combined effects of atmospheric metallicity, rotation period and surface gravity as it is to the equilibrium temperature. We therefore expect a large scatter along the general trend of increasing heat redistribution with temperature. As a consequence, a large sample of planets will be needed to measure any trends in the hot Jupiter population by observation, unless one can control some of these parameters through careful target selection.

We find that the relationship between phase curve offset and orbital period is non-monotonic within our models. With an offset that generally increases with rotation period for slowly rotating planets and decreases with rotation period for fast rotating planets. We propose that the turning point is due to the balance between the rotational circulation (i.e. the jet) and the overturning circulation, and can be characterised by the thermal Rossby number of the atmosphere (see Equation \ref{eq:rot}). This relationship breaks down for low equilibrium temperature models which are in a different circulation regime than the hotter planets.

We further find that phase curve offset and phase curve amplitude do not necessarily track each other. This is because the rotational and divergent part of the atmosphere both advect heat from the day to the night-side of the planet, but have counteracting effect on the hot spot offset. This competition between the rotational and divergent effects create two clear regimes within our models. For those with low periods ($Ro_{T}\leq1$) the amplitude increases when offset decreases, and those with long periods ($Ro_{T}>1$) the opposite is found. This produces a scatter in the relationship between the offset and amplitude.

We find that the combined effects of the heat redistribution and phase shift also impacts the temperature contrast between the planetary limbs. As heat redistribution becomes increasingly less efficient at higher temperatures, the evening limb becomes increasingly hotter than the morning limb, increasing the east/west temperature contrast. For the hottest planets, however, the reduced hot spot shift reduces the limb temperature asymmetries. We predict the largest limb-to-limb temperature differences for planets with $T_{\rm eq}\approx 2000K$.

By comparing our spectra to the re-analysed Spitzer/IRAC secondary eclipse data from \citet{2023arXiv230103639D}, we propose a novel mechanism that explains quantitatively the abrupt increase in the brightness temperature seen in this population: the appearance of TiO/VO in the atmosphere at $T_{\rm eq}\approx 1800K$ reduces the efficiency of heat transport and creates a sudden brightening of the flux in the Spitzer band-pass.

Further comparing our models to the spectral variation of the water feature strength observed within the HST/WFC3 band-pass \citep{2021NatAs...5.1224M}, we find that most observation between 1400-2200K are poorly matched by our grid of models. We suggest that this could be due to the partial depletion of TiO/VO due to non local condensation processes. We leave for future work whether an atmosphere partially depleted of TiO/VO can indeed fit both Hubble and Spitzer population level trends. 

Finally, by looking at observed phase curve parameters, we find that the intrinsic scatter of our model grid aligns well with the observed spread in the data. We also find that the trend that phase curve amplitude increases with equilibrium temperature is well matched by our models. However, we do not predict the possible sudden jump towards lower phase curve amplitudes for planets with period larger than 1 day. We also find that the offset trends in both equilibrium temperature and period are well matched with our models, although our model is unable to reproduce any of the observed negative offsets.

Overall, this GCM library is a powerful tool that can be used to investigate the the effects of individual parameters, and a combination of parameters, on key observable features. There are many further avenues to explore using these models, including a deeper look into the dynamics of hot Jupiter atmospheres. Not only this but the parameter space spanned by the grid lends it to being used as model data base in the future, from which fast interpolated PT structure and spectra can be calculated.

\section*{Acknowledgements}
We would like to thank Xi Zhang, Xianyu Tan and Thaddeus Komacek for useful conversations pertaining to this work. We further thank the referee for their insightful comments. A.R acknowledges the support of the United Kingdom’s Science and Technology Facilities Council for funding this research. This work used the DiRAC@Durham facility managed by the Institute for Computational Cosmology on behalf of the STFC DiRAC HPC Facility (www.dirac.ac.uk). The equipment was funded by BEIS capital funding via STFC capital grants ST/P002293/1 and ST/R002371/1, Durham University and STFC operations grant ST/R000832/1. DiRAC is part of the National e-Infrastructure.
\section*{Data Availability}
\begin{figure*}
\includegraphics[width=2\columnwidth]{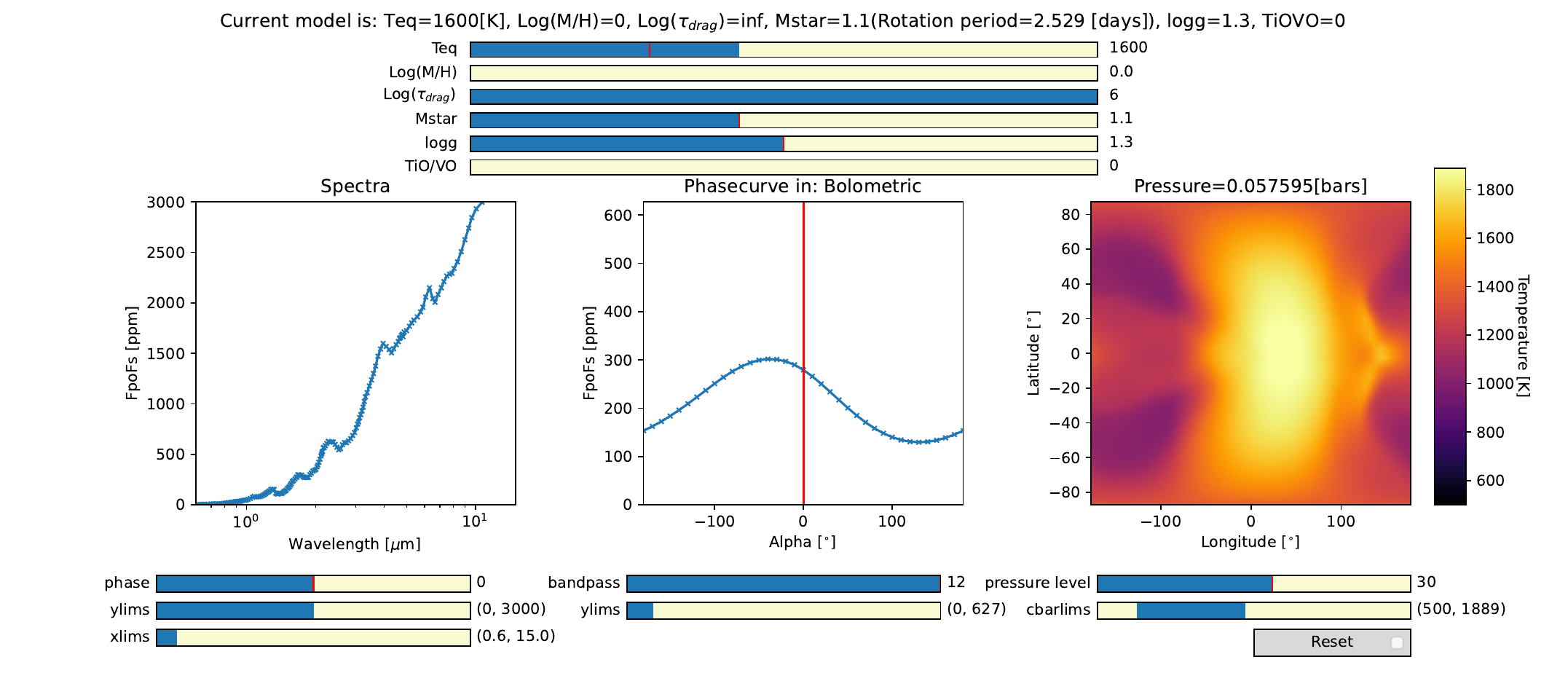}
    \caption{Example of the grid visualization tool for displaying spectra, phase curves and the thermal structure of models in the GCM grid. This tool along with the grid data can be found at at \url{https://zenodo.org/doi/10.5281/zenodo.10785320} and \url{https://3dsim.oca.eu/hot-jupiters-3d-models}.}
    \label{fig:sliders}
\end{figure*}

The grid of modelled spectra, phase curves and thermal structures are made available to the community, together with a python code allowing for visualization the grid properties (see Figure \ref{fig:sliders} for an example), at \url{https://zenodo.org/doi/10.5281/zenodo.10785320} and \url{https://3dsim.oca.eu/hot-jupiters-3d-models}. We provide an additional set of 171 models with varying drag timescale of $10^{3}$, $10^{4}$ and $10^{5}$s that are not discussed in the current publication.

\bibliographystyle{mnras}
\bibliography{paper} 

\begin{thebibliography}{}
\makeatletter
\relax
\def\mn@urlcharsother{\let\do\@makeother \do\$\do\&\do\#\do\^\do\_\do\%\do\~}
\def\mn@doi{\begingroup\mn@urlcharsother \@ifnextchar [ {\mn@doi@} {\mn@doi@[]}}
\def\mn@doi@[#1]#2{\def\@tempa{#1}\ifx\@tempa\@empty \href {http://dx.doi.org/#2} {doi:#2}\else \href {http://dx.doi.org/#2} {#1}\fi \endgroup}
\def\mn@eprint#1#2{\mn@eprint@#1:#2::\@nil}
\def\mn@eprint@arXiv#1{\href {http://arxiv.org/abs/#1} {{\tt arXiv:#1}}}
\def\mn@eprint@dblp#1{\href {http://dblp.uni-trier.de/rec/bibtex/#1.xml} {dblp:#1}}
\def\mn@eprint@#1:#2:#3:#4\@nil{\def\@tempa {#1}\def\@tempb {#2}\def\@tempc {#3}\ifx \@tempc \@empty \let \@tempc \@tempb \let \@tempb \@tempa \fi \ifx \@tempb \@empty \def\@tempb {arXiv}\fi \@ifundefined {mn@eprint@\@tempb}{\@tempb:\@tempc}{\expandafter \expandafter \csname mn@eprint@\@tempb\endcsname \expandafter{\@tempc}}}

\bibitem[\protect\citeauthoryear{{Adcroft}, {Campin}, {Hill}  \& {Marshall}}{{Adcroft} et~al.}{2004}]{2004MWRv..132.2845A}
{Adcroft} A.,  {Campin} J.-M.,  {Hill} C.,   {Marshall} J.,  2004, \mn@doi [Monthly Weather Review] {10.1175/MWR2823.1}, \href {https://ui.adsabs.harvard.edu/abs/2004MWRv..132.2845A} {132, 2845}

\bibitem[\protect\citeauthoryear{{Arcangeli} et~al.,}{{Arcangeli} et~al.}{2018}]{2018ApJ...855L..30A}
{Arcangeli} J.,  et~al., 2018, \mn@doi [\apjl] {10.3847/2041-8213/aab272}, \href {https://ui.adsabs.harvard.edu/abs/2018ApJ...855L..30A} {855, L30}

\bibitem[\protect\citeauthoryear{{Baeyens}, {Decin}, {Carone}, {Venot}, {Ag{\'u}ndez}  \& {Molli{\`e}re}}{{Baeyens} et~al.}{2021}]{2021MNRAS.505.5603B}
{Baeyens} R.,  {Decin} L.,  {Carone} L.,  {Venot} O.,  {Ag{\'u}ndez} M.,   {Molli{\`e}re} P.,  2021, \mn@doi [\mnras] {10.1093/mnras/stab1310}, \href {https://ui.adsabs.harvard.edu/abs/2021MNRAS.505.5603B} {505, 5603}

\bibitem[\protect\citeauthoryear{{Baeyens}, {Konings}, {Venot}, {Carone}  \& {Decin}}{{Baeyens} et~al.}{2022}]{2022MNRAS.512.4877B}
{Baeyens} R.,  {Konings} T.,  {Venot} O.,  {Carone} L.,   {Decin} L.,  2022, \mn@doi [\mnras] {10.1093/mnras/stac809}, \href {https://ui.adsabs.harvard.edu/abs/2022MNRAS.512.4877B} {512, 4877}

\bibitem[\protect\citeauthoryear{{Baraffe}, {Chabrier}, {Fortney}  \& {Sotin}}{{Baraffe} et~al.}{2014}]{2014prpl.conf..763B}
{Baraffe} I.,  {Chabrier} G.,  {Fortney} J.,   {Sotin} C.,  2014, in {Beuther} H.,  {Klessen} R.~S.,  {Dullemond} C.~P.,   {Henning} T.,  eds, Protostars and Planets VI. pp 763--786 (\mn@eprint {arXiv} {1401.4738}), \mn@doi{10.2458/azu_uapress_9780816531240-ch033}

\bibitem[\protect\citeauthoryear{{Baxter} et~al.,}{{Baxter} et~al.}{2020}]{2020A&A...639A..36B}
{Baxter} C.,  et~al., 2020, \mn@doi [\aap] {10.1051/0004-6361/201937394}, \href {https://ui.adsabs.harvard.edu/abs/2020A&A...639A..36B} {639, A36}

\bibitem[\protect\citeauthoryear{{Beatty}, {Madhusudhan}, {Tsiaras}, {Zhao}, {Gilliland}, {Knutson}, {Shporer}  \& {Wright}}{{Beatty} et~al.}{2017}]{2017AJ....154..158B}
{Beatty} T.~G.,  {Madhusudhan} N.,  {Tsiaras} A.,  {Zhao} M.,  {Gilliland} R.~L.,  {Knutson} H.~A.,  {Shporer} A.,   {Wright} J.~T.,  2017, \mn@doi [\aj] {10.3847/1538-3881/aa899b}, \href {https://ui.adsabs.harvard.edu/abs/2017AJ....154..158B} {154, 158}

\bibitem[\protect\citeauthoryear{{Beatty}, {Marley}, {Gaudi}, {Col{\'o}n}, {Fortney}  \& {Showman}}{{Beatty} et~al.}{2019}]{2019AJ....158..166B}
{Beatty} T.~G.,  {Marley} M.~S.,  {Gaudi} B.~S.,  {Col{\'o}n} K.~D.,  {Fortney} J.~J.,   {Showman} A.~P.,  2019, \mn@doi [\aj] {10.3847/1538-3881/ab33fc}, \href {https://ui.adsabs.harvard.edu/abs/2019AJ....158..166B} {158, 166}

\bibitem[\protect\citeauthoryear{{Bell} \& {Cowan}}{{Bell} \& {Cowan}}{2018}]{2018ApJ...857L..20B}
{Bell} T.~J.,  {Cowan} N.~B.,  2018, \mn@doi [\apjl] {10.3847/2041-8213/aabcc8}, \href {https://ui.adsabs.harvard.edu/abs/2018ApJ...857L..20B} {857, L20}

\bibitem[\protect\citeauthoryear{{Bell} et~al.,}{{Bell} et~al.}{2021}]{2021MNRAS.504.3316B}
{Bell} T.~J.,  et~al., 2021, \mn@doi [\mnras] {10.1093/mnras/stab1027}, \href {https://ui.adsabs.harvard.edu/abs/2021MNRAS.504.3316B} {504, 3316}

\bibitem[\protect\citeauthoryear{{Beltz}, {Rauscher}, {Brogi}  \& {Kempton}}{{Beltz} et~al.}{2021}]{2021AJ....161....1B}
{Beltz} H.,  {Rauscher} E.,  {Brogi} M.,   {Kempton} E. M.~R.,  2021, \mn@doi [\aj] {10.3847/1538-3881/abb67b}, \href {https://ui.adsabs.harvard.edu/abs/2021AJ....161....1B} {161, 1}

\bibitem[\protect\citeauthoryear{{Beltz}, {Rauscher}, {Roman}  \& {Guilliat}}{{Beltz} et~al.}{2022}]{2022AJ....163...35B}
{Beltz} H.,  {Rauscher} E.,  {Roman} M.~T.,   {Guilliat} A.,  2022, \mn@doi [\aj] {10.3847/1538-3881/ac3746}, \href {https://ui.adsabs.harvard.edu/abs/2022AJ....163...35B} {163, 35}

\bibitem[\protect\citeauthoryear{{Bitsch}, {Schneider}  \& {Kreidberg}}{{Bitsch} et~al.}{2022}]{2022A&A...665A.138B}
{Bitsch} B.,  {Schneider} A.~D.,   {Kreidberg} L.,  2022, \mn@doi [\aap] {10.1051/0004-6361/202243345}, \href {https://ui.adsabs.harvard.edu/abs/2022A&A...665A.138B} {665, A138}

\bibitem[\protect\citeauthoryear{{Blecic}, {Dobbs-Dixon}  \& {Greene}}{{Blecic} et~al.}{2017}]{2017ApJ...848..127B}
{Blecic} J.,  {Dobbs-Dixon} I.,   {Greene} T.,  2017, \mn@doi [\apj] {10.3847/1538-4357/aa8171}, \href {https://ui.adsabs.harvard.edu/abs/2017ApJ...848..127B} {848, 127}

\bibitem[\protect\citeauthoryear{{Brogi}, {de Kok}, {Albrecht}, {Snellen}, {Birkby}  \& {Schwarz}}{{Brogi} et~al.}{2016}]{2016ApJ...817..106B}
{Brogi} M.,  {de Kok} R.~J.,  {Albrecht} S.,  {Snellen} I.~A.~G.,  {Birkby} J.~L.,   {Schwarz} H.,  2016, \mn@doi [\apj] {10.3847/0004-637X/817/2/106}, \href {https://ui.adsabs.harvard.edu/abs/2016ApJ...817..106B} {817, 106}

\bibitem[\protect\citeauthoryear{{Caldas}, {Leconte}, {Selsis}, {Waldmann}, {Bord{\'e}}, {Rocchetto}  \& {Charnay}}{{Caldas} et~al.}{2019}]{2019A&A...623A.161C}
{Caldas} A.,  {Leconte} J.,  {Selsis} F.,  {Waldmann} I.~P.,  {Bord{\'e}} P.,  {Rocchetto} M.,   {Charnay} B.,  2019, \mn@doi [\aap] {10.1051/0004-6361/201834384}, \href {https://ui.adsabs.harvard.edu/abs/2019A&A...623A.161C} {623, A161}

\bibitem[\protect\citeauthoryear{{Charnay}, {Meadows}, {Misra}, {Leconte}  \& {Arney}}{{Charnay} et~al.}{2015}]{2015ApJ...813L...1C}
{Charnay} B.,  {Meadows} V.,  {Misra} A.,  {Leconte} J.,   {Arney} G.,  2015, \mn@doi [\apjl] {10.1088/2041-8205/813/1/L1}, \href {https://ui.adsabs.harvard.edu/abs/2015ApJ...813L...1C} {813, L1}

\bibitem[\protect\citeauthoryear{{Cont} et~al.,}{{Cont} et~al.}{2022}]{2022A&A...668A..53C}
{Cont} D.,  et~al., 2022, \mn@doi [\aap] {10.1051/0004-6361/202244277}, \href {https://ui.adsabs.harvard.edu/abs/2022A&A...668A..53C} {668, A53}

\bibitem[\protect\citeauthoryear{{Cooper} \& {Showman}}{{Cooper} \& {Showman}}{2005}]{2005ApJ...629L..45C}
{Cooper} C.~S.,  {Showman} A.~P.,  2005, \mn@doi [\apjl] {10.1086/444354}, \href {https://ui.adsabs.harvard.edu/abs/2005ApJ...629L..45C} {629, L45}

\bibitem[\protect\citeauthoryear{{Cooper} \& {Showman}}{{Cooper} \& {Showman}}{2006}]{2006ApJ...649.1048C}
{Cooper} C.~S.,  {Showman} A.~P.,  2006, \mn@doi [\apj] {10.1086/506312}, \href {https://ui.adsabs.harvard.edu/abs/2006ApJ...649.1048C} {649, 1048}

\bibitem[\protect\citeauthoryear{{Coulombe} et~al.,}{{Coulombe} et~al.}{2023}]{2023arXiv230108192C}
{Coulombe} L.-P.,  et~al., 2023, \mn@doi [arXiv e-prints] {10.48550/arXiv.2301.08192}, \href {https://ui.adsabs.harvard.edu/abs/2023arXiv230108192C} {p. arXiv:2301.08192}

\bibitem[\protect\citeauthoryear{{Cowan} \& {Agol}}{{Cowan} \& {Agol}}{2011}]{2011ApJ...729...54C}
{Cowan} N.~B.,  {Agol} E.,  2011, \mn@doi [\apj] {10.1088/0004-637X/729/1/54}, \href {https://ui.adsabs.harvard.edu/abs/2011ApJ...729...54C} {729, 54}

\bibitem[\protect\citeauthoryear{{Cowan}, {Machalek}, {Croll}, {Shekhtman}, {Burrows}, {Deming}, {Greene}  \& {Hora}}{{Cowan} et~al.}{2012}]{2012ApJ...747...82C}
{Cowan} N.~B.,  {Machalek} P.,  {Croll} B.,  {Shekhtman} L.~M.,  {Burrows} A.,  {Deming} D.,  {Greene} T.,   {Hora} J.~L.,  2012, \mn@doi [\apj] {10.1088/0004-637X/747/1/82}, \href {https://ui.adsabs.harvard.edu/abs/2012ApJ...747...82C} {747, 82}

\bibitem[\protect\citeauthoryear{{Cridland}, {Eistrup}  \& {van Dishoeck}}{{Cridland} et~al.}{2019}]{2019A&A...627A.127C}
{Cridland} A.~J.,  {Eistrup} C.,   {van Dishoeck} E.~F.,  2019, \mn@doi [\aap] {10.1051/0004-6361/201834378}, \href {https://ui.adsabs.harvard.edu/abs/2019A&A...627A.127C} {627, A127}

\bibitem[\protect\citeauthoryear{{Dang} et~al.,}{{Dang} et~al.}{2018}]{2018NatAs...2..220D}
{Dang} L.,  et~al., 2018, \mn@doi [Nature Astronomy] {10.1038/s41550-017-0351-6}, \href {https://ui.adsabs.harvard.edu/abs/2018NatAs...2..220D} {2, 220}

\bibitem[\protect\citeauthoryear{{Deming}, {Line}, {Knutson}, {Crossfield}, {Kempton}, {Komacek}, {Wallack}  \& {Fu}}{{Deming} et~al.}{2023}]{2023arXiv230103639D}
{Deming} D.,  {Line} M.~R.,  {Knutson} H.~A.,  {Crossfield} I. J.~M.,  {Kempton} E. M.~R.,  {Komacek} T.~D.,  {Wallack} N.~L.,   {Fu} G.,  2023, \mn@doi [arXiv e-prints] {10.48550/arXiv.2301.03639}, \href {https://ui.adsabs.harvard.edu/abs/2023arXiv230103639D} {p. arXiv:2301.03639}

\bibitem[\protect\citeauthoryear{{Dobbs-Dixon} \& {Agol}}{{Dobbs-Dixon} \& {Agol}}{2013}]{2013MNRAS.435.3159D}
{Dobbs-Dixon} I.,  {Agol} E.,  2013, \mn@doi [\mnras] {10.1093/mnras/stt1509}, \href {https://ui.adsabs.harvard.edu/abs/2013MNRAS.435.3159D} {435, 3159}

\bibitem[\protect\citeauthoryear{{Dobbs-Dixon} \& {Cowan}}{{Dobbs-Dixon} \& {Cowan}}{2017}]{2017ApJ...851L..26D}
{Dobbs-Dixon} I.,  {Cowan} N.~B.,  2017, \mn@doi [\apjl] {10.3847/2041-8213/aa9bec}, \href {https://ui.adsabs.harvard.edu/abs/2017ApJ...851L..26D} {851, L26}

\bibitem[\protect\citeauthoryear{{Drummond}, {Tremblin}, {Baraffe}, {Amundsen}, {Mayne}, {Venot}  \& {Goyal}}{{Drummond} et~al.}{2016}]{2016A&A...594A..69D}
{Drummond} B.,  {Tremblin} P.,  {Baraffe} I.,  {Amundsen} D.~S.,  {Mayne} N.~J.,  {Venot} O.,   {Goyal} J.,  2016, \mn@doi [\aap] {10.1051/0004-6361/201628799}, \href {https://ui.adsabs.harvard.edu/abs/2016A&A...594A..69D} {594, A69}

\bibitem[\protect\citeauthoryear{{Drummond}, {Mayne}, {Baraffe}, {Tremblin}, {Manners}, {Amundsen}, {Goyal}  \& {Acreman}}{{Drummond} et~al.}{2018a}]{2018A&A...612A.105D}
{Drummond} B.,  {Mayne} N.~J.,  {Baraffe} I.,  {Tremblin} P.,  {Manners} J.,  {Amundsen} D.~S.,  {Goyal} J.,   {Acreman} D.,  2018a, \mn@doi [\aap] {10.1051/0004-6361/201732010}, \href {https://ui.adsabs.harvard.edu/abs/2018A&A...612A.105D} {612, A105}

\bibitem[\protect\citeauthoryear{{Drummond} et~al.,}{{Drummond} et~al.}{2018b}]{2018ApJ...855L..31D}
{Drummond} B.,  et~al., 2018b, \mn@doi [\apjl] {10.3847/2041-8213/aab209}, \href {https://ui.adsabs.harvard.edu/abs/2018ApJ...855L..31D} {855, L31}

\bibitem[\protect\citeauthoryear{{Ehrenreich} et~al.,}{{Ehrenreich} et~al.}{2020}]{2020Natur.580..597E}
{Ehrenreich} D.,  et~al., 2020, \mn@doi [\nat] {10.1038/s41586-020-2107-1}, \href {https://ui.adsabs.harvard.edu/abs/2020Natur.580..597E} {580, 597}

\bibitem[\protect\citeauthoryear{{Esteves}, {De Mooij}  \& {Jayawardhana}}{{Esteves} et~al.}{2015}]{2015ApJ...804..150E}
{Esteves} L.~J.,  {De Mooij} E. J.~W.,   {Jayawardhana} R.,  2015, \mn@doi [\apj] {10.1088/0004-637X/804/2/150}, \href {https://ui.adsabs.harvard.edu/abs/2015ApJ...804..150E} {804, 150}

\bibitem[\protect\citeauthoryear{{Evans} et~al.,}{{Evans} et~al.}{2016}]{2016ApJ...822L...4E}
{Evans} T.~M.,  et~al., 2016, \mn@doi [\apjl] {10.3847/2041-8205/822/1/L4}, \href {https://ui.adsabs.harvard.edu/abs/2016ApJ...822L...4E} {822, L4}

\bibitem[\protect\citeauthoryear{{Feng}, {Line}, {Fortney}, {Stevenson}, {Bean}, {Kreidberg}  \& {Parmentier}}{{Feng} et~al.}{2016}]{2016ApJ...829...52F}
{Feng} Y.~K.,  {Line} M.~R.,  {Fortney} J.~J.,  {Stevenson} K.~B.,  {Bean} J.,  {Kreidberg} L.,   {Parmentier} V.,  2016, \mn@doi [\apj] {10.3847/0004-637X/829/1/52}, \href {https://ui.adsabs.harvard.edu/abs/2016ApJ...829...52F} {829, 52}

\bibitem[\protect\citeauthoryear{{Fortney}}{{Fortney}}{2005}]{2005MNRAS.364..649F}
{Fortney} J.~J.,  2005, \mn@doi [\mnras] {10.1111/j.1365-2966.2005.09587.x}, \href {https://ui.adsabs.harvard.edu/abs/2005MNRAS.364..649F} {364, 649}

\bibitem[\protect\citeauthoryear{{Fortney} \& {Nettelmann}}{{Fortney} \& {Nettelmann}}{2010}]{2010SSRv..152..423F}
{Fortney} J.~J.,  {Nettelmann} N.,  2010, \mn@doi [\ssr] {10.1007/s11214-009-9582-x}, \href {https://ui.adsabs.harvard.edu/abs/2010SSRv..152..423F} {152, 423}

\bibitem[\protect\citeauthoryear{{Fortney}, {Marley}, {Lodders}, {Saumon}  \& {Freedman}}{{Fortney} et~al.}{2005}]{2005ApJ...627L..69F}
{Fortney} J.~J.,  {Marley} M.~S.,  {Lodders} K.,  {Saumon} D.,   {Freedman} R.,  2005, \mn@doi [\apjl] {10.1086/431952}, \href {https://ui.adsabs.harvard.edu/abs/2005ApJ...627L..69F} {627, L69}

\bibitem[\protect\citeauthoryear{{Fortney}, {Cooper}, {Showman}, {Marley}  \& {Freedman}}{{Fortney} et~al.}{2006}]{2006ApJ...652..746F}
{Fortney} J.~J.,  {Cooper} C.~S.,  {Showman} A.~P.,  {Marley} M.~S.,   {Freedman} R.~S.,  2006, \mn@doi [\apj] {10.1086/508442}, \href {https://ui.adsabs.harvard.edu/abs/2006ApJ...652..746F} {652, 746}

\bibitem[\protect\citeauthoryear{{Fortney}, {Lodders}, {Marley}  \& {Freedman}}{{Fortney} et~al.}{2008}]{2008ApJ...678.1419F}
{Fortney} J.~J.,  {Lodders} K.,  {Marley} M.~S.,   {Freedman} R.~S.,  2008, \mn@doi [\apj] {10.1086/528370}, \href {https://ui.adsabs.harvard.edu/abs/2008ApJ...678.1419F} {678, 1419}

\bibitem[\protect\citeauthoryear{{Fortney}, {Mordasini}, {Nettelmann}, {Kempton}, {Greene}  \& {Zahnle}}{{Fortney} et~al.}{2013}]{2013ApJ...775...80F}
{Fortney} J.~J.,  {Mordasini} C.,  {Nettelmann} N.,  {Kempton} E. M.~R.,  {Greene} T.~P.,   {Zahnle} K.,  2013, \mn@doi [\apj] {10.1088/0004-637X/775/1/80}, \href {https://ui.adsabs.harvard.edu/abs/2013ApJ...775...80F} {775, 80}

\bibitem[\protect\citeauthoryear{{Freedman}, {Marley}  \& {Lodders}}{{Freedman} et~al.}{2008}]{2008ApJS..174..504F}
{Freedman} R.~S.,  {Marley} M.~S.,   {Lodders} K.,  2008, \mn@doi [\apjs] {10.1086/521793}, \href {https://ui.adsabs.harvard.edu/abs/2008ApJS..174..504F} {174, 504}

\bibitem[\protect\citeauthoryear{{Freedman}, {Lustig-Yaeger}, {Fortney}, {Lupu}, {Marley}  \& {Lodders}}{{Freedman} et~al.}{2014}]{2014ApJS..214...25F}
{Freedman} R.~S.,  {Lustig-Yaeger} J.,  {Fortney} J.~J.,  {Lupu} R.~E.,  {Marley} M.~S.,   {Lodders} K.,  2014, \mn@doi [\apjs] {10.1088/0067-0049/214/2/25}, \href {https://ui.adsabs.harvard.edu/abs/2014ApJS..214...25F} {214, 25}

\bibitem[\protect\citeauthoryear{{Fu} et~al.,}{{Fu} et~al.}{2022}]{2022ApJ...925L...3F}
{Fu} G.,  et~al., 2022, \mn@doi [\apjl] {10.3847/2041-8213/ac4968}, \href {https://ui.adsabs.harvard.edu/abs/2022ApJ...925L...3F} {925, L3}

\bibitem[\protect\citeauthoryear{{Garhart} et~al.,}{{Garhart} et~al.}{2020}]{2020AJ....159..137G}
{Garhart} E.,  et~al., 2020, \mn@doi [\aj] {10.3847/1538-3881/ab6cff}, \href {https://ui.adsabs.harvard.edu/abs/2020AJ....159..137G} {159, 137}

\bibitem[\protect\citeauthoryear{{Goody} \& {Yung}}{{Goody} \& {Yung}}{1989}]{1989artb.book.....G}
{Goody} R.~M.,  {Yung} Y.~L.,  1989, {Atmospheric radiation : theoretical basis}

\bibitem[\protect\citeauthoryear{{Gordon} \& {McBride}}{{Gordon} \& {McBride}}{1994}]{GMB1994}
{Gordon} S.,  {McBride} B.~J.,  1994, NASA Reference Publication, 1311

\bibitem[\protect\citeauthoryear{{Goyal} et~al.,}{{Goyal} et~al.}{2020}]{2020MNRAS.498.4680G}
{Goyal} J.~M.,  et~al., 2020, \mn@doi [\mnras] {10.1093/mnras/staa2300}, \href {https://ui.adsabs.harvard.edu/abs/2020MNRAS.498.4680G} {498, 4680}

\bibitem[\protect\citeauthoryear{{Goyal}, {Lewis}, {Wakeford}, {MacDonald}  \& {Mayne}}{{Goyal} et~al.}{2021}]{2021ApJ...923..242G}
{Goyal} J.~M.,  {Lewis} N.~K.,  {Wakeford} H.~R.,  {MacDonald} R.~J.,   {Mayne} N.~J.,  2021, \mn@doi [\apj] {10.3847/1538-4357/ac27b2}, \href {https://ui.adsabs.harvard.edu/abs/2021ApJ...923..242G} {923, 242}

\bibitem[\protect\citeauthoryear{{Greene}, {Line}, {Montero}, {Fortney}, {Lustig-Yaeger}  \& {Luther}}{{Greene} et~al.}{2016}]{2016ApJ...817...17G}
{Greene} T.~P.,  {Line} M.~R.,  {Montero} C.,  {Fortney} J.~J.,  {Lustig-Yaeger} J.,   {Luther} K.,  2016, \mn@doi [\apj] {10.3847/0004-637X/817/1/17}, \href {https://ui.adsabs.harvard.edu/abs/2016ApJ...817...17G} {817, 17}

\bibitem[\protect\citeauthoryear{{Hammond} \& {Lewis}}{{Hammond} \& {Lewis}}{2021}]{2021PNAS..11822705H}
{Hammond} M.,  {Lewis} N.~T.,  2021, \mn@doi [Proceedings of the National Academy of Science] {10.1073/pnas.2022705118}, \href {https://ui.adsabs.harvard.edu/abs/2021PNAS..11822705H} {118, e2022705118}

\bibitem[\protect\citeauthoryear{{Hammond} \& {Pierrehumbert}}{{Hammond} \& {Pierrehumbert}}{2018}]{2018ApJ...869...65H}
{Hammond} M.,  {Pierrehumbert} R.~T.,  2018, \mn@doi [\apj] {10.3847/1538-4357/aaec03}, \href {https://ui.adsabs.harvard.edu/abs/2018ApJ...869...65H} {869, 65}

\bibitem[\protect\citeauthoryear{{Hansen}, {Schwartz}  \& {Cowan}}{{Hansen} et~al.}{2014}]{2014MNRAS.444.3632H}
{Hansen} C.~J.,  {Schwartz} J.~C.,   {Cowan} N.~B.,  2014, \mn@doi [\mnras] {10.1093/mnras/stu1699}, \href {https://ui.adsabs.harvard.edu/abs/2014MNRAS.444.3632H} {444, 3632}

\bibitem[\protect\citeauthoryear{{Hauschildt}, {Allard}  \& {Baron}}{{Hauschildt} et~al.}{1999}]{1999ApJ...512..377H}
{Hauschildt} P.~H.,  {Allard} F.,   {Baron} E.,  1999, \mn@doi [\apj] {10.1086/306745}, \href {https://ui.adsabs.harvard.edu/abs/1999ApJ...512..377H} {512, 377}

\bibitem[\protect\citeauthoryear{{Hellier} et~al.,}{{Hellier} et~al.}{2009}]{2009Natur.460.1098H}
{Hellier} C.,  et~al., 2009, \mn@doi [\nat] {10.1038/nature08245}, \href {https://ui.adsabs.harvard.edu/abs/2009Natur.460.1098H} {460, 1098}

\bibitem[\protect\citeauthoryear{{Helling} et~al.,}{{Helling} et~al.}{2019}]{2019A&A...631A..79H}
{Helling} C.,  et~al., 2019, \mn@doi [\aap] {10.1051/0004-6361/201935771}, \href {https://ui.adsabs.harvard.edu/abs/2019A&A...631A..79H} {631, A79}

\bibitem[\protect\citeauthoryear{{Heng}, {Frierson}  \& {Phillipps}}{{Heng} et~al.}{2011}]{2011MNRAS.418.2669H}
{Heng} K.,  {Frierson} D. M.~W.,   {Phillipps} P.~J.,  2011, \mn@doi [\mnras] {10.1111/j.1365-2966.2011.19658.x}, \href {https://ui.adsabs.harvard.edu/abs/2011MNRAS.418.2669H} {418, 2669}

\bibitem[\protect\citeauthoryear{{Hindle}, {Bushby}  \& {Rogers}}{{Hindle} et~al.}{2021}]{2021ApJ...922..176H}
{Hindle} A.~W.,  {Bushby} P.~J.,   {Rogers} T.~M.,  2021, \mn@doi [\apj] {10.3847/1538-4357/ac0e2e}, \href {https://ui.adsabs.harvard.edu/abs/2021ApJ...922..176H} {922, 176}

\bibitem[\protect\citeauthoryear{{Hubeny}, {Burrows}  \& {Sudarsky}}{{Hubeny} et~al.}{2003}]{2003ApJ...594.1011H}
{Hubeny} I.,  {Burrows} A.,   {Sudarsky} D.,  2003, \mn@doi [\apj] {10.1086/377080}, \href {https://ui.adsabs.harvard.edu/abs/2003ApJ...594.1011H} {594, 1011}

\bibitem[\protect\citeauthoryear{{JWST Transiting Exoplanet Community Early Release Science Team} et~al.,}{{JWST Transiting Exoplanet Community Early Release Science Team} et~al.}{2023}]{2023Natur.614..649J}
{JWST Transiting Exoplanet Community Early Release Science Team} et~al., 2023, \mn@doi [\nat] {10.1038/s41586-022-05269-w}, \href {https://ui.adsabs.harvard.edu/abs/2023Natur.614..649J} {614, 649}

\bibitem[\protect\citeauthoryear{{Kataria}, {Showman}, {Lewis}, {Fortney}, {Marley}  \& {Freedman}}{{Kataria} et~al.}{2013}]{2013ApJ...767...76K}
{Kataria} T.,  {Showman} A.~P.,  {Lewis} N.~K.,  {Fortney} J.~J.,  {Marley} M.~S.,   {Freedman} R.~S.,  2013, \mn@doi [\apj] {10.1088/0004-637X/767/1/76}, \href {https://ui.adsabs.harvard.edu/abs/2013ApJ...767...76K} {767, 76}

\bibitem[\protect\citeauthoryear{{Kataria}, {Showman}, {Fortney}, {Stevenson}, {Line}, {Kreidberg}, {Bean}  \& {D{\'e}sert}}{{Kataria} et~al.}{2015}]{2015ApJ...801...86K}
{Kataria} T.,  {Showman} A.~P.,  {Fortney} J.~J.,  {Stevenson} K.~B.,  {Line} M.~R.,  {Kreidberg} L.,  {Bean} J.~L.,   {D{\'e}sert} J.-M.,  2015, \mn@doi [\apj] {10.1088/0004-637X/801/2/86}, \href {https://ui.adsabs.harvard.edu/abs/2015ApJ...801...86K} {801, 86}

\bibitem[\protect\citeauthoryear{{Kataria}, {Sing}, {Lewis}, {Visscher}, {Showman}, {Fortney}  \& {Marley}}{{Kataria} et~al.}{2016}]{2016ApJ...821....9K}
{Kataria} T.,  {Sing} D.~K.,  {Lewis} N.~K.,  {Visscher} C.,  {Showman} A.~P.,  {Fortney} J.~J.,   {Marley} M.~S.,  2016, \mn@doi [\apj] {10.3847/0004-637X/821/1/9}, \href {https://ui.adsabs.harvard.edu/abs/2016ApJ...821....9K} {821, 9}

\bibitem[\protect\citeauthoryear{{Keles}}{{Keles}}{2021}]{2021MNRAS.502.1456K}
{Keles} E.,  2021, \mn@doi [\mnras] {10.1093/mnras/stab099}, \href {https://ui.adsabs.harvard.edu/abs/2021MNRAS.502.1456K} {502, 1456}

\bibitem[\protect\citeauthoryear{{Kempton}, {Perna}  \& {Heng}}{{Kempton} et~al.}{2014}]{2014ApJ...795...24K}
{Kempton} E. M.~R.,  {Perna} R.,   {Heng} K.,  2014, \mn@doi [\apj] {10.1088/0004-637X/795/1/24}, \href {https://ui.adsabs.harvard.edu/abs/2014ApJ...795...24K} {795, 24}

\bibitem[\protect\citeauthoryear{{Kesseli} \& {Snellen}}{{Kesseli} \& {Snellen}}{2021}]{2021ApJ...908L..17K}
{Kesseli} A.~Y.,  {Snellen} I.~A.~G.,  2021, \mn@doi [\apjl] {10.3847/2041-8213/abe047}, \href {https://ui.adsabs.harvard.edu/abs/2021ApJ...908L..17K} {908, L17}

\bibitem[\protect\citeauthoryear{{Kesseli}, {Snellen}, {Casasayas-Barris}, {Molli{\`e}re}  \& {S{\'a}nchez-L{\'o}pez}}{{Kesseli} et~al.}{2022}]{2022AJ....163..107K}
{Kesseli} A.~Y.,  {Snellen} I.~A.~G.,  {Casasayas-Barris} N.,  {Molli{\`e}re} P.,   {S{\'a}nchez-L{\'o}pez} A.,  2022, \mn@doi [\aj] {10.3847/1538-3881/ac4336}, \href {https://ui.adsabs.harvard.edu/abs/2022AJ....163..107K} {163, 107}

\bibitem[\protect\citeauthoryear{{Knutson} et~al.,}{{Knutson} et~al.}{2007}]{2007Natur.447..183K}
{Knutson} H.~A.,  et~al., 2007, \mn@doi [\nat] {10.1038/nature05782}, \href {https://ui.adsabs.harvard.edu/abs/2007Natur.447..183K} {447, 183}

\bibitem[\protect\citeauthoryear{{Koll} \& {Komacek}}{{Koll} \& {Komacek}}{2018}]{2018ApJ...853..133K}
{Koll} D. D.~B.,  {Komacek} T.~D.,  2018, \mn@doi [\apj] {10.3847/1538-4357/aaa3de}, \href {https://ui.adsabs.harvard.edu/abs/2018ApJ...853..133K} {853, 133}

\bibitem[\protect\citeauthoryear{{Komacek} \& {Showman}}{{Komacek} \& {Showman}}{2016}]{2016ApJ...821...16K}
{Komacek} T.~D.,  {Showman} A.~P.,  2016, \mn@doi [\apj] {10.3847/0004-637X/821/1/16}, \href {https://ui.adsabs.harvard.edu/abs/2016ApJ...821...16K} {821, 16}

\bibitem[\protect\citeauthoryear{{Komacek} \& {Tan}}{{Komacek} \& {Tan}}{2018}]{2018RNAAS...2...36K}
{Komacek} T.~D.,  {Tan} X.,  2018, \mn@doi [Research Notes of the American Astronomical Society] {10.3847/2515-5172/aac5e7}, \href {https://ui.adsabs.harvard.edu/abs/2018RNAAS...2...36K} {2, 36}

\bibitem[\protect\citeauthoryear{{Komacek}, {Showman}  \& {Tan}}{{Komacek} et~al.}{2017}]{2017ApJ...835..198K}
{Komacek} T.~D.,  {Showman} A.~P.,   {Tan} X.,  2017, \mn@doi [\apj] {10.3847/1538-4357/835/2/198}, \href {https://ui.adsabs.harvard.edu/abs/2017ApJ...835..198K} {835, 198}

\bibitem[\protect\citeauthoryear{{Komacek}, {Gao}, {Thorngren}, {May}  \& {Tan}}{{Komacek} et~al.}{2022}]{2022ApJ...941L..40K}
{Komacek} T.~D.,  {Gao} P.,  {Thorngren} D.~P.,  {May} E.~M.,   {Tan} X.,  2022, \mn@doi [\apjl] {10.3847/2041-8213/aca975}, \href {https://ui.adsabs.harvard.edu/abs/2022ApJ...941L..40K} {941, L40}

\bibitem[\protect\citeauthoryear{{Kreidberg} et~al.,}{{Kreidberg} et~al.}{2014}]{2014ApJ...793L..27K}
{Kreidberg} L.,  et~al., 2014, \mn@doi [\apjl] {10.1088/2041-8205/793/2/L27}, \href {https://ui.adsabs.harvard.edu/abs/2014ApJ...793L..27K} {793, L27}

\bibitem[\protect\citeauthoryear{{Lacis} \& {Oinas}}{{Lacis} \& {Oinas}}{1991}]{1991JGR....96.9027L}
{Lacis} A.~A.,  {Oinas} V.,  1991, \mn@doi [\jgr] {10.1029/90JD01945}, \href {https://ui.adsabs.harvard.edu/abs/1991JGR....96.9027L} {96, 9027}

\bibitem[\protect\citeauthoryear{{Lacy} \& {Burrows}}{{Lacy} \& {Burrows}}{2020}]{2020ApJ...905..131L}
{Lacy} B.~I.,  {Burrows} A.,  2020, \mn@doi [\apj] {10.3847/1538-4357/abc01c}, \href {https://ui.adsabs.harvard.edu/abs/2020ApJ...905..131L} {905, 131}

\bibitem[\protect\citeauthoryear{{Lewis} \& {Hammond}}{{Lewis} \& {Hammond}}{2022}]{2022ApJ...941..171L}
{Lewis} N.~T.,  {Hammond} M.,  2022, \mn@doi [\apj] {10.3847/1538-4357/ac8fed}, \href {https://ui.adsabs.harvard.edu/abs/2022ApJ...941..171L} {941, 171}

\bibitem[\protect\citeauthoryear{{Lewis}, {Parmentier}, {Kataria}, {de Wit}, {Showman}, {Fortney}  \& {Marley}}{{Lewis} et~al.}{2017}]{2017arXiv170600466L}
{Lewis} N.~K.,  {Parmentier} V.,  {Kataria} T.,  {de Wit} J.,  {Showman} A.~P.,  {Fortney} J.~J.,   {Marley} M.~S.,  2017, arXiv e-prints, \href {https://ui.adsabs.harvard.edu/abs/2017arXiv170600466L} {p. arXiv:1706.00466}

\bibitem[\protect\citeauthoryear{{Line} \& {Parmentier}}{{Line} \& {Parmentier}}{2016}]{2016ApJ...820...78L}
{Line} M.~R.,  {Parmentier} V.,  2016, \mn@doi [\apj] {10.3847/0004-637X/820/1/78}, \href {https://ui.adsabs.harvard.edu/abs/2016ApJ...820...78L} {820, 78}

\bibitem[\protect\citeauthoryear{{Line} et~al.,}{{Line} et~al.}{2016}]{2016AJ....152..203L}
{Line} M.~R.,  et~al., 2016, \mn@doi [\aj] {10.3847/0004-6256/152/6/203}, \href {https://ui.adsabs.harvard.edu/abs/2016AJ....152..203L} {152, 203}

\bibitem[\protect\citeauthoryear{{Lodders}, {Palme}  \& {Gail}}{{Lodders} et~al.}{2009}]{2009LanB...4B..712L}
{Lodders} K.,  {Palme} H.,   {Gail} H.~P.,  2009, \mn@doi [Landolt B\&ouml;rnstein] {10.1007/978-3-540-88055-4_34}, \href {https://ui.adsabs.harvard.edu/abs/2009LanB...4B..712L} {4B, 712}

\bibitem[\protect\citeauthoryear{{Lothringer} \& {Barman}}{{Lothringer} \& {Barman}}{2019}]{2019ApJ...876...69L}
{Lothringer} J.~D.,  {Barman} T.,  2019, \mn@doi [\apj] {10.3847/1538-4357/ab1485}, \href {https://ui.adsabs.harvard.edu/abs/2019ApJ...876...69L} {876, 69}

\bibitem[\protect\citeauthoryear{{Lothringer}, {Barman}  \& {Koskinen}}{{Lothringer} et~al.}{2018}]{2018ApJ...866...27L}
{Lothringer} J.~D.,  {Barman} T.,   {Koskinen} T.,  2018, \mn@doi [\apj] {10.3847/1538-4357/aadd9e}, \href {https://ui.adsabs.harvard.edu/abs/2018ApJ...866...27L} {866, 27}

\bibitem[\protect\citeauthoryear{{Lubow}, {Tout}  \& {Livio}}{{Lubow} et~al.}{1997}]{1997ApJ...484..866L}
{Lubow} S.~H.,  {Tout} C.~A.,   {Livio} M.,  1997, \mn@doi [\apj] {10.1086/304369}, \href {https://ui.adsabs.harvard.edu/abs/1997ApJ...484..866L} {484, 866}

\bibitem[\protect\citeauthoryear{{MacDonald} \& {Madhusudhan}}{{MacDonald} \& {Madhusudhan}}{2017}]{2017MNRAS.469.1979M}
{MacDonald} R.~J.,  {Madhusudhan} N.,  2017, \mn@doi [\mnras] {10.1093/mnras/stx804}, \href {https://ui.adsabs.harvard.edu/abs/2017MNRAS.469.1979M} {469, 1979}

\bibitem[\protect\citeauthoryear{{MacDonald}, {Goyal}  \& {Lewis}}{{MacDonald} et~al.}{2020}]{2020ApJ...893L..43M}
{MacDonald} R.~J.,  {Goyal} J.~M.,   {Lewis} N.~K.,  2020, \mn@doi [\apjl] {10.3847/2041-8213/ab8238}, \href {https://ui.adsabs.harvard.edu/abs/2020ApJ...893L..43M} {893, L43}

\bibitem[\protect\citeauthoryear{{Maciejewski} et~al.,}{{Maciejewski} et~al.}{2016}]{2016AcA....66...55M}
{Maciejewski} G.,  et~al., 2016, \actaa, \href {https://ui.adsabs.harvard.edu/abs/2016AcA....66...55M} {66, 55}

\bibitem[\protect\citeauthoryear{{Madhusudhan}, {Amin}  \& {Kennedy}}{{Madhusudhan} et~al.}{2014}]{2014ApJ...794L..12M}
{Madhusudhan} N.,  {Amin} M.~A.,   {Kennedy} G.~M.,  2014, \mn@doi [\apjl] {10.1088/2041-8205/794/1/L12}, \href {https://ui.adsabs.harvard.edu/abs/2014ApJ...794L..12M} {794, L12}

\bibitem[\protect\citeauthoryear{{Mansfield} et~al.,}{{Mansfield} et~al.}{2021}]{2021NatAs...5.1224M}
{Mansfield} M.,  et~al., 2021, \mn@doi [Nature Astronomy] {10.1038/s41550-021-01455-4}, \href {https://ui.adsabs.harvard.edu/abs/2021NatAs...5.1224M} {5, 1224}

\bibitem[\protect\citeauthoryear{{Mansfield} et~al.,}{{Mansfield} et~al.}{2022}]{2022AJ....163..261M}
{Mansfield} M.,  et~al., 2022, \mn@doi [\aj] {10.3847/1538-3881/ac658f}, \href {https://ui.adsabs.harvard.edu/abs/2022AJ....163..261M} {163, 261}

\bibitem[\protect\citeauthoryear{{Marley} \& {McKay}}{{Marley} \& {McKay}}{1999}]{1999Icar..138..268M}
{Marley} M.~S.,  {McKay} C.~P.,  1999, \mn@doi [\icarus] {10.1006/icar.1998.6071}, \href {https://ui.adsabs.harvard.edu/abs/1999Icar..138..268M} {138, 268}

\bibitem[\protect\citeauthoryear{{Marley}, {Saumon}, {Fortney}, {Morley}, {Lupu}, {Freedman}  \& {Visscher}}{{Marley} et~al.}{2017}]{2017AAS...23031507M}
{Marley} M.~S.,  {Saumon} D.,  {Fortney} J.~J.,  {Morley} C.,  {Lupu} R.~E.,  {Freedman} R.,   {Visscher} C.,  2017, in American Astronomical Society Meeting Abstracts \#230. p. 315.07

\bibitem[\protect\citeauthoryear{{May} \& {Stevenson}}{{May} \& {Stevenson}}{2020}]{2020AJ....160..140M}
{May} E.~M.,  {Stevenson} K.~B.,  2020, \mn@doi [\aj] {10.3847/1538-3881/aba833}, \href {https://ui.adsabs.harvard.edu/abs/2020AJ....160..140M} {160, 140}

\bibitem[\protect\citeauthoryear{{May} et~al.,}{{May} et~al.}{2022}]{2022arXiv220315059M}
{May} E.,  et~al., 2022, arXiv e-prints, \href {https://ui.adsabs.harvard.edu/abs/2022arXiv220315059M} {p. arXiv:2203.15059}

\bibitem[\protect\citeauthoryear{{Mayne} et~al.,}{{Mayne} et~al.}{2014}]{2014A&A...561A...1M}
{Mayne} N.~J.,  et~al., 2014, \mn@doi [\aap] {10.1051/0004-6361/201322174}, \href {https://ui.adsabs.harvard.edu/abs/2014A&A...561A...1M} {561, A1}

\bibitem[\protect\citeauthoryear{{Mayne} et~al.,}{{Mayne} et~al.}{2017}]{2017A&A...604A..79M}
{Mayne} N.~J.,  et~al., 2017, \mn@doi [\aap] {10.1051/0004-6361/201730465}, \href {https://ui.adsabs.harvard.edu/abs/2017A&A...604A..79M} {604, A79}

\bibitem[\protect\citeauthoryear{{McKay}, {Pollack}  \& {Courtin}}{{McKay} et~al.}{1989}]{1989Icar...80...23M}
{McKay} C.~P.,  {Pollack} J.~B.,   {Courtin} R.,  1989, \mn@doi [\icarus] {10.1016/0019-1035(89)90160-7}, \href {https://ui.adsabs.harvard.edu/abs/1989Icar...80...23M} {80, 23}

\bibitem[\protect\citeauthoryear{{Mendon{\c{c}}a}}{{Mendon{\c{c}}a}}{2020}]{2020MNRAS.491.1456M}
{Mendon{\c{c}}a} J.~M.,  2020, \mn@doi [\mnras] {10.1093/mnras/stz3050}, \href {https://ui.adsabs.harvard.edu/abs/2020MNRAS.491.1456M} {491, 1456}

\bibitem[\protect\citeauthoryear{{Mendon{\c{c}}a}, {Malik}, {Demory}  \& {Heng}}{{Mendon{\c{c}}a} et~al.}{2018a}]{2018AJ....155..150M}
{Mendon{\c{c}}a} J.~M.,  {Malik} M.,  {Demory} B.-O.,   {Heng} K.,  2018a, \mn@doi [\aj] {10.3847/1538-3881/aaaebc}, \href {https://ui.adsabs.harvard.edu/abs/2018AJ....155..150M} {155, 150}

\bibitem[\protect\citeauthoryear{{Mendon{\c{c}}a}, {Tsai}, {Malik}, {Grimm}  \& {Heng}}{{Mendon{\c{c}}a} et~al.}{2018b}]{2018ApJ...869..107M}
{Mendon{\c{c}}a} J.~M.,  {Tsai} S.-m.,  {Malik} M.,  {Grimm} S.~L.,   {Heng} K.,  2018b, \mn@doi [\apj] {10.3847/1538-4357/aaed23}, \href {https://ui.adsabs.harvard.edu/abs/2018ApJ...869..107M} {869, 107}

\bibitem[\protect\citeauthoryear{{Menou} \& {Rauscher}}{{Menou} \& {Rauscher}}{2009}]{2009ApJ...700..887M}
{Menou} K.,  {Rauscher} E.,  2009, \mn@doi [\apj] {10.1088/0004-637X/700/1/887}, \href {https://ui.adsabs.harvard.edu/abs/2009ApJ...700..887M} {700, 887}

\bibitem[\protect\citeauthoryear{{Merritt} et~al.,}{{Merritt} et~al.}{2020}]{2020A&A...636A.117M}
{Merritt} S.~R.,  et~al., 2020, \mn@doi [\aap] {10.1051/0004-6361/201937409}, \href {https://ui.adsabs.harvard.edu/abs/2020A&A...636A.117M} {636, A117}

\bibitem[\protect\citeauthoryear{{Mikal-Evans}, {Sing}, {Kataria}, {Wakeford}, {Mayne}, {Lewis}, {Barstow}  \& {Spake}}{{Mikal-Evans} et~al.}{2020}]{2020MNRAS.496.1638M}
{Mikal-Evans} T.,  {Sing} D.~K.,  {Kataria} T.,  {Wakeford} H.~R.,  {Mayne} N.~J.,  {Lewis} N.~K.,  {Barstow} J.~K.,   {Spake} J.~J.,  2020, \mn@doi [\mnras] {10.1093/mnras/staa1628}, \href {https://ui.adsabs.harvard.edu/abs/2020MNRAS.496.1638M} {496, 1638}

\bibitem[\protect\citeauthoryear{{Miller-Ricci Kempton} \& {Rauscher}}{{Miller-Ricci Kempton} \& {Rauscher}}{2012}]{2012ApJ...751..117M}
{Miller-Ricci Kempton} E.,  {Rauscher} E.,  2012, \mn@doi [\apj] {10.1088/0004-637X/751/2/117}, \href {https://ui.adsabs.harvard.edu/abs/2012ApJ...751..117M} {751, 117}

\bibitem[\protect\citeauthoryear{{Mitchell} \& {Vallis}}{{Mitchell} \& {Vallis}}{2010}]{2010JGRE..11512008M}
{Mitchell} J.~L.,  {Vallis} G.~K.,  2010, \mn@doi [Journal of Geophysical Research (Planets)] {10.1029/2010JE003587}, \href {https://ui.adsabs.harvard.edu/abs/2010JGRE..11512008M} {115, E12008}

\bibitem[\protect\citeauthoryear{{Molaverdikhani}, {Henning}  \& {Molli{\`e}re}}{{Molaverdikhani} et~al.}{2019}]{2019ApJ...883..194M}
{Molaverdikhani} K.,  {Henning} T.,   {Molli{\`e}re} P.,  2019, \mn@doi [\apj] {10.3847/1538-4357/ab3e30}, \href {https://ui.adsabs.harvard.edu/abs/2019ApJ...883..194M} {883, 194}

\bibitem[\protect\citeauthoryear{{Molli{\`e}re}, {van Boekel}, {Dullemond}, {Henning}  \& {Mordasini}}{{Molli{\`e}re} et~al.}{2015}]{2015ApJ...813...47M}
{Molli{\`e}re} P.,  {van Boekel} R.,  {Dullemond} C.,  {Henning} T.,   {Mordasini} C.,  2015, \mn@doi [\apj] {10.1088/0004-637X/813/1/47}, \href {https://ui.adsabs.harvard.edu/abs/2015ApJ...813...47M} {813, 47}

\bibitem[\protect\citeauthoryear{{Moses} et~al.,}{{Moses} et~al.}{2013}]{2013ApJ...777...34M}
{Moses} J.~I.,  et~al., 2013, \mn@doi [\apj] {10.1088/0004-637X/777/1/34}, \href {https://ui.adsabs.harvard.edu/abs/2013ApJ...777...34M} {777, 34}

\bibitem[\protect\citeauthoryear{{{\"O}berg}, {Murray-Clay}  \& {Bergin}}{{{\"O}berg} et~al.}{2011}]{2011ApJ...743L..16O}
{{\"O}berg} K.~I.,  {Murray-Clay} R.,   {Bergin} E.~A.,  2011, \mn@doi [\apjl] {10.1088/2041-8205/743/1/L16}, \href {https://ui.adsabs.harvard.edu/abs/2011ApJ...743L..16O} {743, L16}

\bibitem[\protect\citeauthoryear{{Parmentier} \& {Crossfield}}{{Parmentier} \& {Crossfield}}{2018}]{2018haex.bookE.116P}
{Parmentier} V.,  {Crossfield} I. J.~M.,  2018, in {Deeg} H.~J.,  {Belmonte} J.~A.,  eds, , Handbook of Exoplanets.
p.~116, \mn@doi{10.1007/978-3-319-55333-7_116}

\bibitem[\protect\citeauthoryear{{Parmentier} \& {Guillot}}{{Parmentier} \& {Guillot}}{2014}]{2014A&A...562A.133P}
{Parmentier} V.,  {Guillot} T.,  2014, \mn@doi [\aap] {10.1051/0004-6361/201322342}, \href {https://ui.adsabs.harvard.edu/abs/2014A&A...562A.133P} {562, A133}

\bibitem[\protect\citeauthoryear{{Parmentier}, {Showman}  \& {Lian}}{{Parmentier} et~al.}{2013}]{2013A&A...558A..91P}
{Parmentier} V.,  {Showman} A.~P.,   {Lian} Y.,  2013, \mn@doi [\aap] {10.1051/0004-6361/201321132}, \href {https://ui.adsabs.harvard.edu/abs/2013A&A...558A..91P} {558, A91}

\bibitem[\protect\citeauthoryear{{Parmentier}, {Showman}  \& {de Wit}}{{Parmentier} et~al.}{2015a}]{2015ExA....40..481P}
{Parmentier} V.,  {Showman} A.~P.,   {de Wit} J.,  2015a, \mn@doi [Experimental Astronomy] {10.1007/s10686-014-9395-0}, \href {https://ui.adsabs.harvard.edu/abs/2015ExA....40..481P} {40, 481}

\bibitem[\protect\citeauthoryear{{Parmentier}, {Guillot}, {Fortney}  \& {Marley}}{{Parmentier} et~al.}{2015b}]{2015A&A...574A..35P}
{Parmentier} V.,  {Guillot} T.,  {Fortney} J.~J.,   {Marley} M.~S.,  2015b, \mn@doi [\aap] {10.1051/0004-6361/201323127}, \href {https://ui.adsabs.harvard.edu/abs/2015A&A...574A..35P} {574, A35}

\bibitem[\protect\citeauthoryear{{Parmentier}, {Fortney}, {Showman}, {Morley}  \& {Marley}}{{Parmentier} et~al.}{2016}]{2016ApJ...828...22P}
{Parmentier} V.,  {Fortney} J.~J.,  {Showman} A.~P.,  {Morley} C.,   {Marley} M.~S.,  2016, \mn@doi [\apj] {10.3847/0004-637X/828/1/22}, \href {https://ui.adsabs.harvard.edu/abs/2016ApJ...828...22P} {828, 22}

\bibitem[\protect\citeauthoryear{{Parmentier} et~al.,}{{Parmentier} et~al.}{2018}]{2018A&A...617A.110P}
{Parmentier} V.,  et~al., 2018, \mn@doi [\aap] {10.1051/0004-6361/201833059}, \href {https://ui.adsabs.harvard.edu/abs/2018A&A...617A.110P} {617, A110}

\bibitem[\protect\citeauthoryear{{Parmentier}, {Showman}  \& {Fortney}}{{Parmentier} et~al.}{2021}]{2021MNRAS.501...78P}
{Parmentier} V.,  {Showman} A.~P.,   {Fortney} J.~J.,  2021, \mn@doi [\mnras] {10.1093/mnras/staa3418}, \href {https://ui.adsabs.harvard.edu/abs/2021MNRAS.501...78P} {501, 78}

\bibitem[\protect\citeauthoryear{{Pelletier} et~al.,}{{Pelletier} et~al.}{2023}]{2023arXiv230608739P}
{Pelletier} S.,  et~al., 2023, \mn@doi [arXiv e-prints] {10.48550/arXiv.2306.08739}, \href {https://ui.adsabs.harvard.edu/abs/2023arXiv230608739P} {p. arXiv:2306.08739}

\bibitem[\protect\citeauthoryear{{Perez-Becker} \& {Showman}}{{Perez-Becker} \& {Showman}}{2013}]{2013ApJ...776..134P}
{Perez-Becker} D.,  {Showman} A.~P.,  2013, \mn@doi [\apj] {10.1088/0004-637X/776/2/134}, \href {https://ui.adsabs.harvard.edu/abs/2013ApJ...776..134P} {776, 134}

\bibitem[\protect\citeauthoryear{{Perna}, {Heng}  \& {Pont}}{{Perna} et~al.}{2012}]{2012ApJ...751...59P}
{Perna} R.,  {Heng} K.,   {Pont} F.,  2012, \mn@doi [\apj] {10.1088/0004-637X/751/1/59}, \href {https://ui.adsabs.harvard.edu/abs/2012ApJ...751...59P} {751, 59}

\bibitem[\protect\citeauthoryear{{Piette}, {Madhusudhan}, {McKemmish}, {Gandhi}, {Masseron}  \& {Welbanks}}{{Piette} et~al.}{2020}]{2020MNRAS.496.3870P}
{Piette} A. A.~A.,  {Madhusudhan} N.,  {McKemmish} L.~K.,  {Gandhi} S.,  {Masseron} T.,   {Welbanks} L.,  2020, \mn@doi [\mnras] {10.1093/mnras/staa1592}, \href {https://ui.adsabs.harvard.edu/abs/2020MNRAS.496.3870P} {496, 3870}

\bibitem[\protect\citeauthoryear{{Pluriel}, {Zingales}, {Leconte}  \& {Parmentier}}{{Pluriel} et~al.}{2020}]{2020A&A...636A..66P}
{Pluriel} W.,  {Zingales} T.,  {Leconte} J.,   {Parmentier} V.,  2020, \mn@doi [\aap] {10.1051/0004-6361/202037678}, \href {https://ui.adsabs.harvard.edu/abs/2020A&A...636A..66P} {636, A66}

\bibitem[\protect\citeauthoryear{{Powell}, {Zhang}, {Gao}  \& {Parmentier}}{{Powell} et~al.}{2018}]{2018ApJ...860...18P}
{Powell} D.,  {Zhang} X.,  {Gao} P.,   {Parmentier} V.,  2018, \mn@doi [\apj] {10.3847/1538-4357/aac215}, \href {https://ui.adsabs.harvard.edu/abs/2018ApJ...860...18P} {860, 18}

\bibitem[\protect\citeauthoryear{{Powell}, {Louden}, {Kreidberg}, {Zhang}, {Gao}  \& {Parmentier}}{{Powell} et~al.}{2019}]{2019ApJ...887..170P}
{Powell} D.,  {Louden} T.,  {Kreidberg} L.,  {Zhang} X.,  {Gao} P.,   {Parmentier} V.,  2019, \mn@doi [\apj] {10.3847/1538-4357/ab55d9}, \href {https://ui.adsabs.harvard.edu/abs/2019ApJ...887..170P} {887, 170}

\bibitem[\protect\citeauthoryear{{Prinoth} et~al.,}{{Prinoth} et~al.}{2022}]{2022NatAs...6..449P}
{Prinoth} B.,  et~al., 2022, \mn@doi [Nature Astronomy] {10.1038/s41550-021-01581-z}, \href {https://ui.adsabs.harvard.edu/abs/2022NatAs...6..449P} {6, 449}

\bibitem[\protect\citeauthoryear{{Rasio}, {Tout}, {Lubow}  \& {Livio}}{{Rasio} et~al.}{1996}]{1996ApJ...470.1187R}
{Rasio} F.~A.,  {Tout} C.~A.,  {Lubow} S.~H.,   {Livio} M.,  1996, \mn@doi [\apj] {10.1086/177941}, \href {https://ui.adsabs.harvard.edu/abs/1996ApJ...470.1187R} {470, 1187}

\bibitem[\protect\citeauthoryear{{Rauscher} \& {Menou}}{{Rauscher} \& {Menou}}{2012}]{2012ApJ...750...96R}
{Rauscher} E.,  {Menou} K.,  2012, \mn@doi [\apj] {10.1088/0004-637X/750/2/96}, \href {https://ui.adsabs.harvard.edu/abs/2012ApJ...750...96R} {750, 96}

\bibitem[\protect\citeauthoryear{{Robbins-Blanch}, {Kataria}, {Batalha}  \& {Adams}}{{Robbins-Blanch} et~al.}{2022}]{2022ApJ...930...93R}
{Robbins-Blanch} N.,  {Kataria} T.,  {Batalha} N.~E.,   {Adams} D.~J.,  2022, \mn@doi [\apj] {10.3847/1538-4357/ac658c}, \href {https://ui.adsabs.harvard.edu/abs/2022ApJ...930...93R} {930, 93}

\bibitem[\protect\citeauthoryear{{Rogers}}{{Rogers}}{2017}]{2017NatAs...1E.131R}
{Rogers} T.~M.,  2017, \mn@doi [Nature Astronomy] {10.1038/s41550-017-0131}, \href {https://ui.adsabs.harvard.edu/abs/2017NatAs...1E.131R} {1, 0131}

\bibitem[\protect\citeauthoryear{{Rogers} \& {Komacek}}{{Rogers} \& {Komacek}}{2014}]{2014ApJ...794..132R}
{Rogers} T.~M.,  {Komacek} T.~D.,  2014, \mn@doi [\apj] {10.1088/0004-637X/794/2/132}, \href {https://ui.adsabs.harvard.edu/abs/2014ApJ...794..132R} {794, 132}

\bibitem[\protect\citeauthoryear{{Roth}, {Drummond}, {H{\'e}brard}, {Tremblin}, {Goyal}  \& {Mayne}}{{Roth} et~al.}{2021}]{2021MNRAS.505.4515R}
{Roth} A.,  {Drummond} B.,  {H{\'e}brard} E.,  {Tremblin} P.,  {Goyal} J.,   {Mayne} N.,  2021, \mn@doi [\mnras] {10.1093/mnras/stab1256}, \href {https://ui.adsabs.harvard.edu/abs/2021MNRAS.505.4515R} {505, 4515}

\bibitem[\protect\citeauthoryear{{Savel} et~al.,}{{Savel} et~al.}{2022}]{2022ApJ...926...85S}
{Savel} A.~B.,  et~al., 2022, \mn@doi [\apj] {10.3847/1538-4357/ac423f}, \href {https://ui.adsabs.harvard.edu/abs/2022ApJ...926...85S} {926, 85}

\bibitem[\protect\citeauthoryear{{Schwartz} \& {Cowan}}{{Schwartz} \& {Cowan}}{2015}]{2015MNRAS.449.4192S}
{Schwartz} J.~C.,  {Cowan} N.~B.,  2015, \mn@doi [\mnras] {10.1093/mnras/stv470}, \href {https://ui.adsabs.harvard.edu/abs/2015MNRAS.449.4192S} {449, 4192}

\bibitem[\protect\citeauthoryear{{Showman} \& {Guillot}}{{Showman} \& {Guillot}}{2002}]{2002A&A...385..166S}
{Showman} A.~P.,  {Guillot} T.,  2002, \mn@doi [\aap] {10.1051/0004-6361:20020101}, \href {https://ui.adsabs.harvard.edu/abs/2002A&A...385..166S} {385, 166}

\bibitem[\protect\citeauthoryear{{Showman}, {Cooper}, {Fortney}  \& {Marley}}{{Showman} et~al.}{2008}]{2008ApJ...682..559S}
{Showman} A.~P.,  {Cooper} C.~S.,  {Fortney} J.~J.,   {Marley} M.~S.,  2008, \mn@doi [\apj] {10.1086/589325}, \href {https://ui.adsabs.harvard.edu/abs/2008ApJ...682..559S} {682, 559}

\bibitem[\protect\citeauthoryear{{Showman}, {Fortney}, {Lian}, {Marley}, {Freedman}, {Knutson}  \& {Charbonneau}}{{Showman} et~al.}{2009}]{2009ApJ...699..564S}
{Showman} A.~P.,  {Fortney} J.~J.,  {Lian} Y.,  {Marley} M.~S.,  {Freedman} R.~S.,  {Knutson} H.~A.,   {Charbonneau} D.,  2009, \mn@doi [\apj] {10.1088/0004-637X/699/1/564}, \href {https://ui.adsabs.harvard.edu/abs/2009ApJ...699..564S} {699, 564}

\bibitem[\protect\citeauthoryear{{Showman}, {Fortney}, {Lewis}  \& {Shabram}}{{Showman} et~al.}{2013}]{2013ApJ...762...24S}
{Showman} A.~P.,  {Fortney} J.~J.,  {Lewis} N.~K.,   {Shabram} M.,  2013, \mn@doi [\apj] {10.1088/0004-637X/762/1/24}, \href {https://ui.adsabs.harvard.edu/abs/2013ApJ...762...24S} {762, 24}

\bibitem[\protect\citeauthoryear{{Showman}, {Lewis}  \& {Fortney}}{{Showman} et~al.}{2014}]{2014arXiv1411.4731S}
{Showman} A.~P.,  {Lewis} N.~K.,   {Fortney} J.~J.,  2014, \mn@doi [arXiv e-prints] {10.48550/arXiv.1411.4731}, \href {https://ui.adsabs.harvard.edu/abs/2014arXiv1411.4731S} {p. arXiv:1411.4731}

\bibitem[\protect\citeauthoryear{{Showman}, {Lewis}  \& {Fortney}}{{Showman} et~al.}{2015}]{2015ApJ...801...95S}
{Showman} A.~P.,  {Lewis} N.~K.,   {Fortney} J.~J.,  2015, \mn@doi [\apj] {10.1088/0004-637X/801/2/95}, \href {https://ui.adsabs.harvard.edu/abs/2015ApJ...801...95S} {801, 95}

\bibitem[\protect\citeauthoryear{{Skemer} et~al.,}{{Skemer} et~al.}{2016}]{2016ApJ...817..166S}
{Skemer} A.~J.,  et~al., 2016, \mn@doi [\apj] {10.3847/0004-637X/817/2/166}, \href {https://ui.adsabs.harvard.edu/abs/2016ApJ...817..166S} {817, 166}

\bibitem[\protect\citeauthoryear{{Snellen}, {de Kok}, {de Mooij}  \& {Albrecht}}{{Snellen} et~al.}{2010}]{2010Natur.465.1049S}
{Snellen} I. A.~G.,  {de Kok} R.~J.,  {de Mooij} E. J.~W.,   {Albrecht} S.,  2010, \mn@doi [\nat] {10.1038/nature09111}, \href {https://ui.adsabs.harvard.edu/abs/2010Natur.465.1049S} {465, 1049}

\bibitem[\protect\citeauthoryear{{Spiegel}, {Silverio}  \& {Burrows}}{{Spiegel} et~al.}{2009}]{2009ApJ...699.1487S}
{Spiegel} D.~S.,  {Silverio} K.,   {Burrows} A.,  2009, \mn@doi [\apj] {10.1088/0004-637X/699/2/1487}, \href {https://ui.adsabs.harvard.edu/abs/2009ApJ...699.1487S} {699, 1487}

\bibitem[\protect\citeauthoryear{{Steinrueck}, {Parmentier}, {Showman}, {Lothringer}  \& {Lupu}}{{Steinrueck} et~al.}{2019}]{2019ApJ...880...14S}
{Steinrueck} M.~E.,  {Parmentier} V.,  {Showman} A.~P.,  {Lothringer} J.~D.,   {Lupu} R.~E.,  2019, \mn@doi [\apj] {10.3847/1538-4357/ab2598}, \href {https://ui.adsabs.harvard.edu/abs/2019ApJ...880...14S} {880, 14}

\bibitem[\protect\citeauthoryear{{Stevenson} et~al.,}{{Stevenson} et~al.}{2010}]{2010Natur.464.1161S}
{Stevenson} K.~B.,  et~al., 2010, \mn@doi [\nat] {10.1038/nature09013}, \href {https://ui.adsabs.harvard.edu/abs/2010Natur.464.1161S} {464, 1161}

\bibitem[\protect\citeauthoryear{{Tan} \& {Komacek}}{{Tan} \& {Komacek}}{2019}]{2019ESS.....432603T}
{Tan} X.,  {Komacek} T.,  2019, in AAS/Division for Extreme Solar Systems Abstracts. p. 326.03

\bibitem[\protect\citeauthoryear{{Tan}, {Komacek}, {Batalha}, {Deming}, {Lupu}, {Parmentier}  \& {Pierrehumbert}}{{Tan} et~al.}{2024}]{2024MNRAS.528.1016T}
{Tan} X.,  {Komacek} T.~D.,  {Batalha} N.~E.,  {Deming} D.,  {Lupu} R.,  {Parmentier} V.,   {Pierrehumbert} R.~T.,  2024, \mn@doi [\mnras] {10.1093/mnras/stae050}, \href {https://ui.adsabs.harvard.edu/abs/2024MNRAS.528.1016T} {528, 1016}

\bibitem[\protect\citeauthoryear{{Taylor}, {Parmentier}, {Irwin}, {Aigrain}, {Lee}  \& {Krissansen-Totton}}{{Taylor} et~al.}{2020}]{2020MNRAS.493.4342T}
{Taylor} J.,  {Parmentier} V.,  {Irwin} P. G.~J.,  {Aigrain} S.,  {Lee} E. K.~H.,   {Krissansen-Totton} J.,  2020, \mn@doi [\mnras] {10.1093/mnras/staa552}, \href {https://ui.adsabs.harvard.edu/abs/2020MNRAS.493.4342T} {493, 4342}

\bibitem[\protect\citeauthoryear{{Thorngren}, {Gao}  \& {Fortney}}{{Thorngren} et~al.}{2019}]{2019ApJ...884L...6T}
{Thorngren} D.,  {Gao} P.,   {Fortney} J.~J.,  2019, \mn@doi [\apjl] {10.3847/2041-8213/ab43d0}, \href {https://ui.adsabs.harvard.edu/abs/2019ApJ...884L...6T} {884, L6}

\bibitem[\protect\citeauthoryear{{Tinetti} et~al.,}{{Tinetti} et~al.}{2017}]{2017EPSC...11..713T}
{Tinetti} G.,  et~al., 2017, in European Planetary Science Congress. pp EPSC2017--713

\bibitem[\protect\citeauthoryear{{Venot} et~al.,}{{Venot} et~al.}{2020}]{2020ApJ...890..176V}
{Venot} O.,  et~al., 2020, \mn@doi [\apj] {10.3847/1538-4357/ab6a94}, \href {https://ui.adsabs.harvard.edu/abs/2020ApJ...890..176V} {890, 176}

\bibitem[\protect\citeauthoryear{{Visscher}, {Lodders}  \& {Fegley}}{{Visscher} et~al.}{2006}]{2006ApJ...648.1181V}
{Visscher} C.,  {Lodders} K.,   {Fegley} Bruce J.,  2006, \mn@doi [\apj] {10.1086/506245}, \href {https://ui.adsabs.harvard.edu/abs/2006ApJ...648.1181V} {648, 1181}

\bibitem[\protect\citeauthoryear{{Visscher}, {Lodders}  \& {Fegley}}{{Visscher} et~al.}{2010}]{2010ApJ...716.1060V}
{Visscher} C.,  {Lodders} K.,   {Fegley} Bruce J.,  2010, \mn@doi [\apj] {10.1088/0004-637X/716/2/1060}, \href {https://ui.adsabs.harvard.edu/abs/2010ApJ...716.1060V} {716, 1060}

\bibitem[\protect\citeauthoryear{{Wakeford}, {Visscher}, {Lewis}, {Kataria}, {Marley}, {Fortney}  \& {Mandell}}{{Wakeford} et~al.}{2017}]{2017MNRAS.464.4247W}
{Wakeford} H.~R.,  {Visscher} C.,  {Lewis} N.~K.,  {Kataria} T.,  {Marley} M.~S.,  {Fortney} J.~J.,   {Mandell} A.~M.,  2017, \mn@doi [\mnras] {10.1093/mnras/stw2639}, \href {https://ui.adsabs.harvard.edu/abs/2017MNRAS.464.4247W} {464, 4247}

\bibitem[\protect\citeauthoryear{{Wang} \& {Wordsworth}}{{Wang} \& {Wordsworth}}{2020}]{2020ApJ...891....7W}
{Wang} H.,  {Wordsworth} R.,  2020, \mn@doi [\apj] {10.3847/1538-4357/ab6dcc}, \href {https://ui.adsabs.harvard.edu/abs/2020ApJ...891....7W} {891, 7}

\bibitem[\protect\citeauthoryear{{Wardenier}, {Parmentier}, {Lee}, {Line}  \& {Gharib-Nezhad}}{{Wardenier} et~al.}{2021}]{2021MNRAS.506.1258W}
{Wardenier} J.~P.,  {Parmentier} V.,  {Lee} E. K.~H.,  {Line} M.~R.,   {Gharib-Nezhad} E.,  2021, \mn@doi [\mnras] {10.1093/mnras/stab1797}, \href {https://ui.adsabs.harvard.edu/abs/2021MNRAS.506.1258W} {506, 1258}

\bibitem[\protect\citeauthoryear{{Wardenier}, {Parmentier}  \& {Lee}}{{Wardenier} et~al.}{2022}]{2022MNRAS.510..620W}
{Wardenier} J.~P.,  {Parmentier} V.,   {Lee} E. K.~H.,  2022, \mn@doi [\mnras] {10.1093/mnras/stab3432}, \href {https://ui.adsabs.harvard.edu/abs/2022MNRAS.510..620W} {510, 620}

\bibitem[\protect\citeauthoryear{{Welbanks}, {Madhusudhan}, {Allard}, {Hubeny}, {Spiegelman}  \& {Leininger}}{{Welbanks} et~al.}{2019}]{2019ApJ...887L..20W}
{Welbanks} L.,  {Madhusudhan} N.,  {Allard} N.~F.,  {Hubeny} I.,  {Spiegelman} F.,   {Leininger} T.,  2019, \mn@doi [\apjl] {10.3847/2041-8213/ab5a89}, \href {https://ui.adsabs.harvard.edu/abs/2019ApJ...887L..20W} {887, L20}

\bibitem[\protect\citeauthoryear{{Zhang} \& {Showman}}{{Zhang} \& {Showman}}{2017}]{2017ApJ...836...73Z}
{Zhang} X.,  {Showman} A.~P.,  2017, \mn@doi [\apj] {10.3847/1538-4357/836/1/73}, \href {https://ui.adsabs.harvard.edu/abs/2017ApJ...836...73Z} {836, 73}

\bibitem[\protect\citeauthoryear{{Zhang}, {Kempton}  \& {Rauscher}}{{Zhang} et~al.}{2017}]{2017ApJ...851...84Z}
{Zhang} J.,  {Kempton} E. M.~R.,   {Rauscher} E.,  2017, \mn@doi [\apj] {10.3847/1538-4357/aa9891}, \href {https://ui.adsabs.harvard.edu/abs/2017ApJ...851...84Z} {851, 84}

\bibitem[\protect\citeauthoryear{{Zhang} et~al.,}{{Zhang} et~al.}{2018}]{2018AJ....155...83Z}
{Zhang} M.,  et~al., 2018, \mn@doi [\aj] {10.3847/1538-3881/aaa458}, \href {https://ui.adsabs.harvard.edu/abs/2018AJ....155...83Z} {155, 83}

\bibitem[\protect\citeauthoryear{{van Sluijs} et~al.,}{{van Sluijs} et~al.}{2023}]{2023MNRAS.522.2145V}
{van Sluijs} L.,  et~al., 2023, \mn@doi [\mnras] {10.1093/mnras/stad1103}, \href {https://ui.adsabs.harvard.edu/abs/2023MNRAS.522.2145V} {522, 2145}

\makeatother
\end{thebibliography}




\appendix

\section{Zonal-Mean Zonal Wind Speed at Equator}
\label{sec:zmws_appendix}
\begin{figure*}
	\includegraphics[width=2\columnwidth]{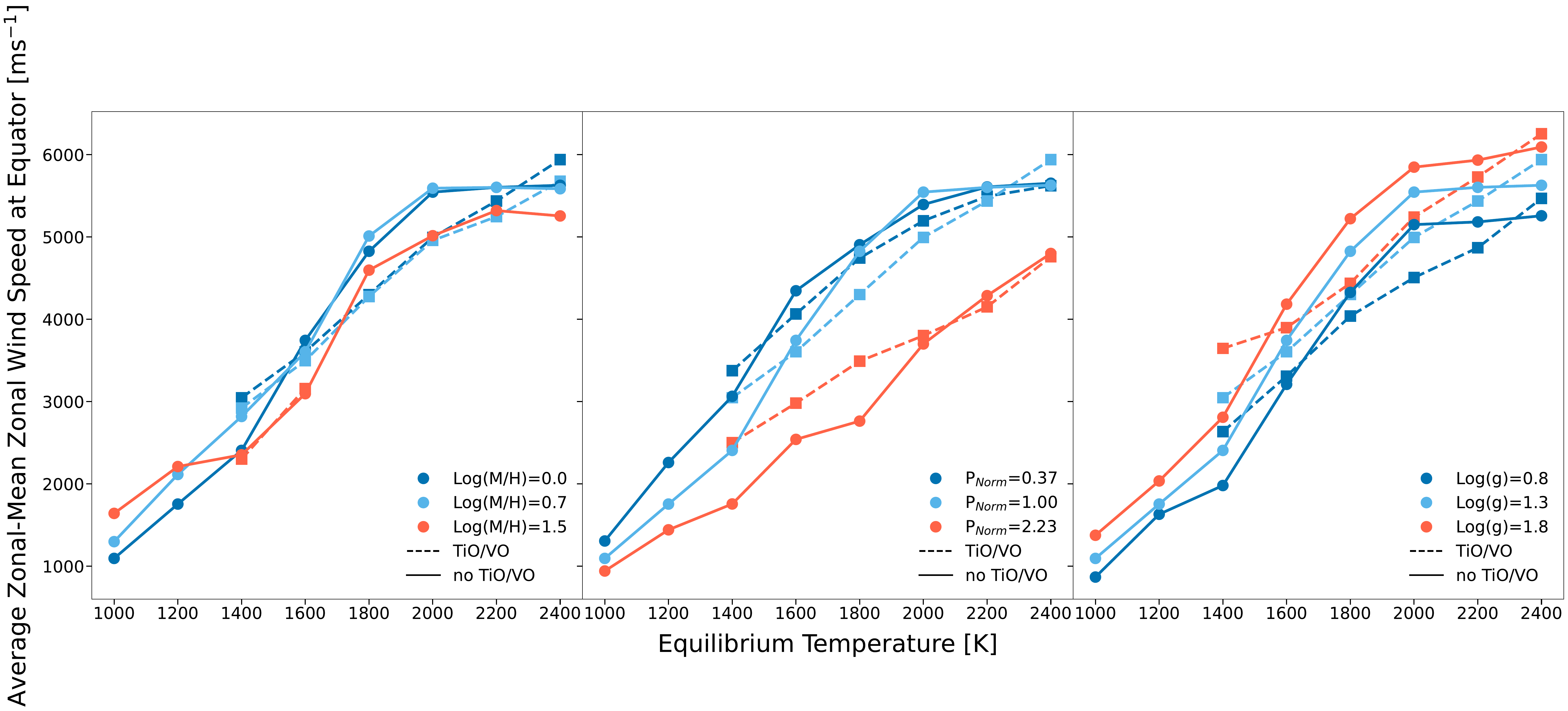}
    \caption{Zonal-mean zonal wind speed at the equator averaged over pressure for a subset of models within the grid. Each panel displays the variation with $T_{\rm eq}$ due to one of: rotation period, surface gravity and metallicity, whilst the other two parameters are kept constant (when constant Log(M/H)=0.0, P$_{\rm Norm}$=1, log(g)=1.3). Cross-terms between parameters are not included. Models excluding and including TiO/VO are displayed with solid and dashed lines respectively. \textbf{Left Panel:} Variation in zonal-mean zonal wind speed at the equator whilst changing metallicity. \textbf{Middle Panel:} Variation in zonal-mean zonal wind speed at the equator with different normalised orbital period. These tracks corresponds to planets tidally-locked around different stellar types. \textbf{Right Panel:} Variation in zonal-mean zonal wind speed at the equator whilst changing surface gravity.}
    \label{fig:zmzw_comp}
\end{figure*}

Figure \ref{fig:zmzw_comp} displays the zonal-mean zonal wind speed at the equator averaged over all pressure levels for different grid parameters.

\section{Amplitude Tracks}
\label{sec:amp_appendix}
\begin{figure*}
	\includegraphics[width=2\columnwidth]{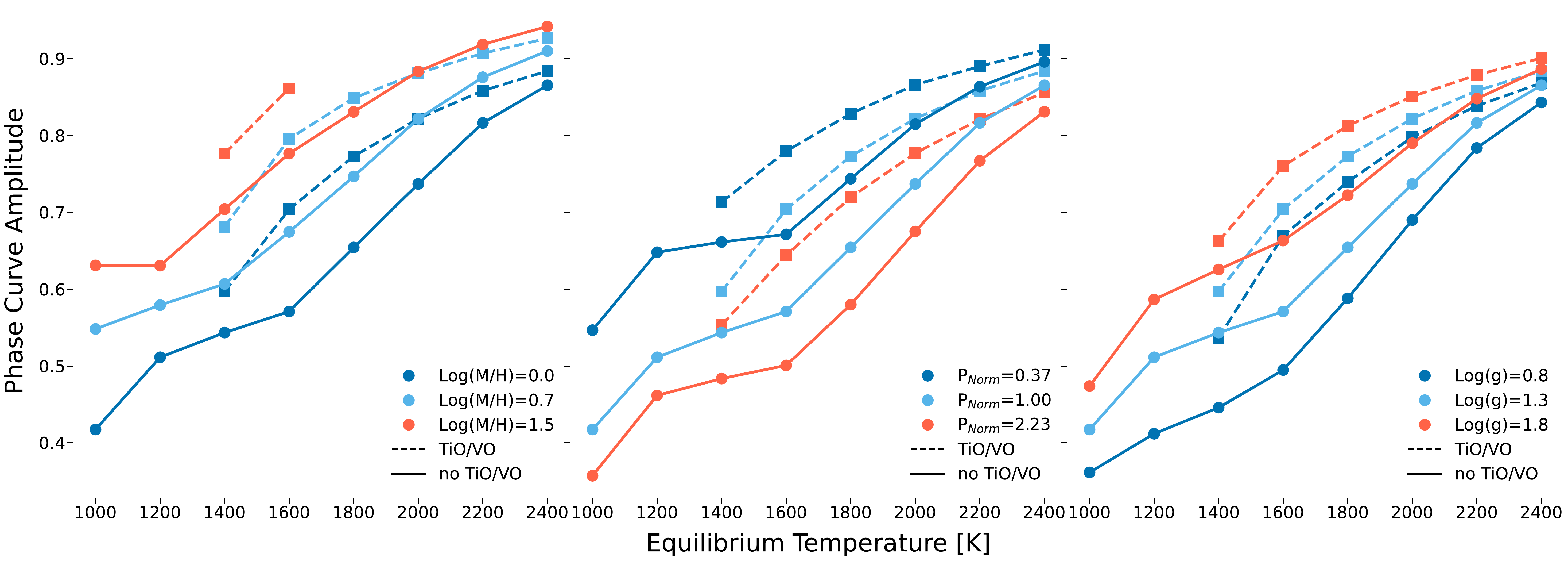}
    \caption{Phase curve amplitude for a subset of models within the grid. Each panel displays the variation in offset with $T_{\rm eq}$ due to one of: rotation period, surface gravity and metallicity, whilst the other two parameters are kept constant (when constant Log(M/H)=0.0, P$_{\rm Norm}$=1, log(g)=1.3). Cross-terms between parameters are not included. Models excluding and including TiO/VO are displayed with solid and dashed lines respectively. \textbf{Left Panel:} Variation in amplitude whilst changing metallicity. \textbf{Middle Panel:} Variation in amplitude with different normalised orbital period. These tracks corresponds to planets tidally-locked around different stellar types. \textbf{Right Panel:} Variation in amplitude whilst changing surface gravity.}
    \label{fig:amp_bol_comp}
\end{figure*}

Figure \ref{fig:amp_bol_comp} displays the phase curve amplitude tracks at different grid parameters, in the same way as Figures \ref{fig:f_bol_tracks} and \ref{fig:offset_bol_comp} do for the redistribution factor and offset respectively. The amplitude follows the same trends as the heat redistribution - increasing with temperature, metallicity and surface gravity, but decreasing with orbital period.

\section{Spitzer 3.6 micron Phase Curve Data Comparison}
\label{sec:spitzer1_offsetandamp}

\begin{figure*}
	\includegraphics[width=2\columnwidth]{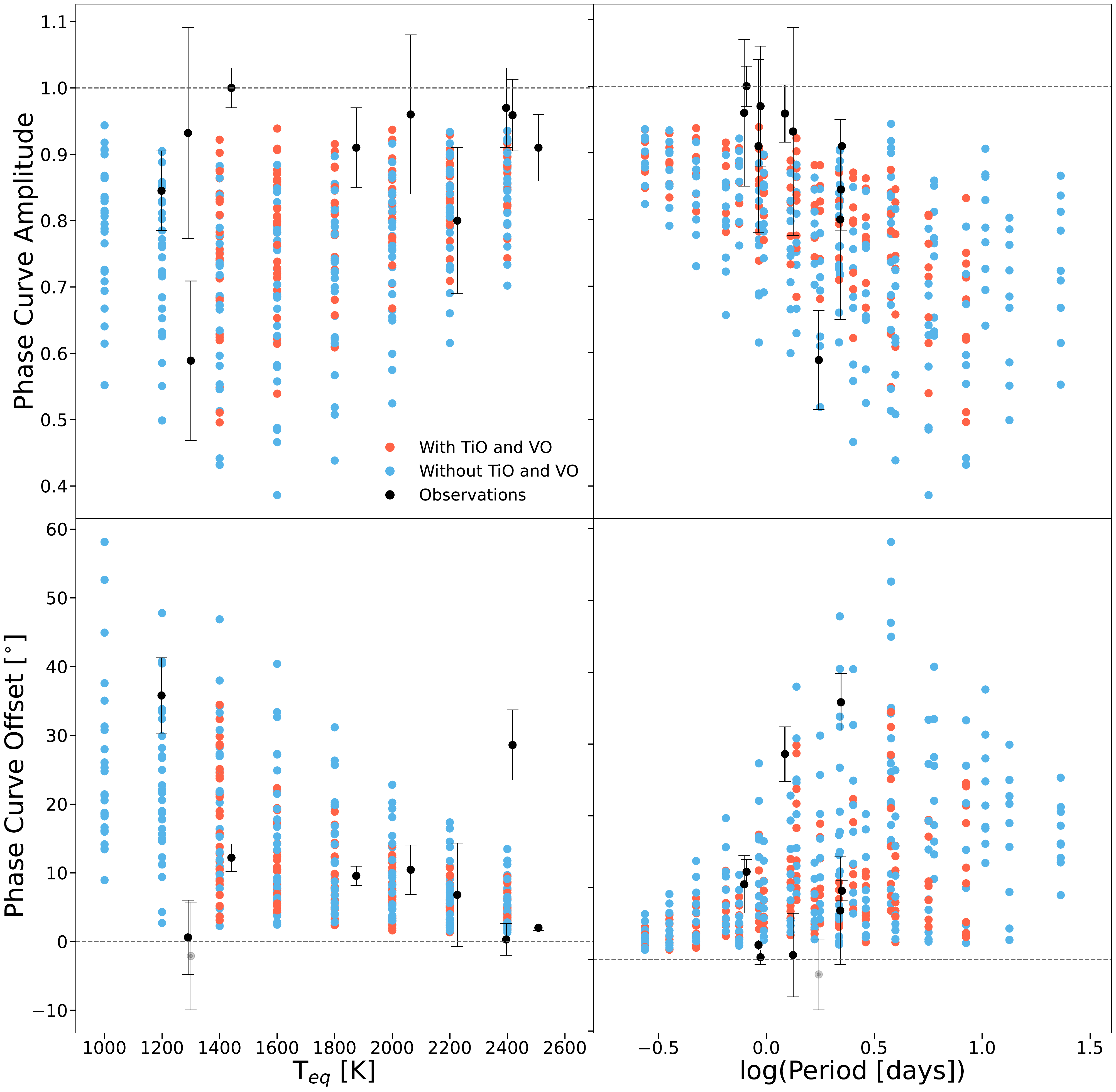}
    \caption{This figure shows a comparison between the phase curve offset and amplitudes in the Spitzer/IRAC $3.6\mu m$ band calculated for the 3D GCM grid and the observational data. The blue points are models without TiO/VO, the red points are the models with TiO/VO and the black points with error-bars are the observed data. \textbf{Left Panels:} shows the offset (lower) and amplitude (upper) vs equilibrium temperature. \textbf{Right Panels:} shows the offset (lower) and amplitude (upper) vs orbital period.}
    \label{fig:pccomp_3.6}
\end{figure*}

Figure \ref{fig:pccomp_3.6} displays a comparison between the model grid and phase curve offsets and amplitudes observed in the Spitzer/IRAC $3.6\mu m$ band-pass. The results here follow much the same trends as the $4.5\mu m$ data found in Section \ref{sec:spitzer2_offsetandamp}, with the trends and scatter in the observational data well produced by the model grid. We also find the noticeable turning point in the phase curve offset for the model grid, but lack the observational data for planets with long enough periods to observe this in the data. In the $3.6 \mu m$ band-pass we also observe a turning point in the amplitude with both increasing equilibrium temperature and period.

\section{Geometric \& Bond Albedo}
\label{sec:albedo}
\begin{figure*}
\includegraphics[width=2\columnwidth]{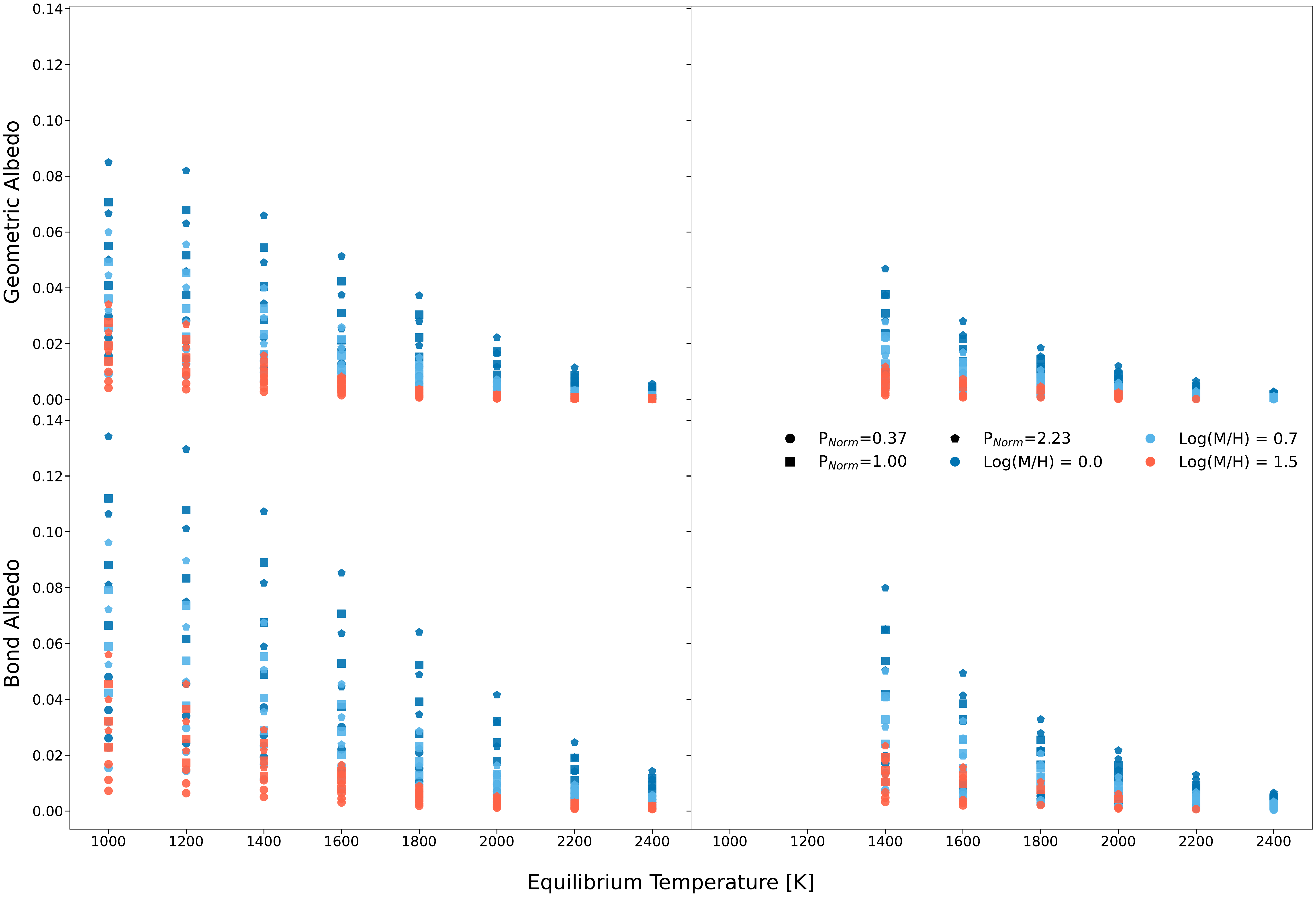}
\caption{Calculated values for the bolometric geometric (\textbf{Top Panels}) and bond (\textbf{Bottom Panels}) albedos from the model grid plotted against the equilibrium temperature. \textbf{Left Panels:} Show models not containing TiO/VO. \textbf{Right Panels:} Show models containing TiO/VO. The shape of each point depicts the orbital period, and the color depicts the metallicity of the model.}
\label{fig:albedo}
\end{figure*}

The bond albedo, $A_{b}$ of a planetary body is a measure of the fraction of incident radiation that is reflected back into space without being absorbed over all wavelengths and viewing angles, because of this it is crucial to working out planetary energy balance. The spherical albedo is this same measurement, but for a singular wavelength. The geometric albedo, related to the spherical albedo by $A_{s}=qA_{g}$ where q is the phase integral, is the albedo measured when the light source is between the observer and the planet. Figure \ref{fig:albedo} displays the bolometric geometric and bond albedos for all the grid models. All models within the grid have a geometric albedo of $\leq0.1$ and a bond albedo of $\leq0.15$. The albedo also decreases at higher equilibrium temperatures. This distribution confirms our expectation that cloudless hot Jupiter models have low albedos.

\section{Model Convergence}
\label{sec:convergence}
\begin{figure}
	\includegraphics[width=\columnwidth]{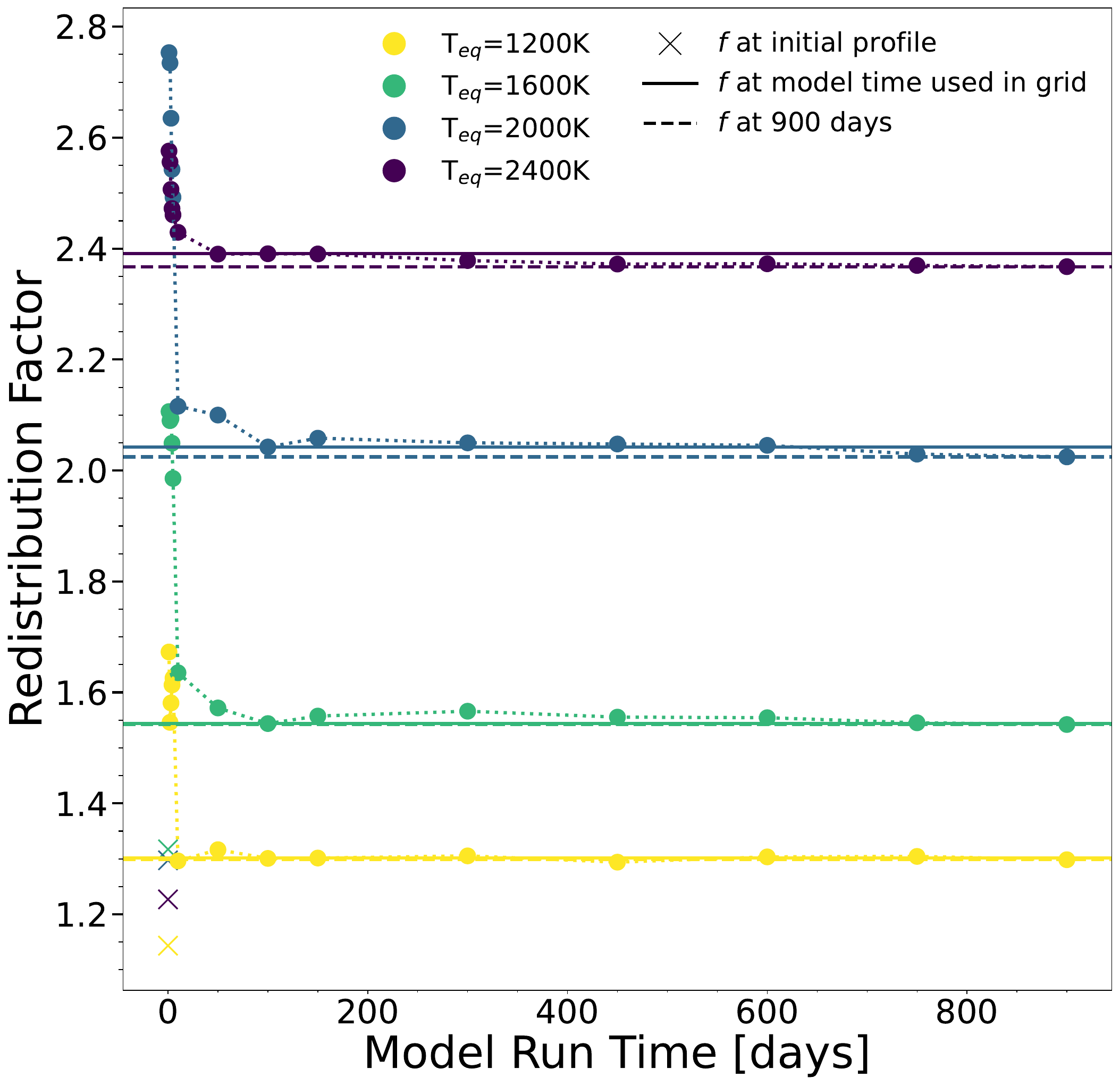}
    \caption{Redistribution factor as a function of model run time for a range of equilibrium temperatures in the model grid. The dashed lines show the value after 900 days and the solid lines the value at the integration time within the grid (150 days for models with equilibrium temperatures above 1400K and 300 days for models below). The points marked with a cross show the redistribution factor of the initial profile.}
    \label{fig:conv}
\end{figure}

As mentioned in Section \ref{sec:methods_dynamics}, as our models only evolve for between 150-300 days (depending on the equilibrium temperature) they are not in a fully converged state. The primary causes for incomplete model convergence can be categorised as systematic flux lost due to truncation of the stellar spectrum or parameter dependant flux transformed into kinetic energy to drive winds in the model. To fully converge the deep atmosphere would take potentially tens of thousands of CPU hours (see Figure 3 in \citet{2020MNRAS.491.1456M}), making the costs and time for running the grid completely unreasonable and seemingly unnecessary. However, as the radiative timescale at the photosphere is much smaller than that of the deep atmosphere, the photospheric level will converge substantially faster. For this study, as we are primarily interested in the trends due to planetary parameters rather than specific atmospheric characterisation, it is important that these quantities have reaches a steady state within our model run time rather than 'full convergence' being achieved. To test this, we allow models at four equilibrium temperatures to evolve for 900 days. We then sample the redistribution factor (see Section \ref{sec:redist}) at a number of different time steps to check if it has achieved steady state. This can be seen in Figure \ref{fig:conv}. From this we find that the redistribution factor has negligible change after the 150 day mark, suggesting that we have in fact reached a pseudo steady-state at the photospheric pressure level in these models.


\bsp	
\label{lastpage}
\end{document}